\documentclass[twocolumn,aps,pra,longbibliography,superscriptaddress,nofootinbib,10pt]{revtex4-1}
\usepackage[latin1]{inputenc}
\usepackage{graphicx,dcolumn,bm}
\usepackage{subfigure}
\usepackage{multirow}
\usepackage{array}
\usepackage{arydshln}
\usepackage{color}
\usepackage[colorlinks,linkcolor=blue,citecolor=blue,hyperindex]{hyperref}
\usepackage{amsfonts}
\usepackage{amssymb}
\usepackage{amsmath}
\usepackage{latexsym}
\usepackage{simplewick}
\usepackage{epstopdf}
\usepackage{multirow}
\usepackage{float}
\usepackage[table]{xcolor}
\usepackage{multibib}
\usepackage{mathrsfs}
\usepackage[sans]{dsfont}
\usepackage[mathscr]{eucal}

\usepackage[normalem]{ulem}
\usepackage{epsfig}
\definecolor{Blue}{rgb}{0.0,0.0,1}
\definecolor{Red}{rgb}{1,0.0,0.0}
\definecolor{Green}{rgb}{0,0.5,0.0}
\usepackage{tikz}
\usetikzlibrary{decorations.pathmorphing}
\usetikzlibrary{arrows}
\usetikzlibrary{intersections,shapes.arrows}
\usetikzlibrary{calc}
\usetikzlibrary{quotes,angles}
\usepackage{nicefrac}
\usepackage{pgfplots}
\usepgfplotslibrary{fillbetween}
\pgfplotsset{compat=1.13,colormap={violetnew}{rgb=(0.293416, 0.0574044, 0.529412) rgb=(0.394818,0.233715,0.671945) rgb =(0.49622,0.410025,0.814477) rgb=(0.588672,0.567494,0.910066) rgb=(0.663226,0.687282,0.911765) rgb=(0.73778,0.807069,0.913465) rgb=(0.807267,0.861883,0.894034) rgb=(0.874222,0.884211,0.864039) rgb=(0.941176, 0.906538, 0.834043)}}
\usepgfplotslibrary{groupplots} 
\usepgfplotslibrary[groupplots] 
\usetikzlibrary{pgfplots.groupplots} 
\usetikzlibrary[pgfplots.groupplots] 
\usepgfplotslibrary{statistics}
\usepackage{pgfplotstable}
\tikzset{jumpdot/.style={mark=*,solid},excl/.append style={jumpdot,fill=white},incl/.append style={jumpdot,fill=black}}
\begin{document}

\title{Mixedness timescale in non-Hermitian quantum systems}
\author{Diego Paiva Pires}
\affiliation{Departamento de F\'{i}sica, Universidade Federal do Maranh\~{a}o, Campus Universit\'{a}rio do Bacanga, 65080-805, S\~{a}o Lu\'{i}s, Maranh\~{a}o, Brazil}
\author{Tommaso Macr\`{i}}
\affiliation{Departamento de F\'{i}sica Te\'{o}rica e Experimental, Universidade Federal do Rio Grande do Norte, 59072-970 Natal, Rio Grande do Norte, Brazil}
\affiliation{ITAMP, Harvard-Smithsonian Center for Astrophysics, Cambridge, Massachusetts 02138, USA}

\begin{abstract}
We discuss the short-time perturbative expansion of the linear entropy for finite-dimensional quantum systems whose dynamics can be effectively described by a non-Hermitian Hamiltonian. We derive a timescale for the degree of mixedness for an input state undergoing non-Hermitian dynamics and specialize these results in the case of a driven-dissipative two-level system. Next, we derive a timescale for the growth of mixedness for bipartite quantum systems that depends on the effective non-Hermitian Hamiltonian. In the Hermitian limit, this result recovers the perturbative expansion for coherence loss in Hermitian systems, while it provides an entanglement timescale for initial pure and uncorrelated states. To illustrate these findings, we consider the many-body transverse-field $XY$ Hamiltonian coupled to an imaginary all-to-all Ising model. We find that the non-Hermitian Hamiltonian enhances the short-time dynamics of the linear entropy for the considered input states. Overall, each timescale depends on minimal ingredients such as the probe state and the non-Hermitian Hamiltonian of the system, and its evaluation requires low computational cost. Our results find applications to non-Hermitian quantum sensing, quantum thermodynamics of non-Hermitian systems, and $\mathcal{PT}$-symmetric quantum field theory.
\end{abstract}

\maketitle


\section{Introduction}
\label{sec:00000000000}

The study of non-Hermitian systems~\cite{PhysRevLett.109.230405,arXiv:2201.05367} has paved the way for recent developments across the subjects of quantum sensing~\cite{PhysRevLett.125.240506,PhysRevLett.127.026404}, $\mathcal{PT}$ symmetry and exceptional points~\cite{Zyablovsky_2016,PhysRevLett.119.190401,PhysRevLett.127.186602,Yuto_Ashida2020,Bergholtz2021,PhysRevResearch.2.033022,PhysRevResearch.3.013015,PhysRevA.106.023309}, linear response theory~\cite{PhysRevLett.128.016802}, quantum many-body systems~\cite{Korff_2007,Korff_2008,Castro_Alvaredo_2009,NatCommun_8_15791_2017,Xiao2019,PhysRevB.101.121109,PhysRevLett.123.090603,PhysRevLett.125.260601,ProgTheorExpPhys_12_2020,arXiv:2201.09895,arXiv:2210.11937}, skin effect~\cite{PhysRevB.102.205118,arXiv:2201.10318,PhysRevResearch.4.033041}, bulk-edge correspondence~\cite{PhysRevLett.121.086803,PhysRevLett.125.226402}, phase transitions~\cite{PhysRevB.103.155417,PhysRevA.105.022219}, and the quantum boomerang effect in localized systems~\cite{arXiv:2206.02922,PhysRevX.12.011035,PhysRevA.99.023629,PhysRevB.106.L060301}, to cite a few. 

Recent theoretical achievements discussed the effects of postselection on the dynamics of open quantum systems, thus reconciling the approaches of effective non-Hermitian Hamiltonians and Liouvillian superoperators~\cite{PhysRevA.100.062131,PhysRevA.101.062112}. In this setting, several works have addressed the role of non-Hermitian features on legitimate quantum mechanical signatures, e.g., quantum coherence and entanglement~\cite{PhysRevB.104.155141,PhysRevA.104.052405,PhysRevLett.126.170503,PhysRevB.105.L121115,PhysRevResearch.4.033022}. For example, it has been shown that quantum coherence can be characterized under the framework of multiple quantum coherences~\cite{PhysRevB.105.075103,PhysRevA.99.052354,PhysRevA.102.012429}. In turn, the dynamics of entanglement in non-Hermitian systems has been widely probed with entropic measures, but their evaluation generally involves the full spectral decomposition of the state driven by the non-Hermitian Hamiltonian~\cite{JStatMech_033102_2016,QIP_10_234_2017,arXiv:2105.09793,PhysRevB.105.075126}. This can be a challenging computational task, especially for interacting quantum many-body systems.

To overcome this issue, the onset growth of mixedness at earlier times of the dynamics can be addressed through the so-called linear entropy, a useful information-theoretic quantifier that is related to the second-order R\'{e}nyi entropy and quantum purity~\cite{PhysRevX.7.031011,NJP_20_053058_2018}. Remarkably, those quantities have been experimentally probed in optical lattices~\cite{20159_arxiv_1509.01160,PhysRevLett.120.050406,Science_353_6301_794}, and trapped ion setups~\cite{PhysRevA.98.052334,Brydges260}. In the Hermitian case, it is known that the short-time perturbative expansion of the linear entropy implies a universal timescale for the entanglement dynamics of inte\-rac\-ting bipartite systems with initial pure state~\cite{PhysRevD.97.066008,PhysRevA.97.022317,PhysRevA.71.042321}. It is worth mentioning that this timescale is inversely proportional to the fluctuations of the coupling between subsystems~\cite{NEMES1986367,RevModPhys.75.715}. Importantly, this result also assigns a timescale for the decoherence mechanism in subsystems of such composite quantum systems~\cite{PhysRevLett.77.207,PhysRevLett.70.1187,JPhysChemLett_17_8_2017}. We also mention the study of the growth of entanglement through a perturbative expansion of the entanglement negativity~\cite{PhysRevA.99.012322,arXiv:2209.03976}, and also quantum fidelities~\cite{PhysRevA.56.4466,GORIN200633}. To the best of our knowledge, despite these remarkable achievements in the Hermitian setting, deriving an analogous timescale for non-Hermitian systems remains a gap to be filled.

Here we address timescales for the growth of the li\-near entropy for finite-dimensional quantum systems described by effective non-Hermitian Hamiltonians. The physical system is initialized in a quantum state which can be chosen as either a pure or mixed one, possibly an entangled state or even an uncorrelated one. We investigate the short-time perturbative expansion of the linear entropy for a given input state driven by a general non-Hermitian Hamiltonian. In this setting, up to the se\-cond order in time, the onset growth of mixedness of the evolved state is governed by two competing timescales that are intrinsically related to the anti-Hermitian part of the non-Hermitian Hamiltonian. In particular, we specialize these results in the case of a driven non-Hermitian two-level system and discuss the mixedness of a single-qubit state. 

Next, focusing on the reduced dynamics of bipartite systems described by non-Hermitian Hamiltonians, we derive the short-time perturbative expansion of the linear entropy for a given evolved marginal state of the composite system. In the Hermitian limit, these results recover the perturbative expansion for coherence loss in Hermitian systems~\cite{PhysRevLett.77.207}. In particular, for initial pure and uncorrelated states, we find the lowest order entanglement timescale for quantum systems described by Hermitian Hamiltonians addressed in Refs.~\cite{PhysRevD.97.066008,PhysRevA.97.022317}. To illustrate these findings, we consider a paradigmatic many-body non-Hermitian Hamiltonian, and present analytical calculations and numerical simulations to support our theo\-re\-ti\-cal predictions. We verify that, unlike the Hermitian case, the non-Hermitian Hamiltonian is responsible for an enhancement in the short-time dynamics of the linear entropy for the multiparticle states that have been considered.

Overall, our results can be of relevance to both the communities of photonics~\cite{Zyablovsky_2016,Nat_Phot_11_752_2017}, and also atomic, molecular and optical physics~\cite{PhysRevX.8.031079,Bergholtz2021}. Our findings might be useful, for example, to understand the interplay of quantum speed limits and the dynamics generated by non-Hermitian Hamiltonians, the latter providing an effective description for the dynamics of open quantum systems whose dynamics satisfies a Lindblad type master equation~\cite{PhysRevLett.110.050403,PhysRevA.97.022109}. In this context, the aforementioned timescales can find applications in the study of quantum state transfer protocols in dissipative two-level systems, particularly with regard to the search of non-Hermitian Hamiltonians related to optimal speed limits~\cite{PhysRevA.104.052620}. In addition, non-Hermitian mixedness timescales can be of interest in the study of topology and localization signatures in many-body systems, for example, in the investigation of the non-Hermitian skin effect by means of inverse participation ratio~\cite{PhysRevB.102.205118,arXiv:2201.10318,PhysRevResearch.4.033041}. Furthermore, it may also be useful in the study of thermalization dynamics in non-Hermitian many-body systems~\cite{PhysRevLett.128.213901,arXiv:2202.08638}.

The paper is organized as follows. In Sec.~\ref{sec:00000000001}, we briefly review useful properties regarding the linear entropy. In Sec.~\ref{sec:00000000002} we investigate the short-time perturbative expansion of the linear entropy for finite-dimensional quantum systems whose dynamics are driven by a non-Hermitian Hamiltonian. In Sec.~\ref{sec:00000000002A}, we illustrate our findings by means of the two-level system. In Sec.~\ref{sec:00000000003} we derive a mixedness timescale for bipartite quantum systems evolving under the action of a given non-Hermitian Hamiltonian. In Sec.~\ref{sec:00000000003A}, we specialize these results to the case of two initially uncorrelated subsystems. In addition, Sec.~\ref{sec:00000000003B} addresses the case of a many-body system with non-Hermitian Hamiltonian describing the transverse-field $XY$ model perturbed by a fully connected Ising Hamiltonian with imaginary exchange coupling. Finally, in Sec.~\ref{sec:00000000004} we summarize our conclusions.


\section{Linear entropy}
\label{sec:00000000001}

In this section, we review the main properties of li\-near entropy, i.e., a versatile information-theoretic measure that quantifies the degree of mixedness of a given state~\cite{Schlosshauer_book}. Linear entropy has been used to witness multipartite entanglement~\cite{PhysRep474_1_2009,RevModPhys.81.865,PhysRevA.68.022318,JOptB_6_S730_2004,PhysRevA.75.032301}. We point out that Ref.~\cite{PhysRevLett.93.110501} addresses a separability criterion for multipartite entangled states of bosons that is based on purity, thus being related to the linear entropy. In addition, Refs.~\cite{PhysRevA.54.1838,doi:10.1063/1.1498490} discuss entanglement criteria for bipartite mixed states based on the so-called conditional Tsallis entropy. In this regard, by setting the second-order Tsallis entropy, one obtains an entanglement criterion for linear entropy. In addition, similar criteria based on conditional R\'{e}nyi entropy have been addressed in Refs.~\cite{HORODECKI1996377,PhysRevLett.79.5194,PhysRevA.60.3461,VIDIELLABARRANCO1999335,ABE2001157,PhysRevA.63.042104,TERHAL2002313}. Let us consider a quantum system with finite-dimensional Hilbert space $\mathcal{H}$, with $d = \dim\mathcal{H}$. The space of quantum states $\mathcal{S} \subset \mathcal{H}$ is a convex set of Hermitian, positive semidefinite, trace-one, $d\times d$ matrices, i.e., $\mathcal{S} = \{\rho \in \mathcal{H} \mid {\rho^{\dagger}} = \rho,~\rho\geq 0,~\text{Tr}(\rho) = 1\}$. The normalized linear entropy of the quantum state $\rho$ is defined as~\cite{PhysRevA.70.052309}
\begin{equation}
\label{eq:00000000001}
{S_L}(\rho) := \frac{d}{d - 1}[1 - f(\rho)] ~,
\end{equation}
where $f(\rho) =  \text{Tr}({\rho^2})$ stands for the quantum purity. The latter quantity is bounded as $1/d \leq  f(\rho) \leq 1$, which implies that the linear entropy ranges as $0 \leq {S_L}(\rho) \leq 1$ for all $\rho \in \mathcal{S}$. In addition, given the spectral decomposition $\rho = {\sum_j}\, {p_j}|j\rangle \langle{j}|$ in terms of the basis of states ${\{|j\rangle\}_{j = 1,\ldots, d}}$, with $0 \leq {p_j} \leq 1$ and ${\sum_j}\, {p_j} = 1$, one readily concludes that ${S_L}(\rho) = [d/(d - 1)][1 - {\sum_j}\, {p_j^2}]$. Importantly, it has been shown that quantum states with linear entropy satisfying the lower bound ${S_L}(\rho) \geq d(d - 2)/{(d - 1)^2}$ are separable~\cite{PhysRevA.58.883}.

Linear entropy remains invariant under unitary transformations over the input state, i.e., ${S_L}(V\rho{V^{\dagger}}) = {S_L}(\rho)$, with $V{V^{\dagger}} = {V^{\dagger}}V = \mathbb{I}$ and for all $\rho \in \mathcal{S}$. It is related to the second-order R\'{e}nyi entropy ${S_2}(\rho)$~\cite{Renyi1961,10.1063.1.4838856}, also known as collision entropy~\cite{PhysRevA.85.012108}, and thus becomes ${S_L}(\rho) = [d/(d - 1)][1 - e^{-{S_2}(\rho)}]$. Furthermore, Eq.~\eqref{eq:00000000001} is also written as ${S_L}(\rho) = [d/(d - 1)] {{\text{H}}_{2}}(\rho)$, with ${{\text{H}}_{2}}(\rho)$ being the second-order Tsallis entropy~\cite{Tsallis1988}. Interestingly, the linear entropy is also connected with the quantum Fisher information. To see this, let $\Lambda \in \mathcal{H}$ be a given Hermitian operator which generates the unitary evolution ${\rho_{\varphi}} := {\mathcal{U}_{\varphi}}(\rho) = {e^{- i \varphi \Lambda}} \rho \, {e^{ i \varphi \Lambda}}$ imprinting a phase shift $\varphi$ on the initial quantum state $\rho \in \mathcal{S}$ of a finite-dimensional quantum system. In this case, given the task of estimating the parameter $\varphi$ for such a unitary encoding protocol, it can be proved that the linear entropy satisfies the lower bound ${S_L}(\rho) \geq [2d/(d - 1)][{\langle{\Lambda^2}\rangle_{\rho}}  - {\langle{\Lambda}\rangle^2_{\rho}} - (1/4)F(\rho,\Lambda)]/{[{\lambda_{\text{max}}}(\Lambda) - {\lambda_{\text{min}}}(\Lambda)]^2}$, with $\langle \bullet \rangle_{\rho} = \text{Tr}(\bullet \rho)$, where $F(\rho, \Lambda) := F({\rho_{\varphi}})$ is the quantum Fisher information (QFI) respective to evolved state $\rho_{\varphi}$, while ${\lambda_{\text{max}}}(\Lambda)$ and ${\lambda_{\text{min}}}(\Lambda)$ are the largest and smallest eigenvalues of $\Lambda$, respectively~\cite{arXiv:1701.07461}. On the one hand, for a given initial mixed state $\rho = {\sum_j} {p_j}| j \rangle\langle{j}|$, the QFI is written as $F(\rho,\Lambda) = 2\, {\sum_{j,k}} \, [{({p_j} - {p_k})^2}/({p_j} + {p_k})] \, {|\langle{j}|\Lambda|{k}\rangle|^2}$~\cite{IntJQuantInf_125_7_2009,10.1116_1.5119961}. On the other hand, QFI reduces to the squared variance of the generator $\Lambda$ for an input pure state, i.e., one gets $F(\rho,\Lambda) = 4(\langle{\psi}|{\Lambda^2}|\psi\rangle - {\langle{\psi}|{\Lambda}|\psi\rangle^2})$ for $\rho = |\psi\rangle\langle\psi |$. In turn, the latter case implies that the aforementioned lower bound saturates to ${S_L}(|\psi\rangle\langle\psi |) = 0$, which is expected for any pure state. To the best of our knowledge, the latter bound holds for unitary evolutions governed by Hermitian operators. Hence, it should no longer apply to the case of nonunitary evolutions generated by effective non-Hermitian Hamiltonians.

Recently, linear entropy has also been discussed for systems described by non-Hermitian Hamiltonians~\cite{JStatMech_033102_2016,Entropy_18120451,QIP_10_234_2017}. In this regard, at least two approaches can be highlighted. The first one consists of equipping the Hilbert space with a generalized inner product structure, called metric operator~\cite{Bender2007,JPhysA_42_5_055303,PhysScr_82_038110}. In the second case, one introduces a renormalized density operator whose dynamics is governed by modified Heisenberg equations of motion~\cite{PhysRevLett.109.230405,PhysRevA.42.1467,doi:10.1142,EPJD_253_69_2015}. Throughout this paper, we will follow this last perspective, which in turn finds applications in a broader context, ranging from dissipative systems~\cite{arXiv:2201.05367} to topological phases in non-Hermitian systems~\cite{PhysRevX.8.031079,PhysRevLett.119.190401,Bergholtz2021}, also including recent studies in localization~\cite{arXiv:2206.02922} and cri\-ti\-ca\-lity in non-Hermitian many-body systems~\cite{NatCommun_8_15791_2017,Xiao2019,PhysRevB.101.121109,PhysRevLett.123.090603,PhysRevLett.125.260601,ProgTheorExpPhys_12_2020,arXiv:2201.09895,arXiv:2210.11937}.


\section{Mixedness timescale for non-Hermitian systems}
\label{sec:00000000002}

We consider a quantum system with finite-dimensional Hilbert space $\mathcal{H}$, with $d = \text{dim}\mathcal{H}$, which is initialized in the state ${\rho_0} \in \mathcal{S}$ [see Sec.~\ref{sec:00000000001}]. This input state undergoes a nonunitary evolution governed by the time-independent non-Hermitian Hamiltonian $H = {H_1} + i{H_2}$, where ${H_1} = ({1}/{2})(H + {H^{\dagger}})$, and ${H_2} = ({1}/{2i})(H - {H^{\dagger}})$. In turn, the noncommuting observables $H_1$ and $H_2$ stand for the Hermitian and anti-Hermitian parts of the Hamiltonian $H$, res\-pec\-tively. In the remainder of the paper, we set $\hbar = 1$. The dynamics of the normalized time-dependent density matrix ${\widetilde{\rho}_t} = {\rho_t}/{\text{Tr}({\rho_t})}$ fulfills the equation of motion~\cite{PhysRevA.42.1467,doi:10.1142,EPJD_253_69_2015}
\begin{equation}
\label{eq:00000000002}
\frac{d{\widetilde{\rho}_t}}{dt} = - {i} \, [{H_1}, {\widetilde{\rho}_t}] + \{ {H_2} , {\widetilde{\rho}_t}\}  - {2} \, \text{Tr}({\widetilde{\rho}_t}{H_2}){\widetilde{\rho}_t}  ~,
\end{equation}
which in turn represents a completely positive and trace-preserving operation. Physically, the nonunitary evolution described in Eq.~\eqref{eq:00000000002} maps a given physical state to another physical state. In this sense, given the input density matrix $\rho_0$, the time-dependent evolved state $\widetilde{\rho}_t$ must also be a density matrix, which is expected to be (i) Hermitian, ${\widetilde{\rho}_t^{\,\dagger}} = {\widetilde{\rho}_t}$ for all $t \geq 0$; (ii) positive semidefinite, ${\widetilde{\rho}_t} \geq 0$ for all $t \geq 0$; and (iii) normalized, $\text{Tr}({\widetilde{\rho}_t}) = 1$ for all $t \geq 0$, which in turn ensures the probability conservation. In this regard, note that the last term in the right-hand side of Eq.~\eqref{eq:00000000002} is related to the conservation of probability. In view of dissipative systems, for example, the dynamics generated by non-Hermitian Hamiltonians is related to the conditioning of postselection on measurement outcomes, and discarding of quantum jumps~\cite{PhysRevLett.109.230405,arXiv:2201.05367,PhysRevA.100.062131,PhysRevA.101.062112}. We point out that the dynamical map drives a nonunitary evolution under which the state ${\widetilde{\rho}_t}$ exhibits a time-dependent mixedness. In the Hermitian setting, the mixedness stands as a conserved quantity for any quantum state undergoing a unitary evolution generated by a Hermitian Hamiltonian.

Here, we choose the normalized linear entropy ${S_L}({\widetilde{\rho}_t}) = [d/(d - 1)][1 - f({\widetilde{\rho}_t})]$ as a useful quantum information-theoretic quantifier to probe the mixedness of state ${\widetilde{\rho}_t}$, with $f({\widetilde{\rho}_t}) = \text{Tr}({\widetilde{\rho}_t^{\, 2}})$. As discussed in Sec.~\ref{sec:00000000001}, for the case of unitary evolutions generated by Hermitian Hamiltonians, both the purity and the linear entropy remain invariant for all $t > 0$. However, for the nonunitary dynamics dictated by non-Hermitian Hamiltonians, the linear entropy becomes a time-dependent quantity, and its evaluation requires the full spectral decomposition of the evolved state. This task has a high computational cost for many-body quantum systems. 

In this section, we are interested in the short-time perturbative expansion of ${S_L}({\widetilde{\rho}_t})$ to understand the initial growth of the mixedness of the evolved state ${\widetilde{\rho}_t}$ [see Eq.~\eqref{eq:00000000002}]. The Taylor expansion of the linear entropy up to second order in $t$ around $t = 0$ yields
\begin{equation}
\label{eq:00000000003}
{S_L}({\widetilde{\rho}_t}) \approx {S_L}({\rho_0}) - \frac{d}{d - 1}\left(\frac{t}{T_1} + \frac{t^2}{T_2^2}\right) + O({t^3}) ~,
\end{equation}
where we define
\begin{equation}
\label{eq:00000000004}
{T_1^{-1}} := {4} \, {\text{cov}_{\rho_0}}({\rho_0},{H_2}) ~,
\end{equation}
and
\begin{align}
\label{eq:00000000005}
&{T_2^{-2}} :=  - {4} f({\rho_0})\, {\text{var}_{\rho_0}}({H_2}) - {8} \, {\langle{H_2}\rangle_{\rho_0}} \, {\text{cov}_{\rho_0}}({\rho_0},{H_2}) \nonumber\\
&+ {8} \, {\text{cov}_{\rho_0}}({H_2},{\rho_0}{H_2}) - {2i} \, {\text{cov}_{\rho_0}}({\rho_0},[{H_2},{H_1}]) ~,
\end{align}
with $\langle \bullet \rangle_{\rho_0} = \text{Tr}({\rho_0} \, \bullet)$ being the expectation value at time $t = 0$, while ${\text{cov}_A}(B,C) = ({1}/{2})\text{Tr}(A\{B,C\}) - \text{Tr}(AB)\text{Tr}(AC)$ defines the covariance functional. In particular, for $B = C$, note that the covariance reduces to the variance ${\text{var}_A}(B) \equiv {\text{cov}_A}(B,B) = \text{Tr}(A{B^2}) - {\text{Tr}(AB)^2}$. We point out that Eqs.~\eqref{eq:00000000004} and~\eqref{eq:00000000005} are related to the first-order and second-order derivatives of the quantum purity $f({\widetilde{\rho}_t})$ at the vicinity of $t = 0$, respectively, with ${[ {f^{(1)}}({\widetilde{\rho}_t}) ]_{t = 0}} = 1/{T_1}$ and $(1/2){[ {f^{(2)}}({\widetilde{\rho}_t}) ]_{t = 0}} = 1/{T_2^2}$, where the $n$th-order derivative of the quantum purity becomes
\begin{equation}
\label{eq:00000000007}
 {f^{(n)}}({\widetilde{\rho}_t}) = {\sum_{k = 0}^n} \, \frac{n!}{{(n - k)!} \, {k!}} \, \text{Tr}\left( \frac{{d^k}{\widetilde{\rho}_t}}{d{t^k}} \,\frac{{d^{n - k}} {\widetilde{\rho}_t}}{d{t^{n - k}}} \right) ~.
\end{equation}

\begin{figure*}[!th]
\begin{center}
\includegraphics[scale=0.9]{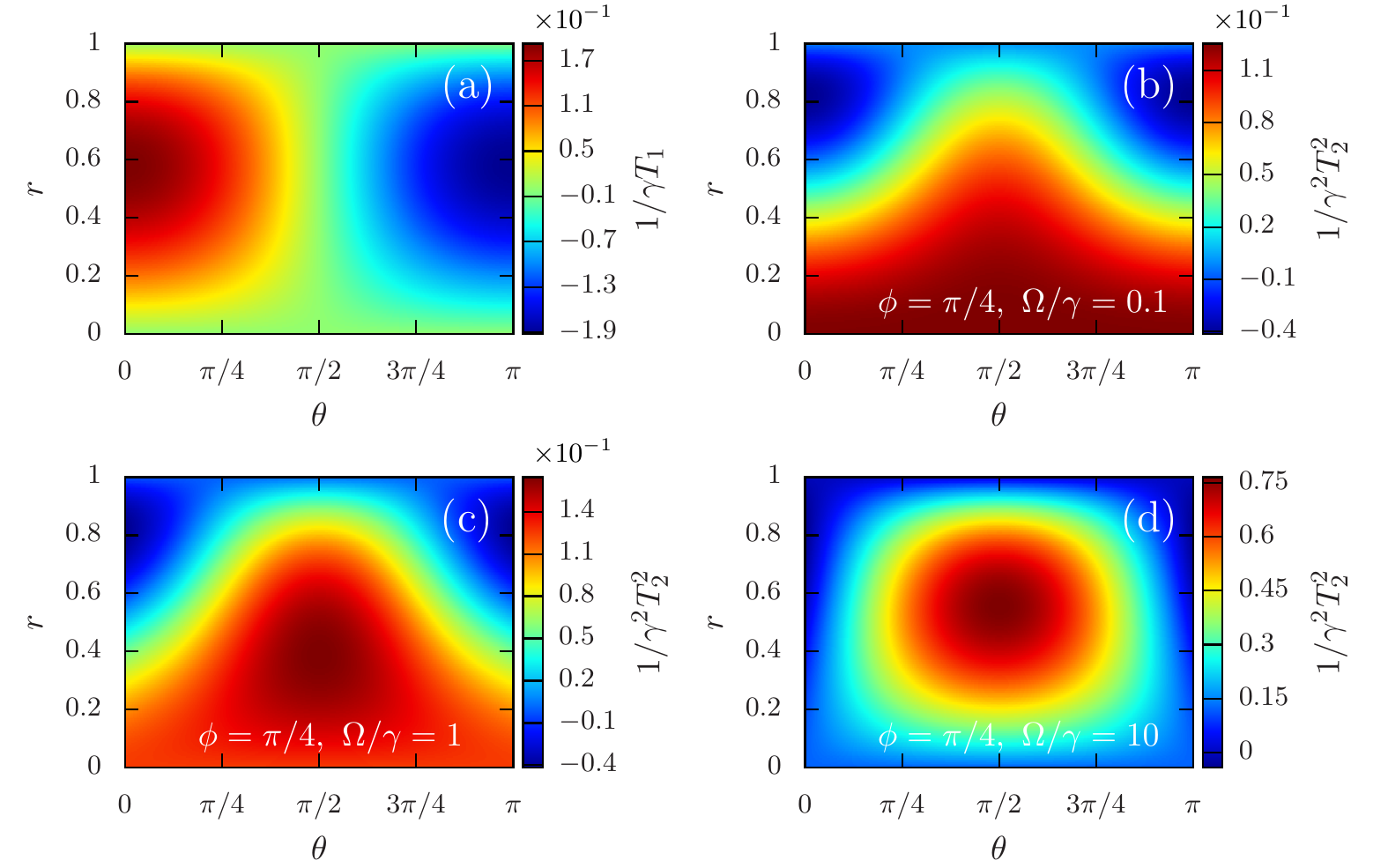}
\caption{(Color online) Density plot of the dimensionless quantities $1/\gamma{T_1}$ [see Eq.~\eqref{eq:00000000010}] and $1/{\gamma^2}{T_2^2}$ [see Eq.~\eqref{eq:00000000011}], as a function of the mixing parameter $r\in [0,1]$ and the azimuthal angle $\theta \in [0,\pi]$, respective to the perturbative expansion of the linear entropy for the driven dissipative two-level system around [see Eq.~\eqref{eq:00000000003}]. In panels (b)--(d), we set the polar angle $\phi = \pi/4$, and consider the ratio $\Omega/\gamma = 0.1$ (b), $\Omega/\gamma = 1$ (c), and $\Omega/\gamma = 0.1$ (d).}
\label{fig:FIG01}
\end{center}
\end{figure*}

Equation~\eqref{eq:00000000003} is the first main result of the paper. We point out that Eq.~\eqref{eq:00000000004}, which in turn is related to the first-order derivative ${[ {f^{(1)}}({\widetilde{\rho}_t}) ]_{t = 0}}$ of the purity, has been previously investigated in the context of gain-loss systems~\cite{PhysRevLett.109.230405}, the quantum-classical description of non-Hermitian systems~\cite{Entropy_18120451}, and the dynamical instability of pure states~\cite{EPJD_253_69_2015}. In turn, Eq.~\eqref{eq:00000000005} is related to the second-order time derivative of the purity around $t = 0$, and also stands as a main result. The coefficients $|1/{T_1}|$ and $|1/{T_2}|$ provide timescales for the linear entropy at earlier times of the dynamics, thus predicting the initial growth of the mixedness of the evolved state of the quantum system. Importantly, they can be evaluated once the input state $\rho_0$ and the Hamiltonian $H = {H_1} + i {H_2}$ of the system have been specified. Note that $1/T_1$ and $1/T_2$ depend on the fluctuations of the obser\-va\-ble $H_2$ that are captured by its covariance respective to the input state. In particular, choosing $H_2$ a zero-valued operator, Eqs.~\eqref{eq:00000000004} and~\eqref{eq:00000000005} vanish, and one gets that ${S_L}({\widetilde{\rho}_t}) \approx {S_L}({\rho_0})$. In fact, this is expected since the linear entropy remains invariant for any quantum state undergoing a unitary evolution ge\-ne\-ra\-ted by a Hermitian operator.

In addition, for any initial pure state with ${\rho_0^2} = {\rho_0} = |\psi_0\rangle\langle{\psi_0}|$, and $f({\rho_0}) = \text{Tr}({\rho_0^2}) = 1$, one can verify that Eqs.~\eqref{eq:00000000004} and~\eqref{eq:00000000005} imply that $1/{T_1} = 0$ and $1/{T_2^2} = 0$, regardless of the operators $H_1$ and $H_2$. In this setting, it can be proved that the mixedness timescales vanish for any perturbative order within the short-time approximation of the linear entropy. Indeed, for the initial pure state $\rho_0 = |\psi_0\rangle\langle{\psi_0}|$ undergoing the nonunitary dynamics dictated by the non-Hermitian Hamiltonian $H = {H_1} + i{H_2}$, we have shown in Appendix~\ref{sec:00000000005} that the purity of the state $\widetilde{\rho}_t$ is given by $f({\widetilde{\rho}_t}) =1$, i.e., such state remains pure for all $t \geq 0$. Hence, one gets that the linear entropy identically vanishes, i.e., ${S_L}({\widetilde{\rho}_t}) = [{d}/({d - 1})] [1 - f({\widetilde{\rho}_t})] = 0$, which implies that any mixedness timescales become zero. Interestingly, we have shown that the same result is obtained within the so-called ``metric approach''. Finally, we note that the last term on the right-hand side of Eq.~\eqref{eq:00000000005} vanishes for $[{H_1},{H_2}] = 0$, i.e., for two commuting operators $H_1$ and $H_2$.

\begin{figure*}[!th]
\begin{center}
\includegraphics[scale=1.0]{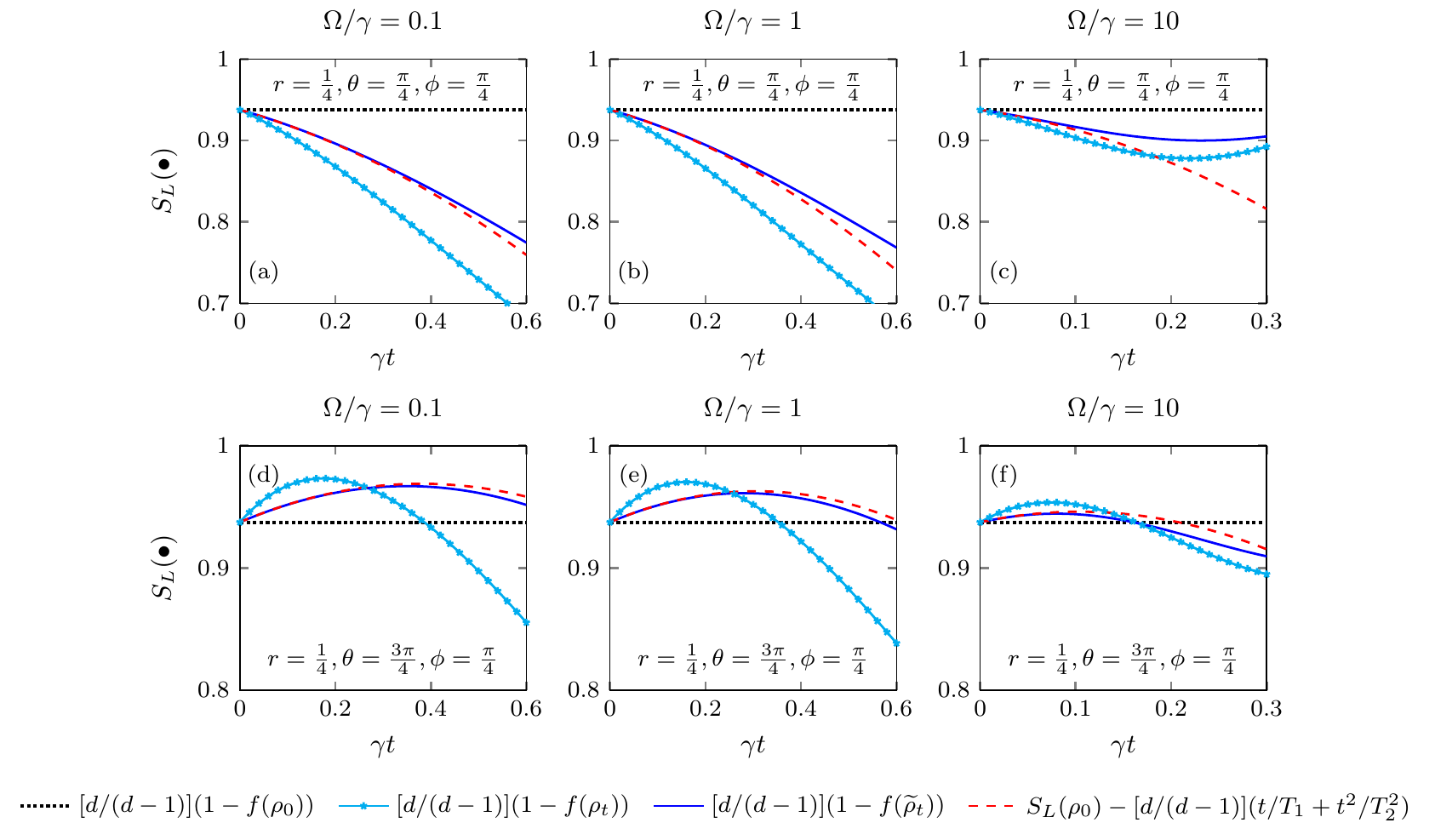}
\caption{(Color online) Plot of the linear entropy ${S_L}({\widetilde{\rho}_t})$, as a function of the dimensionless parameter $\gamma t$, for the driven two-level system described by the Hamiltonian $H = \Delta{|1\rangle}{\langle{1}|} + (\Omega/2)(|{0}\rangle\langle{1}| + |{1}\rangle\langle{0}|)$. Here we choose the ratio $\Delta/\gamma = 0.5$, and set $\Omega/\gamma = 0.1$ (a, d), $\Omega/\gamma = 1$ (b, e), and $\Omega/\gamma = 10$ (c, f). The system is initialized in the single-qubit state $\rho_0 = (1/2)(\mathbb{I} + \vec{r}\cdot\vec{\sigma})$, with $\{r, \theta, \phi\} = \{ 1/4, \pi/4, \pi/4 \}$ (a--c), and $\{r, \theta, \phi\} = \{ 1/4, 3\pi/4, \pi/4 \}$ (d--f). The black dotted line indicates the linear entropy ${S_L}({\rho_0}) = [d/(d - 1)][1 - \text{Tr}({\rho_0^2})]$ respective to the initial state. The cyan dash-dotted line depicts the linear entropy ${S_L}({\rho_t}) = [d/(d - 1)][1 - \text{Tr}({\rho_t^2})]$, where $\rho_t$ satisfies the Markovian master equation in Eq.~\eqref{eq:00000000009}. The blue solid line indicates the linear entropy ${S_L}({\widetilde{\rho}_t}) = [d/(d - 1)][1 - \text{Tr}({\widetilde{\rho}_t^2})]$, where the normalized state ${\widetilde{\rho}_t} = {\rho_t}/\text{Tr}({\rho_t})$ fulfills Eq.~\eqref{eq:00000000002}, with ${H_1} = H$ and $H_2 = - (\gamma/2){|1\rangle}{\langle{1}|}$. The red dashed line represents the linear entropy ${S_L}({\widetilde{\rho}_t}) \approx {S_L}({\rho_0}) - [d/(d - 1)](t/{T_1} + {t^2}/{T_2^2})$ within the short-time approximation [see Eqs.~\eqref{eq:00000000010} and~\eqref{eq:00000000011}].}
\label{fig:FIG02}
\end{center}
\end{figure*}
 

\subsection{Example: Dissipative two-level system}
\label{sec:00000000002A}

To illustrate our findings, we consider a driven two-level system described by the Hamiltonian $H = \Delta{|1\rangle}{\langle{1}|} + (\Omega/2)(|{0}\rangle\langle{1}| + |{1}\rangle\langle{0}|)$, where the two vectors $|0\rangle$ and $|{1}\rangle$ stand for ground and excited states, respectively, with $\Delta$ the energy detuning, and $\Omega$ being their coupling. The system interacts with a zero-temperature thermal reservoir, so that it decays from the excited state $|1\rangle$ to the ground state $|0\rangle$ emitting a photon at a rate $\gamma$. The dynamics is governed by the Markovian master equation 
\begin{equation}
\label{eq:00000000009}
\frac{d{\rho_t}}{dt} = -i({H_{\text{eff}}}{\rho_t} - {\rho_t}{H^{\dagger}_{\text{eff}}}) + \gamma{L}{\rho_t}{L^{\dagger}} ~, 
\end{equation}
where ${H_{\text{eff}}} = H - i(\gamma/2){L^{\dagger}}{L}$ is the effective non-Hermitian Hamiltonian, while $L = |{0}\rangle\langle{1}|$ is the jump operator~\cite{Breuer_Petruccione_book}. In the semiclassical regime, i.e., assuming that the effect of quantum jumps is negligible in the time interval under consideration, an effective description of the master equation can be obtained in terms of the co\-he\-rent nonunitary dissipation of the system, the latter related to the non-Hermitian Hamiltonian ${H_{\text{eff}}}$~\cite{PhysRevA.100.062131}. In this case, by discarding the quantum jump term $\gamma{L}{\rho_t}{L^{\dagger}}$, the dynamics of the system is dictated by the equation $d{\rho_t}/dt \approx -i({H_{\text{eff}}}{\rho_t} - {\rho_t}{H^{\dagger}_{\text{eff}}})$, which no longer des\-cri\-bes a completely positive and trace-preserving evolution. 

To overcome this issue, one introduces the norma\-li\-zed time-dependent density matrix ${\widetilde{\rho}_t} = {\rho_t}/{\text{Tr}({\rho_t})}$, which in turn fulfills Eq.~\eqref{eq:00000000002}, with ${H_1} = H$ and $H_2 = - (\gamma/2){|1\rangle}{\langle{1}|}$. The system is initialized in a single-qubit state $\rho_0 = (1/2)(\mathbb{I} + \vec{r}\cdot\vec{\sigma})$, where $\vec{r} = \{ r\sin\theta\cos\phi, r\sin\theta\sin\phi , r\cos\theta \}$ is the Bloch vector, with $r \in [0,1]$, $\theta \in [0,\pi]$ and $\phi \in [0,2\pi [$, while $\vec{\sigma} = \{{\sigma_x},{\sigma_y},{\sigma_z}\}$ is the vector of Pauli matrices, and $\mathbb{I}$ is the $2\times 2$ identity matrix. We will not show the analytical expressions for the exact linear entropy ${S_L}({\widetilde{\rho}_t})$ of the evolved state as they are cumbersome [see Eq.~\eqref{eq:00000000001}]. However, it is straightforward to obtain the short-time series expansion of ${S_L}({\widetilde{\rho}_t})$ applying Eq.~\eqref{eq:00000000003}, with the linear entropy of the input state as ${S_L}({\rho_0}) = 1 - {r^2}$. Using Eqs.~\eqref{eq:00000000004} and~\eqref{eq:00000000005}, we obtain the following dimensionless coefficients
\begin{equation}
\label{eq:00000000010}
\frac{1}{\gamma \, {T_1}} = \frac{1}{2}(1 - {r^2}) \, r \cos\theta ~,
\end{equation}
and
\begin{equation}
\label{eq:00000000011}
\frac{1}{{\gamma^2} {T_2^2}} = \frac{1}{8}(1 - {r^2})\left(1 - 3{r^2}{\cos^2}\theta + 2 \frac{\Omega}{\gamma} \, r \sin\theta\sin\phi \right) ~,
\end{equation}
respectively. We see that $1/\gamma{T_1}$ is a function of $r$ and $\theta$, while $1/{\gamma^2}{T_2^2}$ depends on the parameters $r$, $\theta$, $\phi$, and $\Omega/\gamma$. We notice that ${S_L}({\rho_0})$, $1/{T_1}$, and $1/{T_2}$ approach zero for any initial single-qubit pure state with $r = 1$, thus implying that the linear entropy ${S_L}({\widetilde{\rho}_t})$ is a vanishing quantity in this case.

In Fig.~\ref{fig:FIG01} we show the plots of the dimensionless quantities $1/\gamma{T_1}$ and $1/{\gamma^2}{T_2^2}$, as a function of the mixing parameter $r$ and the azimuthal angle $\theta$. In Fig.~\ref{fig:FIG01}(a), we see that $1/\gamma{T_1} > 0$ for $\theta \in [0,\pi/2)$ and $0 < r < 1$, while $1/\gamma{T_1} < 0$ for $\theta \in (\pi/2,\pi]$ and $0 < r < 1$. In addition, it follows that $1/{\gamma}{T_1} = 0$ for any chosen initial state with $\theta = \pi/2$ [see Eq.~\eqref{eq:00000000010}]. Next, Figs.~\ref{fig:FIG01}(b)--\ref{fig:FIG01}(d) show the plots of $1/{\gamma^2}{T_2^2}$ in Eq.~\eqref{eq:00000000011}, where we consider the cases $\Omega/\gamma = 0.1$ [see Fig.~\ref{fig:FIG01}(b)], $\Omega/\gamma = 1$ [see Fig.~\ref{fig:FIG01}(c)], and $\Omega/\gamma = 10$ [see Fig.~\ref{fig:FIG01}(d)], also fixing the polar angle $\phi = \pi/4$. On the one hand, for input states with either $\theta = 0$ or $\theta = \pi$, that are all incohe\-rent states respective to the computational basis $\{ |0\rangle, |1\rangle \}$, Eq.~\eqref{eq:00000000011} reduces to ${1}/{{\gamma^2} {T_2^2}} = (1/8)(1 - {r^2})(1 - 3{r^2})$, which is positive for $0 \leq r \leq 1/\sqrt{3}$ [see Figs.~\ref{fig:FIG01}(b)--\ref{fig:FIG01}(d)]. On the other hand, for initial states lying in the equatorial $xy$ plane with $\theta = \pi/2$, one gets that ${1}/{{\gamma^2} {T_2^2}} = (1/8)(1 - {r^2})[ 1 + 2 r (\Omega/\gamma)\sin\phi ]$, which is po\-si\-tive for $0 \leq \phi \leq \pi$ and $0 \leq r \leq 1$. We emphasize that the timescales related to the growth of mixedness can be obtained from the absolute values $|1/\gamma{T_1}|$ and $|1/{\gamma}{T_2}|$.

Figure~\ref{fig:FIG02} shows the plots of the linear entropy ${S_L}({\widetilde{\rho}_t})$, as a function of the dimensionless parameter $\gamma t$, for the aforementioned driven two-level system. The blue solid line refer to the exact linear entropy ${S_L}({\widetilde{\rho}_t})$ [see Eq.~\eqref{eq:00000000001}], while the red dashed line depicts the short-time expansion of this quantity in Eq.~\eqref{eq:00000000003}. We set input states with $\{r, \theta, \phi\} = \{ 1/4, \pi/4, \pi/4 \}$ [see Figs.~\ref{fig:FIG02}(a)--\ref{fig:FIG02}(c)], and $\{r, \theta, \phi\} = \{ 1/4, 3\pi/4, \pi/4 \}$ [see Figs.~\ref{fig:FIG02}(d)--\ref{fig:FIG02}(f)]. In addition, for a fixed ratio $\Delta/\gamma = 0.5$, we consider the cases $\Omega/\gamma = 0.1$ [see Figs.~\ref{fig:FIG02}(a) and~\ref{fig:FIG02}(d)], $\Omega/\gamma = 1$ [see Figs.~\ref{fig:FIG02}(b) and~\ref{fig:FIG02}(e)], and $\Omega/\gamma = 10$ [see Figs.~\ref{fig:FIG02}(c) and~\ref{fig:FIG02}(f)]. In each panel, the black dotted line indicates the linear entropy ${S_L}({\rho_0}) = [d/(d - 1)][1 - \text{Tr}({\rho_0^2})]$ respective to the initial state. We note that the linear entropy remains invariant under unitary evolutions generated by the Hermitian Hamiltonian $H$, i.e., ${S_L}({e^{- i t H}}{\rho_0}{e^{+ i t H}}) = {S_L}({\rho_0})$. The cyan dash-dotted line displays the linear entropy ${S_L}({\rho_t}) = [d/(d - 1)][1 - \text{Tr}({\rho_t^2})]$, with $\rho_t$ satisfying the Markovian master equation in Eq.~\eqref{eq:00000000009}. The blue solid line indicates the linear entropy ${S_L}({\widetilde{\rho}_t}) = [d/(d - 1)][1 - \text{Tr}({\widetilde{\rho}_t^2})]$, where the normalized state ${\widetilde{\rho}_t} = {\rho_t}/\text{Tr}({\rho_t})$ fulfills Eq.~\eqref{eq:00000000002}, with ${H_1} = H$ and $H_2 = - (\gamma/2){|1\rangle}{\langle{1}|}$. The red dashed line represents the linear entropy ${S_L}({\widetilde{\rho}_t}) \approx {S_L}({\rho_0}) - [d/(d - 1)](t/{T_1} + {t^2}/{T_2^2})$ within the short-time approximation, with the timescales $1/{T_1}$ and $1/{T_2^2}$ given in Eqs.~\eqref{eq:00000000010} and~\eqref{eq:00000000011}, respectively. In Appendix~\ref{sec:00000000006}, we compare the timescales $1/{T_1}$, $1/{T_2^2}$ and the first-order and second-order time derivatives of the linear entropy for (i) the state $\rho_t$ satisfying Eq.~\eqref{eq:00000000009}; (ii) and the normalized state $\widetilde{\rho}_t$ fulfilling Eq.~\eqref{eq:00000000002} related to the effective non-Hermitian Hamiltonian.

Overall, Fig.~\ref{fig:FIG02} shows that the short-time approximation of ${S_L}({\widetilde{\rho}_t})$ correctly reproduces its growth at earlier times of the dynamics. We find that, for $0.01 \lesssim  \gamma t \lesssim 0.1$, the relative error between the exact linear entropy [see Eq.~\eqref{eq:00000000001}] and its perturbative expansion [see Eqs.~\eqref{eq:00000000003},~\eqref{eq:00000000010}, and~\eqref{eq:00000000011}] is of order ${10^{-7}} \lesssim {\epsilon} \lesssim {10^{-4}}$ for $\Omega/\gamma = 0.1$ and $\Omega/\gamma = 1$, while it ranges as ${10^{-6}} \lesssim {\epsilon} \lesssim {10^{-3}}$ for $\Omega/\gamma = 10$. However, for $\gamma t \gtrsim 0.1$, we have that the latter result is loose and fails to capture the changes in the eigenvalues of the state ${\widetilde{\rho}_t}$ driven by the non-Hermitian Hamiltonian. We emphasize that one should look to higher orders in its Taylor expansion to accurately predict the mixedness degree of the evolved state for later times.

 
 \section{Mixedness timescale for non-Hermitian bipartite systems}
 \label{sec:00000000003}

In this section, we provide a mixedness timescale for bipartite quantum systems whose dynamics can be effectively described by a non-Hermitian Hamiltonian. In detail, using the linear entropy as a useful measure of mixedness, we investigate its short-time expansion up to the second order in $t$ for certain time-dependent marginal states of the composite system.

We consider a bipartite quantum system with a finite-dimensional Hilbert space ${\mathcal{H}_A} \otimes {\mathcal{H}_B}$ split into the subsystems ${\mathcal{H}_A}$ and ${\mathcal{H}_B}$, with ${d_{A,B}} = \text{dim} \, {\mathcal{H}_{A,B}}$. This com\-po\-si\-te system is initialized in the quantum state ${\rho^{AB}_0}$, which in turn can be chosen either a pure or mixed state, entangled or uncorrelated one, from which the mixed marginal states ${\rho^{A,B}_0} = {\text{Tr}_{B,A}}({\rho^{AB}_0})$ can be obtained. The state ${\rho^{AB}_0}$ undergoes a nonunitary evolution ge\-ne\-ra\-ted by the time-independent non-Hermitian Hamiltonian $H = {H_1} + i{H_2}$, with ${H_1} = ({1}/{2})(H + {H^{\dagger}})$ and ${H_2} = - ({i}/{2})(H - {H^{\dagger}})$ being noncommuting obser\-va\-bles acting over ${\mathcal{H}_A} \otimes {\mathcal{H}_B}$. It is noteworthy that the operators $H_1$ and $H_2$ play the role of the Hermitian and anti-Hermitian parts of $H$, respectively. In this setting, it can be proved that the effective dynamics of subsystem $\mathcal{H}_{A,B}$ is go\-ver\-ned by the equation of motion~\cite{PhysRevA.42.1467,doi:10.1142,EPJD_253_69_2015}
\begin{align}
\label{eq:00000000012}
\frac{d}{dt}{\widetilde{\rho}^{A,B}_t} &= - i \, {\text{Tr}_{B,A}}([{H_1}, {\widetilde{\rho}^{AB}_t}]) + {\text{Tr}_{B,A}}(\{ {H_2} , {\widetilde{\rho}^{AB}_t}\})  \nonumber\\
& - 2 \,{\text{Tr}_{AB}}({\widetilde{\rho}^{AB}_t}{H_2})\,{\widetilde{\rho}^{A,B}_t} ~,
\end{align}
where ${\widetilde{\rho}^{A,B}_t} := {\text{Tr}_{B,A}}({\widetilde{\rho}^{AB}_t})$ stands for the time-dependent reduced density matrices, while ${\widetilde{\rho}^{AB}_t} := {{\rho^{AB}_t}}/{{\text{Tr}_{AB}}({\rho^{AB}_t})}$ is the nor\-ma\-li\-zed state of the whole system.

Without loss of generality, hereafter we will address the dynamics of the marginal state ${\widetilde{\rho}^A_t}$, and investigate the short-time behavior of its linear entropy 
\begin{equation}
\label{eq:00000000013}
{S_L}({\widetilde{\rho}^A_t}) = \frac{d_A}{({d_A} - 1)}(1 - f({\widetilde{\rho}^A_t})) ~, 
\end{equation}
where $f({\widetilde{\rho}^A_t}) = {\text{Tr}_A}[{({\widetilde{\rho}^A_t})^2}]$ is the purity of the aforementioned reduced density matrix. In this case, by per\-for\-ming a Taylor expansion of ${S_L}({\widetilde{\rho}^A_t})$ up to se\-cond order in $t$, around $t = 0$, one gets
\begin{align}
\label{eq:00000000014}
{S_L}({\widetilde{\rho}^A_t}) &\approx {S_L}({\rho_0^A}) - \frac{d_A}{({d_A} - 1)} \left(\frac{1}{T_{1,h}} + \frac{1}{T_{1,nh}}\right)t \nonumber\\
&- \frac{d_A}{({d_A} - 1)}\left(\frac{1}{T_{2,h}^2} + \frac{1}{T_{2,nh}^2}\right){t^2} + O({t^3}) ~,
\end{align}
with ${T_{1,h}}$ and ${T_{1, nh}}$ being coefficients related to the first-order derivative of the linear entropy around $t = 0$, and defined as
\begin{equation}
\label{eq:00000000015}
{T_{1,h}^{-1}} := {2i} \, {\langle {\text{Tr}_{B}}( [ {\rho^{AB}_0} , {H_1} ]) \rangle_{A}} ~,
\end{equation}
and
\begin{equation}
\label{eq:00000000016}
{T_{1,nh}^{-1}} := 2\, {\langle{\text{Tr}_B}(\{ {\rho^{AB}_0} , {H_2} \})\rangle_A} - 4 \, f({\rho_0^A}) {\langle{H_2}\rangle_{AB}} ~,
\end{equation}
while ${T_{2,h}}$ and ${T_{2, nh}}$ arise from the second-order derivative of the linear entropy at the vicinity of $t = 0$ as follows
\begin{align}
\label{eq:00000000017}
{T_{2,h}^{- 2}} &:= - {\langle {\text{Tr}_{B}}([[{\rho^{AB}_0} , {H_1}] , {H_1}]) \rangle_A} \nonumber\\
& - {\text{Tr}_{A}}\left({({\text{Tr}_{B}}([{\rho^{AB}_0} , {H_1}]) )^2}\right) ~,
\end{align}
and
\begin{align}
\label{eq:00000000018}
&{T_{2, nh}^{- 2}} :=\nonumber\\ 
&{\langle {\text{Tr}_{B}}(\{ \{ {\rho^{AB}_0} , {H_2} \} , {H_2} \}) \rangle_A} + {\text{Tr}_{A}}\left({({\text{Tr}_{B}}(\{{\rho^{AB}_0} , {H_2} \} ) )^2}\right) \nonumber\\
&+ i {\langle {\text{Tr}_{B}}(\{ [{\rho^{AB}_0} , {H_1} ] , {H_2} \}) \rangle_A} + i {\langle {\text{Tr}_{B}}( [ \{ {\rho^{AB}_0} , {H_2} \} , {H_1} ]) \rangle_A} \nonumber\\
&- 8 \, {\langle {H_2} \rangle_{AB}} \left( {\langle {\text{Tr}_{B}}( \{ {\rho^{AB}_0} , {H_2} \} ) \rangle_A} + i \, {\langle {\text{Tr}_{B}}([ {\rho^{AB}_0} , {H_1} ]) \rangle_A} \right) \nonumber\\
&+ 2 f({\rho_0^A}) \left( i\, {\langle [{H_2},{H_1}] \rangle_{AB}} - 2 \left({\langle{H_2^2}\rangle_{AB}} - 3\, {\langle{H_2}\rangle_{AB}^2} \right) \right) \nonumber\\ 
&- 2i \, {\text{Tr}_A}\left( {\text{Tr}_{B}}([ {H_1} , {\rho^{AB}_0} ]) \, {\text{Tr}_{B}}(\{ {\rho^{AB}_0} , {H_2} \}) \right) ~.
\end{align}
Here $\langle\bullet\rangle_{\mu} := {\text{Tr}_{\mu}}(\bullet \, {\rho^{\mu}_0})$ defines the expectation value at time $t = 0$, with $\mu = \{A, B, AB\}$. We note that Eqs.~\eqref{eq:00000000015},~\eqref{eq:00000000016},~\eqref{eq:00000000017} and~\eqref{eq:00000000018} were obtained from the first-order and second-order derivatives of the quantum purity $f({\widetilde{\rho}^A_t})$ at the vicinity of $t = 0$, with ${[ {f^{(1)}}({\widetilde{\rho}^A_t}) ]_{t = 0}} = 1/{T_{1,h}} + 1/{T_{1,nh}}$ and $(1/2){[ {f^{(2)}}({\widetilde{\rho}^A_t}) ]_{t = 0}} =1/{T_{2,h}^2} + 1/{T_{2,nh}^2}$.

We point out that Eq.~\eqref{eq:00000000014} [see also Eqs.~\eqref{eq:00000000015}--\eqref{eq:00000000018}] is the second main result of the paper. Overall, we see that the coefficients $|{1/{T_{1,h}}} + {1/{T_{1,nh}}}|$ and ${|{1/{T_{2,h}^2}} + {1/{T_{2,nh}^2}}|}$ represent first-order and second-order timescales in the initial growth of the mixedness dynamics signaled by linear entropy. On the one hand, both the coefficients $T_{1,h}$ and $T_{2,h}$ depend on the initial state of the bipartite system and the Hermitian part ${H_1}$ of the non-Hermitian Hamiltonian. On the other hand, the coefficients $1/{T_{1,nh}}$ and $1/{T_{2 , nh}}$ depend on the anti-Hermitian part ${H_2}$ of the effective non-Hermitian Hamiltonian. In particular, note that the result in Eqs.~\eqref{eq:00000000016} and~\eqref{eq:00000000018} approach zero in the Hermitian limit $H^{\dagger} = H = H_1$, i.e., when one sets $H_2$ as a zero-valued observable, regardless of the observable $H_1$. This means that $1/{T_{1,nh}}$ and $1/{T_{2,nh}}$ assign first-order and second-order nontrivial corrections to the mixedness timescales that are induced by the effective non-Hermitian Hamiltonian.

To gain insights into understanding the results in Eqs.~\eqref{eq:00000000014}--\eqref{eq:00000000018}, in the following, we investigate two cases of interest in view of the nonunitary dynamics of non-Hermitian Hamiltonians. The first case des\-cribes bipartite quantum systems with initial uncorrelated states. The second one addresses a multiparticle system whose non-Hermitian Hamiltonian corresponds to the transverse-field $XY$ model with next-nearest neighbor couplings and a perturbing term given by an all-to-all Ising Hamiltonian with an imaginary exchange coupling.


\subsection{Separable initial pure states}
 \label{sec:00000000003A}

Here we specialize the result in Eq.~\eqref{eq:00000000014} to the particular case of uncorrelated initial pure state ${\rho^{AB}_0} = {\rho^A_0} \otimes {\rho^B_0}$, with ${\rho^A_0}$ and ${\rho^B_0}$ normalized pure marginal states, i.e., ${\text{Tr}_{\mu}}[{({\rho^{\mu}_0})^2}] = {\text{Tr}_{\mu}}({\rho^{\mu}_0}) = 1$ for all $\mu = \{A,B,AB\}$. We consider the non-Hermitian Hamiltonian $H = {H_1} + i{H_2}$ with ${H_1} = {\sum_n}\,{A_n}\otimes{B_n}$ and ${H_2} = {\sum_n}\,{C_n}\otimes{D_n}$, where ${A_n}, {C_n} \in {\mathcal{H}_A}$ and ${B_n}, {D_n} \in {\mathcal{H}_B}$ represent non\-commu\-ting local obser\-va\-bles. In this setting, one can prove that both the coefficients $1/{T_{1,h}} = 0$ [see Eq.~\eqref{eq:00000000015}] and $1/{T_{1,nh}} = 0$ [see Eq.~\eqref{eq:00000000016}] identically vanish, and the linear entropy in Eq.~\eqref{eq:00000000014} becomes
\begin{equation}
\label{eq:00000000022}
{S_L}({\widetilde{\rho}^A_t}) \approx  - \frac{d_A}{({d_A} - 1)}\left(\frac{1}{T_{2,h}^2} + \frac{1}{T_{2,nh}^2} \right){t^2} + O({t^3}) ~,
\end{equation}
with the following nonzero coefficients
\begin{align}
\label{eq:00000000023}
{T_{2,h}^{-2}} &= - 2\, {\sum_{k,l}}\left({\langle{A_k}{A_l}\rangle_A} - {\langle{A_k}\rangle_A}{\langle{A_l}\rangle_A}\right)\nonumber\\
&\times\left({\langle{B_k}{B_l}\rangle_B} - {\langle{B_k}\rangle_B}{\langle{B_l}\rangle_B}\right) ~,
\end{align}
and
\begin{align}
\label{eq:00000000024}
{T_{2,nh}^{-2}} &= - 2\, {\sum_{k,l}}\left({\langle{C_k}{C_l}\rangle_A} - {\langle{C_k}\rangle_A}{\langle{C_l}\rangle_A}\right)\nonumber\\
&\times\left({\langle{D_k}{D_l}\rangle_B} - {\langle{D_k}\rangle_B}{\langle{D_l}\rangle_B}\right) \nonumber\\
& + 4 \, {\sum_{k,l}}\, \text{Im}[ \left({\langle{A_k}{C_l}\rangle_A} - {\langle{A_k}\rangle_A}{\langle{C_l}\rangle_A}\right)\nonumber\\
&\times\left({\langle{B_k}{D_l}\rangle_B} - {\langle{B_k}\rangle_B}{\langle{D_l}\rangle_B}\right)] ~.
\end{align}
Overall, Eq.~\eqref{eq:00000000022} implies that the linear entropy varies quadratically at earlier times of the dynamics. We see that $1/{T_{2,h}}$ is proportional to the so-called correlated quantum uncertainty of observables ${A_n} \in {\mathcal{H}_A}$ and ${B_n} \in {\mathcal{H}_B}$, thus being entirely determined by the expectation values of these operators with respect to the initial marginal states $\rho^{A,B}_0$. It is noteworthy that $|1/{T_{2,h}}|$ assigns a universal timescale for two initially pure subsystems to become entangled by means of the coupling with a Hermitian Hamiltonian $H_1$~\cite{PhysRevLett.77.207,PhysRevD.97.066008,PhysRevA.97.022317}. 

In turn, the coefficient $1/{T_{2,nh}}$ depends on the correlated quantum uncertainty of observables ${C_n} \in {\mathcal{H}_A}$ and ${D_n} \in {\mathcal{H}_B}$, and the imaginary part of cross-correlations of the set of local observables. In particular, one ve\-ri\-fies that $1/{T_{2,nh}}$ vanishes in the Hermitian limit $H = H^{\dagger} = H_1$, i.e., when choosing zero valued obser\-va\-bles $C_n$ and $D_n$. In this case, one finds that Eq.~\eqref{eq:00000000023} recovers the so-called idempotency defect for composite systems described by Hermitian Hamiltonians~\cite{PhysRevLett.77.207}, and constitutes a timescale for the entanglement dynamics of subsystems~\cite{PhysRevD.97.066008,PhysRevA.97.022317}. In this setting, we see that $1/{T_{2,nh}}$ re\-pre\-sents a true signature of the non-Hermitian features of $H$ in the mixedness dynamics. Note that, in addition to the coefficient $1/{T_{2,h}}$, the effective non-Hermitian Hamiltonian induces the factor $1/{T_{2,nh}}$ on the entanglement timescale for initially separable pure states.


\subsection{$1$D quantum many-body systems}
 \label{sec:00000000003B}
 %
\begin{figure*}[!th]
\begin{center}
\includegraphics[scale=1.0]{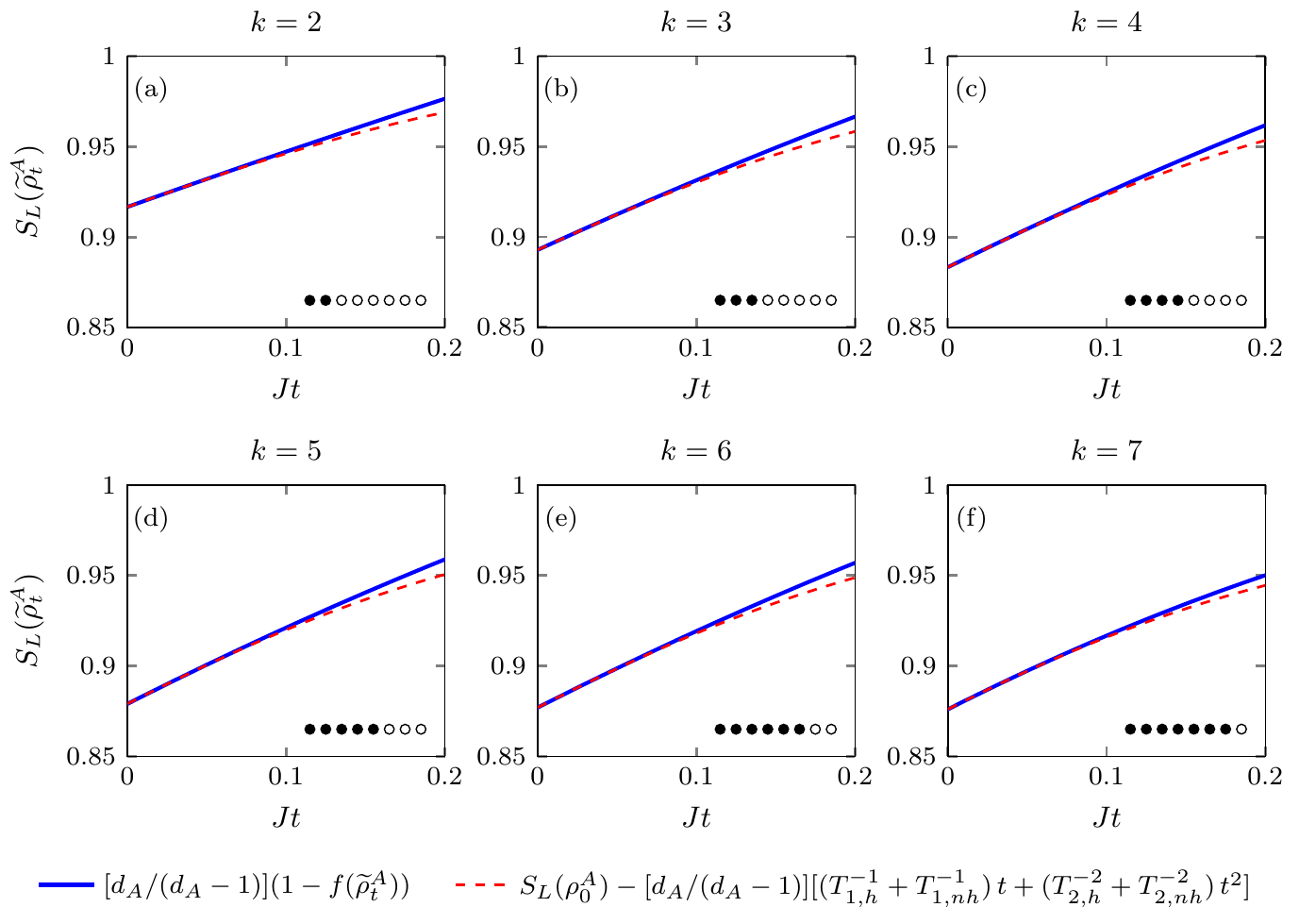}
\caption{(Color online) Plot of the linear entropy ${S_L}({\widetilde{\rho}^A_t})$ for the $k$-particle reduced density matrix ${\widetilde{\rho}_t^A}$, as a function of the dimensionless parameter $Jt$. The nonunitary evolution of subsystem $A$ is governed by Eq.~\eqref{eq:00000000012}, with $H_1$ being the transverse field $XY$ Hamiltonian in Eq.~\eqref{eq:00000000025}, and $H_2$ as the all-to-all Ising model in Eq.~\eqref{eq:00000000026}. The system $A + B$ is initialized in the GHZ mixed state ${\rho^{AB}_0} = ((1 - p)/d) \mathbb{I} + p\,|{\text{GHZ}_N}\rangle\langle{\text{GHZ}_N}|$, where $|{\text{GHZ}_N}\rangle = ({1}/{\sqrt{2}})(\,{|0\rangle^{\otimes N}} + {|1\rangle^{\otimes N}} )$. Here we set $N = 8$, $\gamma = 0.75$, ${J_z}/J = 0.5$, and the mixing parameter $p = 0.5$. The blue solid line corresponds to the exact expression of linear entropy in Eq.~\eqref{eq:00000000013}, and the red dashed line indicates its the short-time perturbative expansion in Eq.~\eqref{eq:00000000014} [see also Eqs.~\eqref{eq:00000000029},~\eqref{eq:00000000030}, and~\eqref{eq:00000000031}].}
\label{fig:FIG03}
\end{center}
\end{figure*}

We set the non-Hermitian Hamiltonian $H = {H_1} + i{H_2}$, where ${H_1}$ describes the transverse-field $XY$ model with open boundary conditions as~\cite{LIEB1961407,PhysRevA.2.1075,PhysRevA.3.786,PhysRevA.3.2137,PhysRevA.4.2331}
\begin{equation}
\label{eq:00000000025}
{H_1} = - J {\sum_{j = 1}^{N - 1}}\left({\gamma_+}{\sigma_j^x}{\sigma_{j + 1}^x} + {\gamma_-}{\sigma_j^y}{\sigma_{j + 1}^y}\right) - h {\sum_{j = 1}^N}\,{\sigma_j^z} ~,
\end{equation}
where $J$ is the coupling constant, $h$ represents the external magnetic field along the $z$ axis, and $\gamma_{\pm} = (1 \pm \gamma)/2$, with $\gamma$ being the anisotropy parameter. For $\gamma = 0$ this Hamiltonian reduces to the isotropic $XX$ model, while for $\gamma = \pm 1$ we recover the Ising model. Furthermore, this model exhibits phase transitions at the isotropic line $\gamma = 0$ ($|h| \leq 1$), and at the critical magnetic field $|h| = 1$. In turn, $H_2$ denotes the many-body fully connected quantum Ising model given by
\begin{equation}
\label{eq:00000000026}
{H_2} = \frac{J_z}{N}\,{\sum_{j < l}}\,{\sigma_j^z}{\sigma_l^z} ~,
\end{equation}
where $J_z$ is the coupling strength, $N$ is the number of spins, and $\{ \sigma^{x,y,z}_s\}_{s = 1,\ldots,N}$ are the Pauli matrices.

\begin{figure*}[!th]
\begin{center}
\includegraphics[scale=1.0]{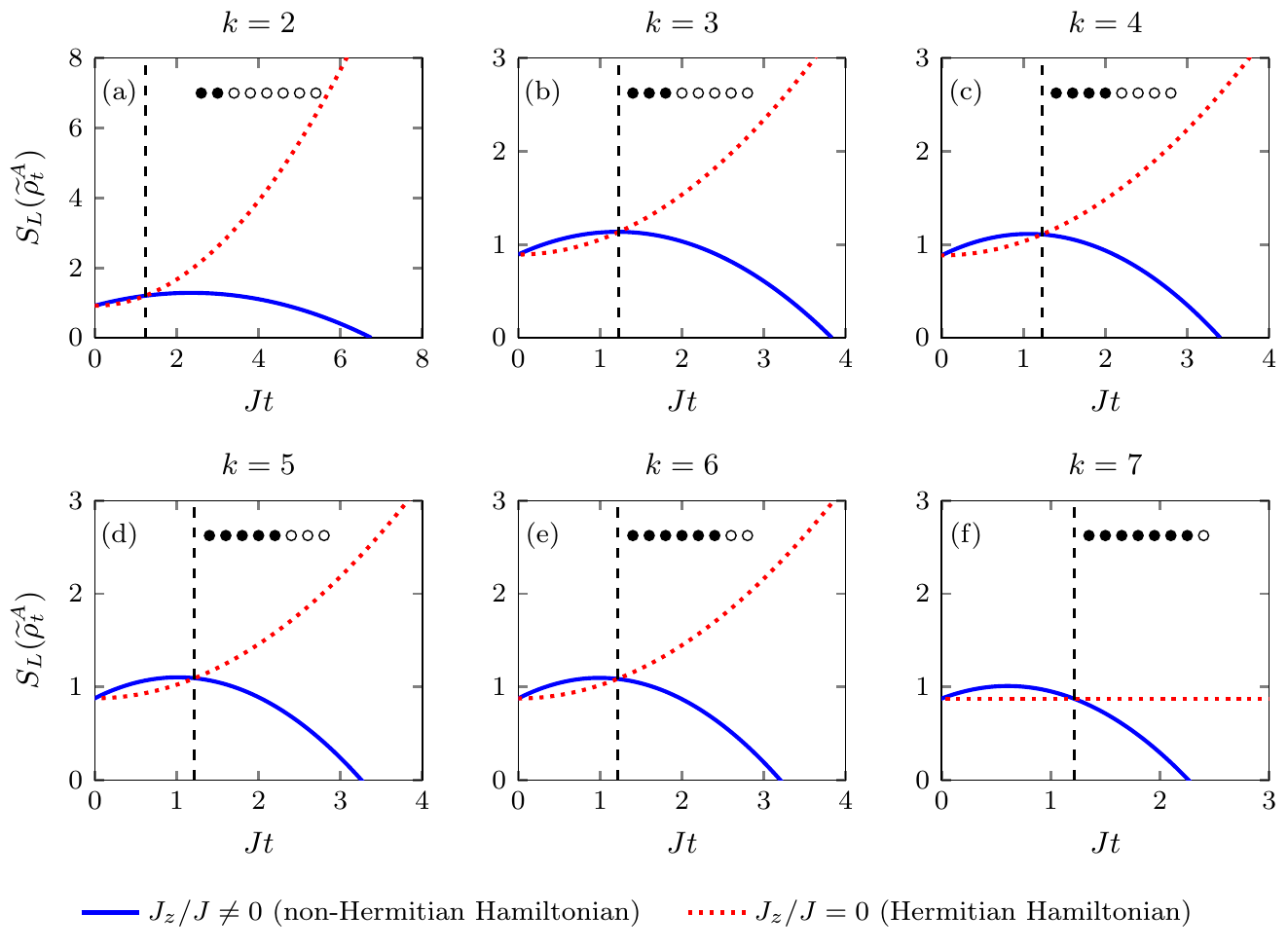}
\caption{(Color online) Plot of the linear entropy ${S_L}({\widetilde{\rho}^A_t})$ in the short-time approximation [see Eq.~\eqref{eq:00000000014}], for the $k$-particle reduced density matrix ${\widetilde{\rho}_t^A}$, as a function of the dimensionless parameter $Jt$. The nonunitary evolution of subsystem $A$ is governed by Eq.~\eqref{eq:00000000012}, with $H_1$ being the transverse field $XY$ Hamiltonian in Eq.~\eqref{eq:00000000025}, and $H_2$ as the all-to-all Ising model in Eq.~\eqref{eq:00000000026}. The system $A + B$ is initialized in the GHZ mixed state ${\rho^{AB}_0} = ((1 - p)/d) \mathbb{I} + p\,|{\text{GHZ}_N}\rangle\langle{\text{GHZ}_N}|$, where $|{\text{GHZ}_N}\rangle = ({1}/{\sqrt{2}})(\,{|0\rangle^{\otimes N}} + {|1\rangle^{\otimes N}} )$. Here we set $N = 8$, $\gamma = 0.75$, and the mixing parameter $p = 0.5$. The blue solid line corresponds to the case ${J_z}/J = 0.5$, and the red dashed line depicts the case ${J_z}/J = 0$.}
\label{fig:FIG04}
\end{center}
\end{figure*}

We consider a bipartition into first sequential $k$ sites ($1,\ldots,k$) as the subsystem $A$, and its complement of sequential $N - k$ sites ($k + 1, \ldots, N$) as subsystem $B$. The system $A + B$ is initialized in the mixed state
\begin{equation}
\label{eq:00000000027}
{\rho^{AB}_0} = \left(\frac{1 - p}{d}\right)\mathbb{I} + p\,|{\text{GHZ}_N}\rangle\langle{\text{GHZ}_N}| ~,
\end{equation}
with $d = 2^N$, $0 \leq p \leq 1$, and $|{\text{GHZ}_N}\rangle$ is the GHZ state of $N$ particles defined as
\begin{equation}
\label{eq:00000000028}
|{\text{GHZ}_N}\rangle = \frac{1}{\sqrt{2}}\left(\,{|0\rangle^{\otimes N}} + {|1\rangle^{\otimes N}}\right) ~,
\end{equation}
and its purity is written as $f({\rho^{AB}_0}) = (1/{2^N})(1 + ({2^N} - 1)\,{p^2})$. Furthermore, one can evaluate the averaged va\-lues ${\langle{H_2}\rangle_{AB}} = ({J_z}/2){(N - 1)p}$ and ${\langle{H_2^2}\rangle_{AB}} = {J_z^2}[(N - 1)/(4N)]( 2 + (N - 2)(N + 1)p)$ of the observable $H_2$ respective to the probe state of system $A + B$. The many-body state ${\rho^{AB}_0}$ undergoes a nonunitary evolution gene\-rated by the non-Hermitian Hamiltonian $H = {H_1} + i{H_2}$, and the subsystem $A$ is described by the reduced $k$-particle state ${\widetilde{\rho}^A_t} = {\text{Tr}_{N - k}}({\widetilde{\rho}^{AB}_t})$ whose dynamics is go\-ver\-ned by Eq.~\eqref{eq:00000000012}. The linear entropy ${S_L}({\widetilde{\rho}^A_t})$ of this marginal state is given in Eq.~\eqref{eq:00000000013}, which in turn reduces to ${S_L}({\rho_0^A}) = [{d_A}/({d_A} - 1)][1 - (1/{2^k})(1 + ({2^{k - 1}} - 1)\, {p^2})]$ at time $t = 0$. The short-time expansion of ${S_L}({\widetilde{\rho}^A_t})$ is given in Eq.~\eqref{eq:00000000014}. In this setting, it is possible to verify that the coefficient $1/{T_{1,h}} = 0$ vanishes [see Eq.~\eqref{eq:00000000015}], while Eq.~\eqref{eq:00000000016} implies the following nonzero contribution
\begin{align}
\label{eq:00000000029}
&{T^{-1}_{1,nh}} =\nonumber\\
& - \frac{{J_z} \, p (1 - p)}{{2^{k - 1}} N}\left( k(k - 1) + N(N - 1)({2^{k - 1}} - 1)p \right) ~.
\end{align}
It is noteworthy that Eq.~\eqref{eq:00000000029} shows that $1/{T_{1,nh}}$ exhibits a polynomial dependence on the mixing parameter $p$, thus being a negative quantity for all $0 < p < 1$, and $k \in \{1,\ldots,N\}$. In particular, it follows that $1/{T_{1,nh}} = 0$ for the initial pure state $|{\text{GHZ}_N}\rangle\langle{\text{GHZ}_N}|$ ($p = 1$) and also for the maximally mixed state $\mathbb{I}/d$ ($p = 0$). We find that $1/{T_{1,nh}}$ is proportional to the coupling strength $J_z$, and identically vanishes in the Hermitian limit (${J_z}/J = 0$).

Next, by using Eq.~\eqref{eq:00000000017}, we obtain 
\begin{equation}
\label{eq:00000000030}
{T^{-2}_{2,h}} = ({\delta_{{d_B},2}} - 1){\gamma^2}{J^2}{p^2} ~, 
\end{equation}
which depends on the coupling $J$, anisotropy parameter $\gamma$, and va\-ni\-shes whenever $\mathcal{H}_B$ is a two-dimensional subspace, i.e., one gets $1/{T_{2,h}} = 0$ for the case ${d_B} = 2$. Finally, by applying Eq.~\eqref{eq:00000000018} and performing lengthy calculations, one obtains the result
\begin{align}
\label{eq:00000000031}
&{T^{-2}_{2,nh}} = - \frac{{J_z^2}}{{2^{k - 1}} \, {N^2}} \left\{ 3{N^2}{(N - 1)^2}(1 - {2^{k - 1}}) \, {p^4} \right. \nonumber\\
&\left. + N(N - 1)[({2^{k - 1}} - 1)(5N(N - 1) - 2) - 4k(k - 1)] \, {p^3} \right. \nonumber\\
&\left. + [ k({k^2}(k - 6) + k + 4) + 2{N}{k}(k - 1)(3N - 1) \right.\nonumber\\
&\left. - 2{N}({2^{k - 1}} - 1)({N^3} - 2{N^2} + 1)] \, {p^2} - 2k(k - 1) \right.\nonumber\\
&\left. - k(k - 1)[k(k - 5) + 2((N - 1)(N + 2) - 1)] \, {p} \right\} ~.
\end{align}
We find that $1/{T^{2}_{2,nh}}$ behaves polynomially with the mi\-xing parameter $p$. In particular, it follows that $1/{T^{2}_{2,nh}} = 0$ for $p = 1$, while for $p = 0$ one obtains that $1/{T^{2}_{2,nh}} = {2^{2 - k}} {J_z^2}k(k - 1)/{N^2}$. Hence, for $N \gg k$, the latter case implies $1/{T^{2}_{2,nh}} \sim {J_z^2}/{N^2}$ for the initial maximally mixed state ($p = 0$), i.e, it scales with the inverse square of the number of particles.

\begin{figure}[!th]
\begin{center}
\includegraphics[scale=0.9]{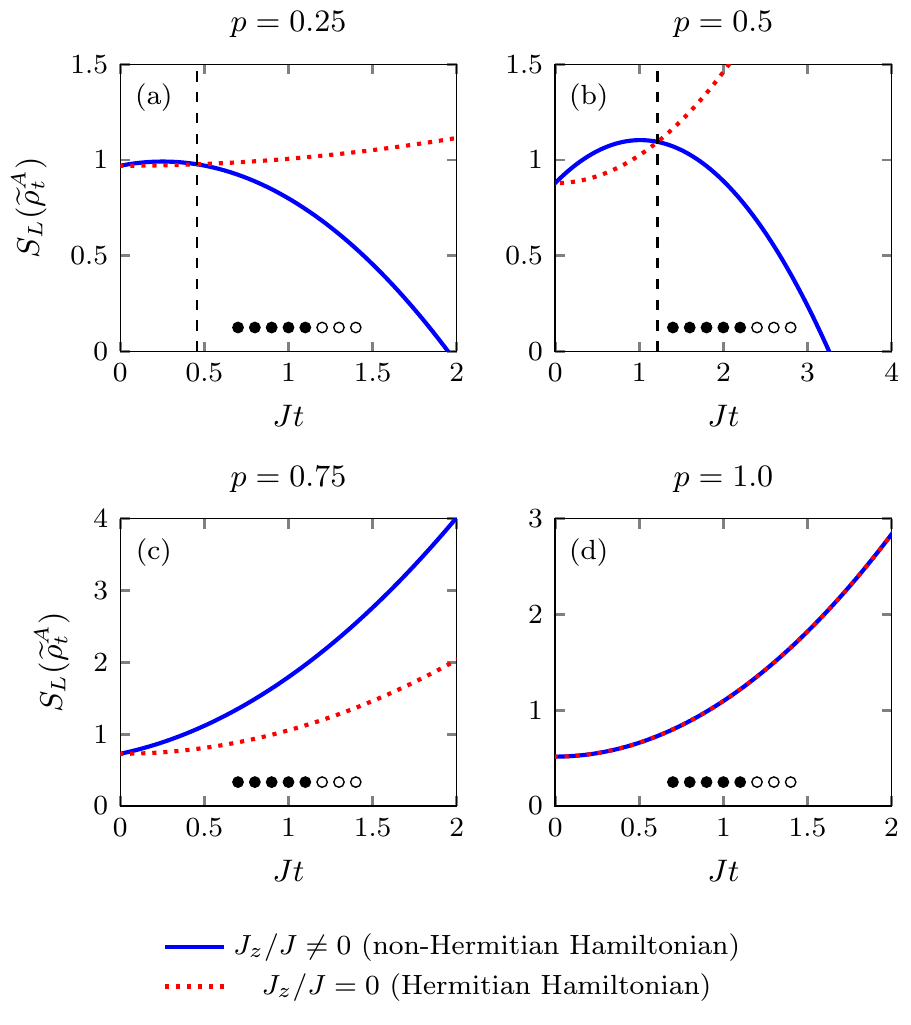}
\caption{(Color online) Plot of the linear entropy ${S_L}({\widetilde{\rho}^A_t})$ [see Eq.~\eqref{eq:00000000013}] in the short-time approximation [see Eq.~\eqref{eq:00000000014}], for the $k$-particle reduced density matrix ${\widetilde{\rho}_t^A}$, as a function of the dimensionless parameter $Jt$. The nonunitary evolution of subsystem $A$ is governed by Eq.~\eqref{eq:00000000012}, with $H_1$ being the transverse field $XY$ Hamiltonian in Eq.~\eqref{eq:00000000025}, and $H_2$ as the all-to-all Ising model in Eq.~\eqref{eq:00000000026}. The system $A + B$ is initialized in the GHZ mixed state ${\rho^{AB}_0} =  ((1 - p)/d) \mathbb{I} + p\,|{\text{GHZ}_N}\rangle\langle{\text{GHZ}_N}|$, where $|{\text{GHZ}_N}\rangle = ({1}/{\sqrt{2}})(\,{|0\rangle^{\otimes N}} + {|1\rangle^{\otimes N}} )$. Here we consider the system size $N = 8$, with $k = 5$, and $\gamma = 0.75$. The blue solid line corresponds to the case ${J_z}/J = 0.5$, and the red dashed line depicts the case ${J_z}/J = 0$.}
\label{fig:FIG05}
\end{center}
\end{figure}

In the following, we will numerically address the short-time dynamics of the linear entropy ${S_L}({\widetilde{\rho}^A_t})$ in Eq.~\eqref{eq:00000000014}. The system $A + B$ is initialized at the GHZ mixed state in Eq.~\eqref{eq:00000000027}, with $H_1$ being the transverse field $XY$ Hamiltonian in Eq.~\eqref{eq:00000000025}, and $H_2$ standing for the all-to-all Ising model in Eq.~\eqref{eq:00000000026}. In this case, bearing in mind that $1/{T_{1,h}} = 0$, we also apply the results in Eqs.~\eqref{eq:00000000029},~\eqref{eq:00000000030},~\eqref{eq:00000000031}. Unless otherwise stated, we set the system size $N = 8$, the anisotropy parameter $\gamma = 0.75$, and the ratio ${J_z}/J = 0.5$.

In Fig.~\ref{fig:FIG03} we plot of the linear entropy ${S_L}({\widetilde{\rho}^A_t})$, as a function of the dimensionless parameter $Jt$, and set the mixing parameter $p = 0.5$. The solid blue line refers to the exact linear entropy ${S_L}({\widetilde{\rho}^A_t})$ [see Eq.~\eqref{eq:00000000013}], while the red dashed line depicts the short-time expansion of this quantity in Eq.~\eqref{eq:00000000014}. We consider the subsystem $A$ with the number of sites $k \in \{2,\ldots,7\}$, and respective dimensions ${d_A} = \{ {2^k} \}_{k = 2,\ldots,7}$. Figures~\ref{fig:FIG03}(a)-\ref{fig:FIG03}(f) show that the short-time expansion of ${S_L}({\widetilde{\rho}^A_t})$ reproduces its growth at early times. In each panel, we find that the relative error between the exact linear entropy [see Eq.~\eqref{eq:00000000013}] and its respective perturbative expansion [see Eqs.~\eqref{eq:00000000014},~\eqref{eq:00000000029},~\eqref{eq:00000000030}, and~\eqref{eq:00000000031}] is of order ${10^{-6}} \lesssim {\epsilon} \lesssim {10^{-3}}$ for $0.01 \lesssim Jt \lesssim 0.1$. Nevertheless, for $Jt \gtrsim 0.1$, we have that the result in Eq.~\eqref{eq:00000000014} becomes loose and it is bounded from above by the exact linear entropy in Eq.~\eqref{eq:00000000013}, thus failing to predict the dynamics of ${S_L}({\widetilde{\rho}^A_t})$ at later times.

Next, Fig.~\ref{fig:FIG04} shows the short-time dynamics of the linear entropy ${S_L}({\widetilde{\rho}^A_t})$ in Eq.~\eqref{eq:00000000014}, for the initial state $\rho_0^{AB}$ with $p = 0.5$ [see Eq.~\eqref{eq:00000000027}]. The blue solid line corresponds to the case ${J_z}/J = 0.5$ (non-Hermitian Hamiltonian), while the red dashed line depicts the case ${J_z}/J = 0$ (Hermitian Hamiltonian). Overall, Figs.~\ref{fig:FIG04}(a)--\ref{fig:FIG04}(e) show that the linear entropy is a concave function for ${J_z}/J \neq 0$, while it is a convex function for ${J_z}/J = 0$. In turn, Fig.~\ref{fig:FIG04}(f) shows that ${S_L}({\widetilde{\rho}^A_t})$ stands as a concave function for ${J_z}/J \neq 0$, while for ${J_z}/J = 0$ it saturates to a fixed value for all $Jt > 0$. To see this, we first note that $1/{T_{1,nh}} = 0$ and $1/{T_{2,nh}} = 0$ for ${J_z} = 0$ [see Eqs.~\eqref{eq:00000000029} and~\eqref{eq:00000000031}, respectively], while one readily obtains that $1/{T_{2,h}} = 0$ as the subsystem $B$ has dimension $d_B = 2$ [see Eq.~\eqref{eq:00000000030}]. Hence, bearing in mind that $1/{T_{1,h}} = 0$, it follows that the linear entropy ${S_L}({\widetilde{\rho}^A_t}) \approx {S_L}({\rho^A_0})$ is time-independent in the short-time approximation, for ${J_z}/J = 0$.

In Fig.~\ref{fig:FIG05} we display the short-time dynamics of the li\-near entropy ${S_L}({\widetilde{\rho}^A_t})$ in Eq.~\eqref{eq:00000000014}, as a function of $Jt$, and consider the subsystem $A$ with $k = 5$ sites. We emphasize that each of the blue solid line corresponds to the case ${J_z}/J = 0.5$ (non-Hermitian Hamiltonian), and the red ones represent the case ${J_z}/J = 0$ (Hermitian Hamiltonian). We set the mixing parameters $p = 0.25$ [see Fig.~\ref{fig:FIG05}(a)], $p = 0.5$ [see Fig.~\ref{fig:FIG05}(b)], $p = 0.75$ [see Fig.~\ref{fig:FIG05}(c)], and $p = 1$ [see Fig.~\ref{fig:FIG05}(d)]. In Figs.~\ref{fig:FIG05}(a) and~\ref{fig:FIG05}(b), one finds that the li\-near entropy is concave whenever ${J_z}/J \neq 0$, while it turns into a convex function in the Hermitian limit with ${J_z}/J = 0$. In Fig.~\ref{fig:FIG05}(c), the linear entropy turn to be a convex function, which is due to the fact that $1/{T_{2,h}} + 1/{T_{2,nh}} < 0$ for $p = 0.75$. Figure~\ref{fig:FIG05}(d) shows that, for $p = 1$, the two linear entropies coincide regardless of the generator $H$. Indeed, we have seen from Eqs.~\eqref{eq:00000000029} and~\eqref{eq:00000000031} that $1/{T_{1,nh}} = 0$ and $1/{T_{2,nh}} = 0$ for initial pure states, respectively. In this case, given that $1/{T_{1,h}} = 0$ and ${d_B} = 8$, the onset growth of the linear entropy satisfies ${S_L}({\widetilde{\rho}^A_t}) \approx (16/31)(1 + 2 {\gamma^2}{J^2}{t^2})+ O({t^3})$ [see Eqs.~\eqref{eq:00000000014} and~\eqref{eq:00000000030}]. 

As a final remark, Figs.~\ref{fig:FIG04} and~\ref{fig:FIG05} show that the non-Hermitian Hamiltonian (${J_z}/J \neq 0$) enhances the short-time dynamics of ${S_L}({\widetilde{\rho}^A_t})$, which bounds from above the respective linear entropy for the Hermitian Hamiltonian (${J_z}/J = 0$). Last, Figs.~\ref{fig:FIG04} and~\ref{fig:FIG05} show a crossover behavior between both the non-Hermitian (${J_z}/J \neq 0$) and Hermitian (${J_z}/J = 0$) cases, but it should be noted that it occurs in a time window that extrapolates the validity of the short-time approximation. Indeed, Fig.~\ref{fig:FIG03} shows that our results find good agreement with the numerical simulation of ${S_L}({\widetilde{\rho}^A_t})$ in Eq.~\eqref{eq:00000000013} for $0 \leq Jt \lesssim 0.1$.


\section{Discussion and Conclusions}
 \label{sec:00000000004}

In this paper, we discuss the timescales related to the onset growth of linear entropy for finite-dimensional quantum systems described by effective non-Hermitian Hamiltonians. We investigate the short-time perturbative expansion of the linear entropy for a given input state driven by a general non-Hermitian Hamiltonian. We emphasize that our approach takes in account initial quantum states that can be either pure or mixed, possibly entangled or even uncorrelated states. Importantly, for bipartite quantum systems initialized in pure and uncorrelated states, our findings recover the results in Refs.~\cite{PhysRevLett.77.207,PhysRevD.97.066008} to the case of nonunitary reduced dynamics driven by Hermitian Hamiltonians.

We address the degree of mixedness of a quantum state that undergoes the nonunitary dynamics generated by an effective non-Hermitian Hamiltonian $H = {H_1} + i{H_2}$ [see Sec.~\ref{sec:00000000002}]. In this setting, Eq.~\eqref{eq:00000000003} stands for the short-time expansion of the linear entropy up to second order in time $t$, around $t = 0$, which in turn depends on the coefficients $1/{T_1}$ and $1/{T_2}$ in Eqs.~\eqref{eq:00000000004} and~\eqref{eq:00000000005}, respectively. Both quantities can be evaluated once the input state and the Hamiltonian $H$ have been specified. We emphasize that Eqs.~\eqref{eq:00000000004} and~\eqref{eq:00000000005} provide two competing timescales in the initial growth of the mixedness of the evolved state at earlier times of the dynamics. In particular, both coefficients vanish whenever the system is initialized in a pure state, regardless of the non-Hermitian part of the Hamiltonian. Moreover, in the Hermitian limit, we have that $1/{T_1} = 0$ and $1/{T_2} = 0$ independently of the initial state of the system. We note that, since the linear entropy defines a conserved quantity for Hermitian quantum systems, it can be proved that any of the coefficients in its perturbative expansion must vanish in this limiting case [see Eq.~\eqref{eq:00000000003}].

We specialize these results to the case of a dissipative non-Hermitian two-level system initialized in a mixed single-qubit state [see Sec.~\ref{sec:00000000002A}]. We find analytical expressions for the coefficients $1/T_1$ [see Eq.~\eqref{eq:00000000010}] and $1/T_2$ [see Eq.~\eqref{eq:00000000011}] in terms of the Bloch sphere parameters. In this case, we compare the exact linear entropy ${S_L}({\rho_t})$ with its aforementioned short-time expansion around $t = 0$. We find good quantitative agreement between these two quantities at earlier times of the dynamics. Of course, for later times one should include higher orders in the Taylor expansion to obtain tighter results for the mixedness of the evolved state.

Next, we investigate the reduced dynamics of composite systems described by non-Hermitian Hamiltonians [see Sec.~\ref{sec:00000000003}]. We derived the short-time perturbative expansion of the linear entropy ${S_L}({\widetilde{\rho}^A_t})$ for a given time-dependent marginal state of a bipartite system [see Eq.~\eqref{eq:00000000014}]. We found that, up to the second order in time $t$, the growth of the linear entropy is governed by the coefficients $1/{T_{1,h}}$ and $1/{T_{1,nh}}$ in Eqs.~\eqref{eq:00000000015} and~\eqref{eq:00000000016}, respectively, and also $1/{T_{2,h}}$ and $1/{T_{2,nh}}$ in Eqs.~\eqref{eq:00000000017} and~\eqref{eq:00000000018}, respectively. On the one hand, one gets that $1/{T_{1,h}}$ and $1/{T_{2,h}}$ depend on $H_1$ and the input state of the system. On the other hand, we have that $1/{T_{1,nh}}$ and $1/{T_{2,nh}}$ depend on $H_2$, thus being intrinsically related to the non-Hermitian features of the Hamiltonian. In the Hermitian limit, i.e., when one sets $H_2$ being a zero-valued operator, we find $1/{T_{1,nh}} = 0$ and $1/{T_{2,nh}} = 0$ for any bipartite system.

In particular, specifying an initial pure and uncorrelated state, we find the vanishing coefficients $1/{T_{1,h}} = 0$ and $1/{T_{1,nh}} = 0$, and the lowest order of the short-time perturbative expansion of the linear entropy ${S_L}({\widetilde{\rho}^A_t})$ depends on $1/{T_{2,h}^2}$ and $1/{T_{2,nh}^2}$ that are given in Eqs.~\eqref{eq:00000000023} and~\eqref{eq:00000000024}, respectively [see Sec.~\ref{sec:00000000003A}]. In the Hermitian limit, $1/{T_{2,nh}^2}$ identically vanishes, and $1/{T_{2,h}^2}$ recovers the perturbative expansion of the idempotency defect measuring the coherence losses for composite systems described by Hermitian Hamiltonians~\cite{PhysRevLett.77.207}. In this setting, we see that $|1/{T_{2,h}}|$ signals the entanglement timescale for quantum systems described by Hermitian Hamiltonians~\cite{PhysRevD.97.066008}. It is noteworthy that this result is also related to the timescale that governs the growth of entanglement for R\'{e}nyi entropies~\cite{PhysRevA.97.022317}.

To illustrate these findings, we investigated the linear entropy of the $k$-particle evolved marginal state for a quantum many-body system described by the transverse-field $XY$ model coupled to the imaginary fully connected Ising Hamiltonian [see Sec.~\ref{sec:00000000003B}]. We found analytical expressions for $1/{T_{1,nh}}$ and $1/{T_{2,nh}^2}$, which in turn scale linearly with the coupling strength of the all-to-all Ising Hamiltonian [see Eqs.~\eqref{eq:00000000029} and~\eqref{eq:00000000031}, respectively]. In addition, it follows that $1/{T_{1,h}}$ vanishes, while $1/{T_{2,h}^2}$ depends on the anisotropy para\-me\-ter of the $XY$ model [see Eq.~\eqref{eq:00000000030}]. We compared the short-time expansion of the linear entropy with its exact numerical simulation [see Fig.~\ref{fig:FIG03}], and discussed its dynamical behavior in both cases of non-Hermitian and Hermitian Hamiltonians [see Figs.~\ref{fig:FIG04} and~\ref{fig:FIG05}]. We find that non-Hermiticity enhances the short-time dynamics of the linear entropy, pro\-vi\-ding an upper bound for the respective linear entropy for Hermitian Hamiltonian.

Our findings provide insightful qualitative and quantitative information about the initial growth of linear entropy at early times. Importantly, the results require low computational cost and their evaluation involves minimal ingredients as the initial state and the non-Hermitian Hamiltonian that governs the nonunitary dynamics. This might be of interest for higher dimensional systems, where evaluating the linear entropy would require the full spectral decomposition of the evolved system. We point out that one could generalize the present discussion in terms of $\alpha$-R\'{e}nyi entropies~\cite{PhysRevA.97.022317}. Furthermore, one can investigate the interplay of the aforementioned timescales and the quantum speed limit for nonunitary evolutions generated by non-Hermitian Hamiltonians~\cite{PhysRevA.106.012403,PhysRevA.104.052223}. We hope to address these questions in further investigations. The results in this paper could find applications in the subjects of non-Hermitian quantum sensing~\cite{Lau2018,PhysRevA.103.042418}, quantum thermodynamics of non-Hermitian systems~\cite{deffner_scirep_23408_2016}, non-Hermitian long-range interacting quantum systems~\cite{arXiv:2109.01063}, and $\mathcal{PT}$-symmetric quantum field theory~\cite{Bender_2020}.


\begin{acknowledgments}
This work was supported by the Brazilian ministries MEC and MCTIC, and the Brazilian funding agencies CNPq, and Coordena\c{c}\~{a}o de Aperfei\c{c}oamento de Pessoal de N\'{i}vel Superior--Brasil (CAPES) (Finance Code 001). D. P. P. acknowledges Funda\c{c}\~{a}o de Amparo \`{a} Pesquisa e ao Desenvolvimento Cient\'{i}fico e Tecnol\'{o}gico do Maranh\~{a}o (FAPEMA). T. M. acknowledges the hospitality of ITAMP-Harvard where part of this work was done. T. M. also acknowledges support from CAPES. This work was supported by the Serrapilheira Institute (Grant No. Serra-1812-27802).
\end{acknowledgments}

\setcounter{equation}{0}
\setcounter{table}{0}
\setcounter{section}{0}
\numberwithin{equation}{section}
\makeatletter
\renewcommand{\thesection}{\Alph{section}} 
\renewcommand{\thesubsection}{\Alph{section}.\arabic{subsection}}
\def\@gobbleappendixname#1\csname thesubsection\endcsname{\Alph{section}.\arabic{subsection}}
\renewcommand{\theequation}{\Alph{section}\arabic{equation}}
\renewcommand{\thefigure}{\arabic{figure}}
\renewcommand{\bibnumfmt}[1]{[#1]}
\renewcommand{\citenumfont}[1]{#1}

\section*{Appendix}


\section{Mixedness for the nonunitary dynamics of initial pure states}
 \label{sec:00000000005}

In this Appendix, we discuss the mixedness timescales for a given initial pure state whose dynamics is governed by an effective non-Hermitian Hamiltonian. Let us consider a finite-dimensional quantum system initialized in the pure state $\rho_0 = |{\psi_0}\rangle\langle{\psi_0}|$, with $\langle{\psi_0}|{\psi_0}\rangle = 1$. In turn, the initial pure state undergoes the nonunitary dynamics generated by a time-independent non-Hermitian Hamiltonian $H = {H_1} + i{H_2}$, with ${H_1^{\dagger}} = {H_1}$ and ${H_2^{\dagger}} = {H_2}$ being Hermitian operators. In this case, the time-dependent normalized density matrix of the system read as
\begin{equation}
\label{eq:00000000032}
{\widetilde{\rho}_t} = \frac{{U_t}{\rho_0}{U_t^{\dagger}}}{\text{Tr}({U_t}{\rho_0}{U_t^{\dagger}})} = \frac{{U_t}|{\psi_0}\rangle\langle{\psi_0}|{U_t^{\dagger}}}{\langle{\psi_0}|{U_t^{\dagger}}{U_t}|{\psi_0}\rangle} ~,
\end{equation}
with ${U_t} = {e^{- i t H}}$ being the nonunitary evolution operator. It can be proved that the normalized state ${\widetilde{\rho}_t}$ in Eq.~\eqref{eq:00000000032} fulfills the differential equation ${d{\widetilde{\rho}_t}}/{dt} = -i [{H_1},{\widetilde{\rho}_t}] + \{{H_2},{\widetilde{\rho}_t} \} - 2 \text{Tr}({\widetilde{\rho}_t} {H_2})\,{\widetilde{\rho}_t}$, which in turn describes a completely positive and trace preserving evolution. In this case, the purity $f({\widetilde{\rho}_t}) = \text{Tr}({\widetilde{\rho}_t^2})$ of the evolved state thus yields
\begin{equation}
\label{eq:00000000033}
f({\widetilde{\rho}_t}) = \frac{\text{Tr}({U_t}|{\psi_0}\rangle\langle{\psi_0}|{U_t^{\dagger}}{U_t}|{\psi_0}\rangle\langle{\psi_0}|{U_t^{\dagger}})}{\langle{\psi_0}|{U_t^{\dagger}}{U_t}|{\psi_0}\rangle^2} = 1 ~,
\end{equation}
where we have used the cyclic property of the trace. Equation~\eqref{eq:00000000033} shows that, for an initial pure state $\rho_0 = |{\psi_0}\rangle\langle{\psi_0}|$, the purity of the evolved normalized state $\widetilde{\rho}_t$ will remain constant, i.e., such state remains pure for all $t \geq 0$. The linear entropy identically vanishes, i.e., ${S_L}({\widetilde{\rho}_t}) = [{d}/({d - 1})] [1 - f({\widetilde{\rho}_t})] = 0$. Hence, for all nonzero positive integer $k \in \mathbb{Z}^+$, the $k$th-order time derivative of both the purity and linear entropy will vanish, i.e., one gets ${d^k}{f}({\widetilde{\rho}_t})/d{t^k} = 0$ and ${d^k}{S_L}({\widetilde{\rho}_t})/d{t^k} = 0$. This result proves that, given an initial pure state undergoing the nonunitary dynamics generated by a non-Hermitian Hamiltonian, the mixedness timescales vanish for any perturbative order within the short-time approximation of the linear entropy. See also Refs.~\cite{PhysRevLett.109.230405,JStatMech_033102_2016,Entropy_18120451,doi:10.1142,EPJD_253_69_2015}.

Next, we show that the same result can obtained when considering the so-called ``metric approach'' for non-Hermitian systems, whose main idea relies on modifying the inner-product structure of the Hilbert space~\cite{Bender2007,PhysScr_82_038110}. Indeed, the Hilbert space is endowed with an inner-product related to the time-dependent operator $G_t$ called ``metric''~\cite{Brody_2013}. Let $H \neq H^{\dagger}$ be the effective time-independent non-Hermitian Hamiltonian governing the nonunitary dynamics of a finite-dimensional quantum system. In this setting, given the initial pure state $\rho_0 = |{\psi_0}\rangle\langle{\psi_0}|$, one gets that the evolved state $|{\psi_t}\rangle = {U_t}|{\psi_0}\rangle = {e^{-i t H}}|{\psi_0}\rangle$ is not properly normalized, i.e., its squared norm $\langle{\psi_t}|{\psi_t}\rangle = \langle{\psi_0}|{U_t}{U_t^{\dagger}}|{\psi_0}\rangle$ stands as a time-dependent quantity (${U_t^{\dagger}} \neq {U_t^{-1}}$). To overcome this issue, one introduces the modified dual vector $({\psi_t}| := \langle{\psi_t}|{G_t}$ that is obtained replacing the conventional Hermitian conjugate, where $G_t$ is a time-dependent, Hermitian, and positive definite operator called metric~\cite{Brody_2013}. In this approach, the inner product $({\psi_t}|{\psi_t}\rangle$ is required to be time-independent, i.e., its time derivative is expected to vanish as $d({\psi_t}|{\psi_t}\rangle/dt = d\langle{\psi_t}| {G_t} |{\psi_t}\rangle/dt = 0$, for all $t \geq 0$. In turn, this constraint implies that the metric operator $G_t$ fulfills the differential equation 
\begin{equation}
\frac{d {G_t}}{d t} = i({G_t}H - {H^\dagger}{G_t}) ~. 
\end{equation}
We point out that, even though $|{\psi_t\rangle}$ exhibits a time-independent squared norm, the vector $|{\psi_t\rangle}$ is no longer normalized to the unity. This motivates to recast the evolved dual state as ${({\psi_t}|{\psi_t}\rangle^{-1}} ({\psi_t}|$, and thus the time-dependent density matrix of the system yields
\begin{equation}
\label{eq:L0000000000000}
{\rho_t} = \frac{|{\psi_t}\rangle({\psi_t}|}{({\psi_t}|{\psi_t}\rangle} = \frac{|{\psi_t}\rangle\langle{\psi_t}|{G_t}}{\langle{\psi_t}|{G_t}|{\psi_t}\rangle} ~.
\end{equation}
In this setting, it can be seen that the purity $f({\rho_t}) = \text{Tr}({\rho_t^2})$ of the evolved state will remain constant and equal to the unity for all $t \geq 0$,
\begin{equation}
\label{eq:L0000000000000A}
f({\rho_t}) = \frac{ \text{Tr}(|{\psi_t}\rangle\langle{\psi_t}|{G_t}|{\psi_t}\rangle\langle{\psi_t}|{G_t})}{\langle{\psi_t}|{G_t}|{\psi_t}\rangle^2} = 1 ~,
\end{equation}
where we have applied the cyclic property of the trace. Equation~\eqref{eq:L0000000000000A} implies that, for any initial pure state, the linear entropy of the evolved state becomes zero, i.e., ${S_L}({\rho_t}) = [{d}/({d - 1})] [1 - f({\rho_t})] = 0$, for all $t \geq 0$. Therefore, given an initial pure state, Eqs.~\eqref{eq:00000000033} and~\eqref{eq:L0000000000000A} show that the purity of its evolved state must be the same regardless of the theoretical framework that was applied to address the nonunitary dynamics generated by a non-Hermitian Hamiltonian.


\section{Timescales for the dissipative two-level system}
 \label{sec:00000000006}
 
In this Appendix, we compare the timescales $1/(\gamma{T_1})$ [see Eq.~\eqref{eq:00000000010}] and $1/({\gamma^2}{T_2^2})$ [see Eq.~\eqref{eq:00000000011}] with both the first-order $-[d{S_L}(\bullet)/dt]_{t = 0}$ and second-order $-[{d^2}{S_L}(\bullet)/d{t^2}]_{t = 0}$ time derivatives around $t = 0$ of the linear entropy, for the dissipative two-level system discussed in Sec.~\ref{sec:00000000002A}. We remind that the system is described by the Hamiltonian $H = \Delta{|1\rangle}{\langle{1}|} + (\Omega/2)(|{0}\rangle\langle{1}| + |{1}\rangle\langle{0}|)$, and one sets the probe single-qubit state $\rho_0 = (1/2)(\mathbb{I} + \vec{r}\cdot\vec{\sigma})$. Hereafter, we choose $\Delta/\gamma = 0.5$, and also set $\{\theta, \phi\} = \{ 3\pi/4, \pi/4 \}$. The overall dynamics of the dissipative systems is governed by the Markovian master equation in Eq.~\eqref{eq:00000000009}. It is noteworthy that by neglecting the effect of quantum jumps, the nonunitary dynamics of the system is recast in terms of the effective non-Hermitian Hamiltonian ${H_{\text{eff}}} = {H_1} + i{H_2}$, with $H_1 = H$ and ${H_2} = - (\gamma/2){|1\rangle}{\langle{1}|}$.

\begin{figure}[!t]
\begin{center}
\includegraphics[scale=1.]{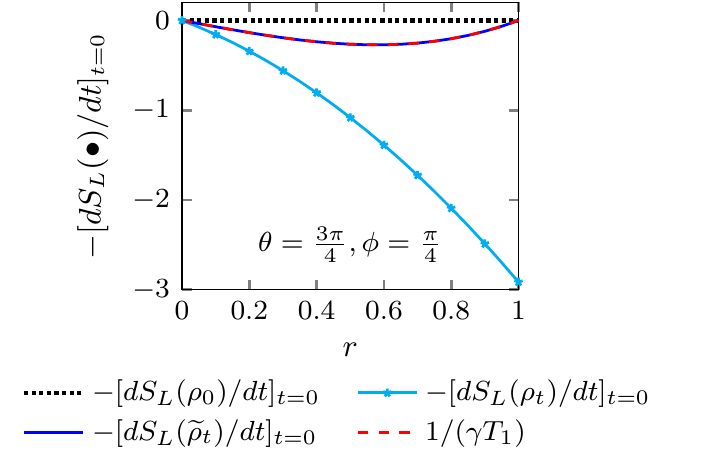}
\caption{(Color online) Plot of the first-order time derivative $-[d{S_L}(\bullet)/dt]_{t = 0}$ around $t = 0$ of the linear entropy, and the timescale $1/\gamma{T_1}$, as a function of the Bloch sphere radius $r$, for the driven two-level system described by the Hamiltonian $H = \Delta{|1\rangle}{\langle{1}|} + (\Omega/2)(|{0}\rangle\langle{1}| + |{1}\rangle\langle{0}|)$. Here we choose the ratio $\Delta/\gamma = 0.5$, and $\Omega/\gamma = 0.1$. The system is initialized in the single-qubit state $\rho_0 = (1/2)(\mathbb{I} + \vec{r}\cdot\vec{\sigma})$, with $\{\theta, \phi\} = \{ 3\pi/4, \pi/4 \}$. The black dotted line indicates $- [d{S_L}({\rho_0})/dt]_{t = 0}$, with ${S_L}({e^{- i t H}}{\rho_0}{e^{+ i t H}}) = {S_L}(\rho_0)$. The cyan dash-dotted line depicts the quantity $-[d{S_L}({\rho_t})/dt]_{t = 0}$, with the state $\rho_t$ evolving under the Markovian master equation in Eq.~\eqref{eq:00000000009}. The blue solid line indicates the first-order time-derivative $- [d{S_L}({\widetilde{\rho}_t})/dt]_{t = 0}$, with the normalized state ${\widetilde{\rho}_t} = {\rho_t}/\text{Tr}({\rho_t})$ satisfying Eq.~\eqref{eq:00000000002}, where ${H_1} = H$ and $H_2 = - (\gamma/2){|1\rangle}{\langle{1}|}$. The red dashed line displays the timescale $1/(\gamma {T_1})$ [see Eq.~\eqref{eq:00000000010}].}
\label{fig:FIG06}
\end{center}
\end{figure}

In Fig.~\ref{fig:FIG06}, we show plots of $-[d{S_L}(\bullet)/dt]_{t = 0}$ and $1/(\gamma{T_1})$, as a function of the Bloch sphere radius $r$. We set $\Omega/\gamma = 0.1$, but we find that the plots are not sensitive to changes in the ratio $\Omega/\gamma$ [see also Fig.~\ref{fig:FIG01}(a)]. The black dotted line indicates the first-order time derivative $-{[d{S_L}({\rho_0})/dt]_{t = 0}}$. Note that this quantity is equal to zero, since the linear entropy ${S_L}({e^{- i t H}}{\rho_0}{e^{+ i t H}}) = {S_L}(\rho_0)$ is time-independent for a unitary evolution generated by the Hermitian Hamiltonian $H$. The cyan dash-dotted line displays the first-order time derivative $-[d{S_L}({\rho_t})/dt]_{t = 0}$, where $\rho_t$ satisfies the Markovian master equation in Eq.~\eqref{eq:00000000009}. The blue solid line depicts the quantity $- [d{S_L}({\widetilde{\rho}_t})/dt]_{t = 0}$, where ${\widetilde{\rho}_t}$ satisfies Eq.~\eqref{eq:00000000002}. The dashed red line displays the timescale $1/(\gamma{T_1})$ in Eq.~\eqref{eq:00000000010}. The quantities $- [d{S_L}({\widetilde{\rho}_t})/dt]_{t = 0}$ and $1/(\gamma{T_1})$ coincide each other.

\begin{figure*}[!t]
\begin{center}
\includegraphics[scale=1.0]{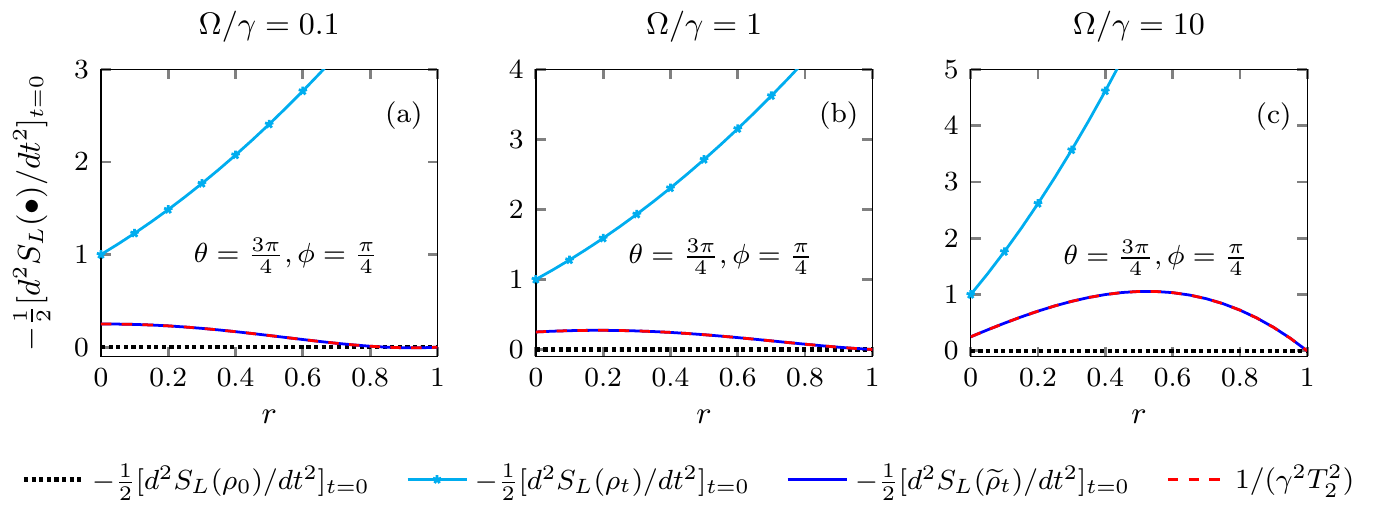}
\caption{(Color online) Plot of the second-order time derivative $-(1/2)[{d^2}{S_L}(\bullet)/d{t^2}]_{t = 0}$ around $t = 0$ of the linear entropy, and the timescale $1/{\gamma^2}{T_2^2}$, as a function of the radius $r$ of the Bloch sphere, for the driven two-level system described by the Hamiltonian $H = \Delta{|1\rangle}{\langle{1}|} + (\Omega/2)(|{0}\rangle\langle{1}| + |{1}\rangle\langle{0}|)$. The system is initialized in the single-qubit state $\rho_0 = (1/2)(\mathbb{I} + \vec{r}\cdot\vec{\sigma})$, with $\{\theta, \phi\} = \{ 3\pi/4, \pi/4 \}$. We set $\Omega/\gamma = 0.1$ (a), $\Omega/\gamma = 1$ (b), and $\Omega/\gamma = 10$ (c). The black dotted line shows the second-order time derivative $-(1/2)[{d^2}{S_L}({\rho_0})/d{t^2}]_{t = 0}$. The cyan dash-dotted line depicts the quantity $-(1/2)[{d^2}{S_L}({\rho_t})/d{t^2}]_{t = 0}$, where the state $\rho_t$ fulfills Eq.~\eqref{eq:00000000009}. The blue solid line indicates the quantity $-(1/2)[{d^2}{S_L}({\widetilde{\rho}_t})/d{t^2}]_{t = 0}$, with the normalized state ${\widetilde{\rho}_t} = {\rho_t}/\text{Tr}({\rho_t})$ satisfying Eq.~\eqref{eq:00000000002}, where ${H_1} = H$ and $H_2 = - (\gamma/2){|1\rangle}{\langle{1}|}$. The red dashed line displays the timescale $1/({\gamma^2} {T_2^2})$ [see Eq.~\eqref{eq:00000000011}].}
\label{fig:FIG07}
\end{center}
\end{figure*}

In Fig.~\ref{fig:FIG07}, we show plots of $-(1/2)[{d^2}{S_L}(\bullet)/d{t^2}]_{t = 0}$ and $1/({\gamma^2}{T_2^2})$, as a function of the Bloch sphere radius $r$. We set $\Omega/\gamma = 0.1$ [see Fig.~\ref{fig:FIG07}(a)], $\Omega/\gamma = 1$ [see Fig.~\ref{fig:FIG07}(b)], and $\Omega/\gamma = 10$ [see Fig.~\ref{fig:FIG07}(c)]. The black dotted line indicates the quantity $-[{d^2}{S_L}({\rho_0})/d{t^2}]_{t = 0}$, which in turn vanishes for the probe state undergoing the unitary evolution generated by $H$. The cyan dash-dotted line shows the quantity $-(1/2)[{d^2}{S_L}({\rho_t})/d{t^2}]_{t = 0}$, with $\rho_t$ satisfying Eq.~\eqref{eq:00000000009}. The blue solid line displays the second-order time derivative $- (1/2)[{d^2}{S_L}({\widetilde{\rho}_t})/d{t^2}]_{t = 0}$, where ${\widetilde{\rho}_t}$ fulfills Eq.~\eqref{eq:00000000002}. The dashed red line represents the timescale $1/({\gamma^2}{T_2^2})$. Note that the quantities $-[{d^2}{S_L}({\widetilde{\rho}_t})/d{t^2}]_{t = 0}$ and $1/({\gamma^2}{T_2^2})$ agree with each other.

In the following, we comment on the differences in the plots of Figs.~\ref{fig:FIG06} and~\ref{fig:FIG07}. On the one hand, the quantities $-[d{S_L}({\rho_t})/dt]_{t = 0}$ and $-(1/2)[{d^2}{S_L}({\rho_t})/d{t^2}]_{t = 0}$ are related to the linear entropy of state $\rho_t$ whose dy\-na\-mics is governed by Eq.~\eqref{eq:00000000009}. On the other hand, note that $-{[d{S_L}({\widetilde{\rho}_t})/dt]_{t = 0}}$ and $- (1/2){[{d^2}{S_L}({\widetilde{\rho}_t})/d{t^2}]_{t = 0}}$ depend on the linear entropy of the normalized state $\widetilde{\rho}_t$ that undergoes the nonunitary effective dynamics generated by the non-Hermitian Hamiltonian ${H_{\text{eff}}} = {H_1} + i{H_2}$, discarding quantum jumps [see Eq.~\eqref{eq:00000000002}]. As expected, these last two time derivatives agree with the timescales $1/({\gamma}{T_1})$ and $1/({\gamma^2}{T_2^2})$, respectively.

\begin{thebibliography}{121}%
\makeatletter
\providecommand \@ifxundefined [1]{%
 \@ifx{#1\undefined}
}%
\providecommand \@ifnum [1]{%
 \ifnum #1\expandafter \@firstoftwo
 \else \expandafter \@secondoftwo
 \fi
}%
\providecommand \@ifx [1]{%
 \ifx #1\expandafter \@firstoftwo
 \else \expandafter \@secondoftwo
 \fi
}%
\providecommand \natexlab [1]{#1}%
\providecommand \enquote  [1]{``#1''}%
\providecommand \bibnamefont  [1]{#1}%
\providecommand \bibfnamefont [1]{#1}%
\providecommand \citenamefont [1]{#1}%
\providecommand \href@noop [0]{\@secondoftwo}%
\providecommand \href [0]{\begingroup \@sanitize@url \@href}%
\providecommand \@href[1]{\@@startlink{#1}\@@href}%
\providecommand \@@href[1]{\endgroup#1\@@endlink}%
\providecommand \@sanitize@url [0]{\catcode `\\12\catcode `\$12\catcode
  `\&12\catcode `\#12\catcode `\^12\catcode `\_12\catcode `\%12\relax}%
\providecommand \@@startlink[1]{}%
\providecommand \@@endlink[0]{}%
\providecommand \url  [0]{\begingroup\@sanitize@url \@url }%
\providecommand \@url [1]{\endgroup\@href {#1}{\urlprefix }}%
\providecommand \urlprefix  [0]{URL }%
\providecommand \Eprint [0]{\href }%
\providecommand \doibase [0]{http://dx.doi.org/}%
\providecommand \selectlanguage [0]{\@gobble}%
\providecommand \bibinfo  [0]{\@secondoftwo}%
\providecommand \bibfield  [0]{\@secondoftwo}%
\providecommand \translation [1]{[#1]}%
\providecommand \BibitemOpen [0]{}%
\providecommand \bibitemStop [0]{}%
\providecommand \bibitemNoStop [0]{.\EOS\space}%
\providecommand \EOS [0]{\spacefactor3000\relax}%
\providecommand \BibitemShut  [1]{\csname bibitem#1\endcsname}%
\let\auto@bib@innerbib\@empty
\bibitem [{\citenamefont {Brody}\ and\ \citenamefont
  {Graefe}(2012)}]{PhysRevLett.109.230405}%
  \BibitemOpen
  \bibfield  {author} {\bibinfo {author} {\bibfnamefont {D.~C.}\ \bibnamefont
  {Brody}}\ and\ \bibinfo {author} {\bibfnamefont {E.-M.}\ \bibnamefont
  {Graefe}},\ }\bibfield  {title} {\enquote {\bibinfo {title} {{M}ixed-{S}tate
  {E}volution in the {P}resence of {G}ain and {L}oss},}\ }\href {\doibase
  10.1103/PhysRevLett.109.230405} {\bibfield  {journal} {\bibinfo  {journal}
  {Phys. Rev. Lett.}\ }\textbf {\bibinfo {volume} {109}},\ \bibinfo {pages}
  {230405} (\bibinfo {year} {2012})}\BibitemShut {NoStop}%
\bibitem [{\citenamefont {Roccati}\ \emph {et~al.}(2022)\citenamefont
  {Roccati}, \citenamefont {Palma}, \citenamefont {Ciccarello},\ and\
  \citenamefont {Bagarello}}]{arXiv:2201.05367}%
  \BibitemOpen
  \bibfield  {author} {\bibinfo {author} {\bibfnamefont {F.}~\bibnamefont
  {Roccati}}, \bibinfo {author} {\bibfnamefont {G.~M.}\ \bibnamefont {Palma}},
  \bibinfo {author} {\bibfnamefont {F.}~\bibnamefont {Ciccarello}}, \ and\
  \bibinfo {author} {\bibfnamefont {F.}~\bibnamefont {Bagarello}},\ }\bibfield
  {title} {\enquote {\bibinfo {title} {Non-{H}ermitian {P}hysics and {M}aster
  {E}quations},}\ }\href {\doibase 10.1142/S1230161222500044} {\bibfield
  {journal} {\bibinfo  {journal} {Open Syst. Inf. Dyn.}\ }\textbf {\bibinfo
  {volume} {29}},\ \bibinfo {pages} {2250004} (\bibinfo {year}
  {2022})}\BibitemShut {NoStop}%
\bibitem [{\citenamefont {Yu}\ \emph {et~al.}(2020)\citenamefont {Yu},
  \citenamefont {Meng}, \citenamefont {Tang}, \citenamefont {Xu}, \citenamefont
  {Wang}, \citenamefont {Yin}, \citenamefont {Ke}, \citenamefont {Liu},
  \citenamefont {Li}, \citenamefont {Yang}, \citenamefont {Chen}, \citenamefont
  {Han}, \citenamefont {Li},\ and\ \citenamefont
  {Guo}}]{PhysRevLett.125.240506}%
  \BibitemOpen
  \bibfield  {author} {\bibinfo {author} {\bibfnamefont {S.}~\bibnamefont
  {Yu}}, \bibinfo {author} {\bibfnamefont {Y.}~\bibnamefont {Meng}}, \bibinfo
  {author} {\bibfnamefont {J.-S.}\ \bibnamefont {Tang}}, \bibinfo {author}
  {\bibfnamefont {X.-Y.}\ \bibnamefont {Xu}}, \bibinfo {author} {\bibfnamefont
  {Y.-T.}\ \bibnamefont {Wang}}, \bibinfo {author} {\bibfnamefont
  {P.}~\bibnamefont {Yin}}, \bibinfo {author} {\bibfnamefont {Z.-J.}\
  \bibnamefont {Ke}}, \bibinfo {author} {\bibfnamefont {W.}~\bibnamefont
  {Liu}}, \bibinfo {author} {\bibfnamefont {Z.-P.}\ \bibnamefont {Li}},
  \bibinfo {author} {\bibfnamefont {Y.-Z.}\ \bibnamefont {Yang}}, \bibinfo
  {author} {\bibfnamefont {G.}~\bibnamefont {Chen}}, \bibinfo {author}
  {\bibfnamefont {Y.-J.}\ \bibnamefont {Han}}, \bibinfo {author} {\bibfnamefont
  {C.-F.}\ \bibnamefont {Li}}, \ and\ \bibinfo {author} {\bibfnamefont {G.-C.}\
  \bibnamefont {Guo}},\ }\bibfield  {title} {\enquote {\bibinfo {title}
  {Experimental {I}nvestigation of {Q}uantum
  $\mathcal{P}\mathcal{T}$-{E}nhanced {S}ensor},}\ }\href {\doibase
  10.1103/PhysRevLett.125.240506} {\bibfield  {journal} {\bibinfo  {journal}
  {Phys. Rev. Lett.}\ }\textbf {\bibinfo {volume} {125}},\ \bibinfo {pages}
  {240506} (\bibinfo {year} {2020})}\BibitemShut {NoStop}%
\bibitem [{\citenamefont {Wang}\ \emph {et~al.}(2021)\citenamefont {Wang},
  \citenamefont {Xiao}, \citenamefont {Budich}, \citenamefont {Yi},\ and\
  \citenamefont {Xue}}]{PhysRevLett.127.026404}%
  \BibitemOpen
  \bibfield  {author} {\bibinfo {author} {\bibfnamefont {K.}~\bibnamefont
  {Wang}}, \bibinfo {author} {\bibfnamefont {L.}~\bibnamefont {Xiao}}, \bibinfo
  {author} {\bibfnamefont {J.~C.}\ \bibnamefont {Budich}}, \bibinfo {author}
  {\bibfnamefont {W.}~\bibnamefont {Yi}}, \ and\ \bibinfo {author}
  {\bibfnamefont {P.}~\bibnamefont {Xue}},\ }\bibfield  {title} {\enquote
  {\bibinfo {title} {Simulating {E}xceptional {N}on-{H}ermitian {M}etals with
  {S}ingle-{P}hoton {I}nterferometry},}\ }\href {\doibase
  10.1103/PhysRevLett.127.026404} {\bibfield  {journal} {\bibinfo  {journal}
  {Phys. Rev. Lett.}\ }\textbf {\bibinfo {volume} {127}},\ \bibinfo {pages}
  {026404} (\bibinfo {year} {2021})}\BibitemShut {NoStop}%
\bibitem [{\citenamefont {Zyablovsky}\ \emph {et~al.}(2016)\citenamefont
  {Zyablovsky}, \citenamefont {Andrianov},\ and\ \citenamefont
  {Pukhov}}]{Zyablovsky_2016}%
  \BibitemOpen
  \bibfield  {author} {\bibinfo {author} {\bibfnamefont {A.~A.}\ \bibnamefont
  {Zyablovsky}}, \bibinfo {author} {\bibfnamefont {E.~S.}\ \bibnamefont
  {Andrianov}}, \ and\ \bibinfo {author} {\bibfnamefont {A.~A.}\ \bibnamefont
  {Pukhov}},\ }\bibfield  {title} {\enquote {\bibinfo {title} {Parametric
  instability of optical non-{H}ermitian systems near the exceptional point},}\
  }\href {\doibase 10.1038/srep29709} {\bibfield  {journal} {\bibinfo
  {journal} {Sci. Rep.}\ }\textbf {\bibinfo {volume} {6}},\ \bibinfo {pages}
  {29709} (\bibinfo {year} {2016})}\BibitemShut {NoStop}%
\bibitem [{\citenamefont {Kawabata}\ \emph {et~al.}(2017)\citenamefont
  {Kawabata}, \citenamefont {Ashida},\ and\ \citenamefont
  {Ueda}}]{PhysRevLett.119.190401}%
  \BibitemOpen
  \bibfield  {author} {\bibinfo {author} {\bibfnamefont {K.}~\bibnamefont
  {Kawabata}}, \bibinfo {author} {\bibfnamefont {Y.}~\bibnamefont {Ashida}}, \
  and\ \bibinfo {author} {\bibfnamefont {M.}~\bibnamefont {Ueda}},\ }\bibfield
  {title} {\enquote {\bibinfo {title} {Information {R}etrieval and
  {C}riticality in {P}arity-{T}ime-{S}ymmetric {S}ystems},}\ }\href {\doibase
  10.1103/PhysRevLett.119.190401} {\bibfield  {journal} {\bibinfo  {journal}
  {Phys. Rev. Lett.}\ }\textbf {\bibinfo {volume} {119}},\ \bibinfo {pages}
  {190401} (\bibinfo {year} {2017})}\BibitemShut {NoStop}%
\bibitem [{\citenamefont {Delplace}\ \emph {et~al.}(2021)\citenamefont
  {Delplace}, \citenamefont {Yoshida},\ and\ \citenamefont
  {Hatsugai}}]{PhysRevLett.127.186602}%
  \BibitemOpen
  \bibfield  {author} {\bibinfo {author} {\bibfnamefont {P.}~\bibnamefont
  {Delplace}}, \bibinfo {author} {\bibfnamefont {T.}~\bibnamefont {Yoshida}}, \
  and\ \bibinfo {author} {\bibfnamefont {Y.}~\bibnamefont {Hatsugai}},\
  }\bibfield  {title} {\enquote {\bibinfo {title} {Symmetry-{P}rotected
  {M}ultifold {E}xceptional {P}oints and {T}heir {T}opological
  {C}haracterization},}\ }\href {\doibase 10.1103/PhysRevLett.127.186602}
  {\bibfield  {journal} {\bibinfo  {journal} {Phys. Rev. Lett.}\ }\textbf
  {\bibinfo {volume} {127}},\ \bibinfo {pages} {186602} (\bibinfo {year}
  {2021})}\BibitemShut {NoStop}%
\bibitem [{\citenamefont {Ashida}\ \emph {et~al.}(2020)\citenamefont {Ashida},
  \citenamefont {Gong},\ and\ \citenamefont {Ueda}}]{Yuto_Ashida2020}%
  \BibitemOpen
  \bibfield  {author} {\bibinfo {author} {\bibfnamefont {Y.}~\bibnamefont
  {Ashida}}, \bibinfo {author} {\bibfnamefont {Z.}~\bibnamefont {Gong}}, \ and\
  \bibinfo {author} {\bibfnamefont {M.}~\bibnamefont {Ueda}},\ }\bibfield
  {title} {\enquote {\bibinfo {title} {Non-{H}ermitian physics},}\ }\href
  {\doibase 10.1080/00018732.2021.1876991} {\bibfield  {journal} {\bibinfo
  {journal} {Adv. Phys.}\ }\textbf {\bibinfo {volume} {69}},\ \bibinfo {pages}
  {249} (\bibinfo {year} {2020})}\BibitemShut {NoStop}%
\bibitem [{\citenamefont {Bergholtz}\ \emph {et~al.}(2021)\citenamefont
  {Bergholtz}, \citenamefont {Budich},\ and\ \citenamefont
  {Kunst}}]{Bergholtz2021}%
  \BibitemOpen
  \bibfield  {author} {\bibinfo {author} {\bibfnamefont {E.~J.}\ \bibnamefont
  {Bergholtz}}, \bibinfo {author} {\bibfnamefont {J.~C.}\ \bibnamefont
  {Budich}}, \ and\ \bibinfo {author} {\bibfnamefont {F.~K.}\ \bibnamefont
  {Kunst}},\ }\bibfield  {title} {\enquote {\bibinfo {title} {Exceptional
  topology of non-{H}ermitian systems},}\ }\href {\doibase
  10.1103/RevModPhys.93.015005} {\bibfield  {journal} {\bibinfo  {journal}
  {Rev. Mod. Phys.}\ }\textbf {\bibinfo {volume} {93}},\ \bibinfo {pages}
  {015005} (\bibinfo {year} {2021})}\BibitemShut {NoStop}%
\bibitem [{\citenamefont {Shackleton}\ and\ \citenamefont
  {Scheurer}(2020)}]{PhysRevResearch.2.033022}%
  \BibitemOpen
  \bibfield  {author} {\bibinfo {author} {\bibfnamefont {H.}~\bibnamefont
  {Shackleton}}\ and\ \bibinfo {author} {\bibfnamefont {M.~S.}\ \bibnamefont
  {Scheurer}},\ }\bibfield  {title} {\enquote {\bibinfo {title} {Protection of
  parity-time symmetry in topological many-body systems: {N}on-{H}ermitian
  toric code and fracton models},}\ }\href {\doibase
  10.1103/PhysRevResearch.2.033022} {\bibfield  {journal} {\bibinfo  {journal}
  {Phys. Rev. Research}\ }\textbf {\bibinfo {volume} {2}},\ \bibinfo {pages}
  {033022} (\bibinfo {year} {2020})}\BibitemShut {NoStop}%
\bibitem [{\citenamefont {Tzeng}\ \emph {et~al.}(2021)\citenamefont {Tzeng},
  \citenamefont {Ju}, \citenamefont {Chen},\ and\ \citenamefont
  {Huang}}]{PhysRevResearch.3.013015}%
  \BibitemOpen
  \bibfield  {author} {\bibinfo {author} {\bibfnamefont {Y.-C.}\ \bibnamefont
  {Tzeng}}, \bibinfo {author} {\bibfnamefont {C.-Y.}\ \bibnamefont {Ju}},
  \bibinfo {author} {\bibfnamefont {G.-Y.}\ \bibnamefont {Chen}}, \ and\
  \bibinfo {author} {\bibfnamefont {W.-M.}\ \bibnamefont {Huang}},\ }\bibfield
  {title} {\enquote {\bibinfo {title} {Hunting for the non-{H}ermitian
  exceptional points with fidelity susceptibility},}\ }\href {\doibase
  10.1103/PhysRevResearch.3.013015} {\bibfield  {journal} {\bibinfo  {journal}
  {Phys. Rev. Research}\ }\textbf {\bibinfo {volume} {3}},\ \bibinfo {pages}
  {013015} (\bibinfo {year} {2021})}\BibitemShut {NoStop}%
\bibitem [{\citenamefont {Louren\ifmmode~\mbox{\c{c}}\else \c{c}\fi{}o}\ \emph
  {et~al.}(2022)\citenamefont {Louren\ifmmode~\mbox{\c{c}}\else \c{c}\fi{}o},
  \citenamefont {Higgins}, \citenamefont {Zhang}, \citenamefont {Hennrich},\
  and\ \citenamefont {Macr\`{\i}}}]{PhysRevA.106.023309}%
  \BibitemOpen
  \bibfield  {author} {\bibinfo {author} {\bibfnamefont {J.~A.~S.}\
  \bibnamefont {Louren\ifmmode~\mbox{\c{c}}\else \c{c}\fi{}o}}, \bibinfo
  {author} {\bibfnamefont {G.}~\bibnamefont {Higgins}}, \bibinfo {author}
  {\bibfnamefont {C.}~\bibnamefont {Zhang}}, \bibinfo {author} {\bibfnamefont
  {M.}~\bibnamefont {Hennrich}}, \ and\ \bibinfo {author} {\bibfnamefont
  {T.}~\bibnamefont {Macr\`{\i}}},\ }\bibfield  {title} {\enquote {\bibinfo
  {title} {Non-{H}ermitian dynamics and $\mathcal{PT}$-symmetry breaking in
  interacting mesoscopic {R}ydberg platforms},}\ }\href {\doibase
  10.1103/PhysRevA.106.023309} {\bibfield  {journal} {\bibinfo  {journal}
  {Phys. Rev. A}\ }\textbf {\bibinfo {volume} {106}},\ \bibinfo {pages}
  {023309} (\bibinfo {year} {2022})}\BibitemShut {NoStop}%
\bibitem [{\citenamefont {Sticlet}\ \emph {et~al.}(2022)\citenamefont
  {Sticlet}, \citenamefont {D\'ora},\ and\ \citenamefont
  {Moca}}]{PhysRevLett.128.016802}%
  \BibitemOpen
  \bibfield  {author} {\bibinfo {author} {\bibfnamefont {D.}~\bibnamefont
  {Sticlet}}, \bibinfo {author} {\bibfnamefont {B.}~\bibnamefont {D\'ora}}, \
  and\ \bibinfo {author} {\bibfnamefont {C.~P.}\ \bibnamefont {Moca}},\
  }\bibfield  {title} {\enquote {\bibinfo {title} {Kubo {F}ormula for
  {N}on-{H}ermitian {S}ystems and {T}achyon {O}ptical {C}onductivity},}\ }\href
  {\doibase 10.1103/PhysRevLett.128.016802} {\bibfield  {journal} {\bibinfo
  {journal} {Phys. Rev. Lett.}\ }\textbf {\bibinfo {volume} {128}},\ \bibinfo
  {pages} {016802} (\bibinfo {year} {2022})}\BibitemShut {NoStop}%
\bibitem [{\citenamefont {Korff}\ and\ \citenamefont
  {Weston}(2007)}]{Korff_2007}%
  \BibitemOpen
  \bibfield  {author} {\bibinfo {author} {\bibfnamefont {C.}~\bibnamefont
  {Korff}}\ and\ \bibinfo {author} {\bibfnamefont {R.}~\bibnamefont {Weston}},\
  }\bibfield  {title} {\enquote {\bibinfo {title} {{PT} symmetry on the
  lattice: {T}he quantum group invariant {XXZ} spin chain},}\ }\href {\doibase
  10.1088/1751-8113/40/30/016} {\bibfield  {journal} {\bibinfo  {journal} {J.
  Phys. A: Math. Theor.}\ }\textbf {\bibinfo {volume} {40}},\ \bibinfo {pages}
  {8845} (\bibinfo {year} {2007})}\BibitemShut {NoStop}%
\bibitem [{\citenamefont {Korff}(2008)}]{Korff_2008}%
  \BibitemOpen
  \bibfield  {author} {\bibinfo {author} {\bibfnamefont {C.}~\bibnamefont
  {Korff}},\ }\bibfield  {title} {\enquote {\bibinfo {title} {{PT} symmetry of
  the non-{H}ermitian {XX} spin-chain: {N}on-local bulk interaction from
  complex boundary fields},}\ }\href {\doibase 10.1088/1751-8113/41/29/295206}
  {\bibfield  {journal} {\bibinfo  {journal} {J. Phys. A: Math. Theor.}\
  }\textbf {\bibinfo {volume} {41}},\ \bibinfo {pages} {295206} (\bibinfo
  {year} {2008})}\BibitemShut {NoStop}%
\bibitem [{\citenamefont {Castro-Alvaredo}\ and\ \citenamefont
  {Fring}(2009)}]{Castro_Alvaredo_2009}%
  \BibitemOpen
  \bibfield  {author} {\bibinfo {author} {\bibfnamefont {O.~A.}\ \bibnamefont
  {Castro-Alvaredo}}\ and\ \bibinfo {author} {\bibfnamefont {A.}~\bibnamefont
  {Fring}},\ }\bibfield  {title} {\enquote {\bibinfo {title} {A spin chain
  model with non-{H}ermitian interaction: {T}he {I}sing quantum spin chain in
  an imaginary field},}\ }\href {\doibase 10.1088/1751-8113/42/46/465211}
  {\bibfield  {journal} {\bibinfo  {journal} {J. Phys. A: Math. Theor.}\
  }\textbf {\bibinfo {volume} {42}},\ \bibinfo {pages} {465211} (\bibinfo
  {year} {2009})}\BibitemShut {NoStop}%
\bibitem [{\citenamefont {Ashida}\ \emph {et~al.}(2017)\citenamefont {Ashida},
  \citenamefont {Furukawa},\ and\ \citenamefont
  {Ueda}}]{NatCommun_8_15791_2017}%
  \BibitemOpen
  \bibfield  {author} {\bibinfo {author} {\bibfnamefont {Y.}~\bibnamefont
  {Ashida}}, \bibinfo {author} {\bibfnamefont {S.}~\bibnamefont {Furukawa}}, \
  and\ \bibinfo {author} {\bibfnamefont {M.}~\bibnamefont {Ueda}},\ }\bibfield
  {title} {\enquote {\bibinfo {title} {Parity-time-symmetric quantum critical
  phenomena},}\ }\href {\doibase 10.1038/ncomms15791} {\bibfield  {journal}
  {\bibinfo  {journal} {Nat. Commun.}\ }\textbf {\bibinfo {volume} {8}},\
  \bibinfo {pages} {15791} (\bibinfo {year} {2017})}\BibitemShut {NoStop}%
\bibitem [{\citenamefont {Xiao}\ \emph {et~al.}(2019)\citenamefont {Xiao},
  \citenamefont {Wang}, \citenamefont {Zhan}, \citenamefont {Bian},
  \citenamefont {Kawabata}, \citenamefont {Ueda}, \citenamefont {Yi},\ and\
  \citenamefont {Xue}}]{Xiao2019}%
  \BibitemOpen
  \bibfield  {author} {\bibinfo {author} {\bibfnamefont {L.}~\bibnamefont
  {Xiao}}, \bibinfo {author} {\bibfnamefont {K.}~\bibnamefont {Wang}}, \bibinfo
  {author} {\bibfnamefont {X.}~\bibnamefont {Zhan}}, \bibinfo {author}
  {\bibfnamefont {Z.}~\bibnamefont {Bian}}, \bibinfo {author} {\bibfnamefont
  {K.}~\bibnamefont {Kawabata}}, \bibinfo {author} {\bibfnamefont
  {M.}~\bibnamefont {Ueda}}, \bibinfo {author} {\bibfnamefont {W.}~\bibnamefont
  {Yi}}, \ and\ \bibinfo {author} {\bibfnamefont {P.}~\bibnamefont {Xue}},\
  }\bibfield  {title} {\enquote {\bibinfo {title} {Observation of {C}ritical
  {P}henomena in {P}arity-{T}ime-{S}ymmetric {Q}uantum {D}ynamics},}\ }\href
  {\doibase 10.1103/PhysRevLett.123.230401} {\bibfield  {journal} {\bibinfo
  {journal} {Phys. Rev. Lett.}\ }\textbf {\bibinfo {volume} {123}},\ \bibinfo
  {pages} {230401} (\bibinfo {year} {2019})}\BibitemShut {NoStop}%
\bibitem [{\citenamefont {Lee}\ \emph {et~al.}(2020)\citenamefont {Lee},
  \citenamefont {Lee},\ and\ \citenamefont {Yang}}]{PhysRevB.101.121109}%
  \BibitemOpen
  \bibfield  {author} {\bibinfo {author} {\bibfnamefont {E.}~\bibnamefont
  {Lee}}, \bibinfo {author} {\bibfnamefont {H.}~\bibnamefont {Lee}}, \ and\
  \bibinfo {author} {\bibfnamefont {B.-J.}\ \bibnamefont {Yang}},\ }\bibfield
  {title} {\enquote {\bibinfo {title} {Many-body approach to non-{H}ermitian
  physics in fermionic systems},}\ }\href {\doibase
  10.1103/PhysRevB.101.121109} {\bibfield  {journal} {\bibinfo  {journal}
  {Phys. Rev. B}\ }\textbf {\bibinfo {volume} {101}},\ \bibinfo {pages}
  {121109(R)} (\bibinfo {year} {2020})}\BibitemShut {NoStop}%
\bibitem [{\citenamefont {Hamazaki}\ \emph {et~al.}(2019)\citenamefont
  {Hamazaki}, \citenamefont {Kawabata},\ and\ \citenamefont
  {Ueda}}]{PhysRevLett.123.090603}%
  \BibitemOpen
  \bibfield  {author} {\bibinfo {author} {\bibfnamefont {R.}~\bibnamefont
  {Hamazaki}}, \bibinfo {author} {\bibfnamefont {K.}~\bibnamefont {Kawabata}},
  \ and\ \bibinfo {author} {\bibfnamefont {M.}~\bibnamefont {Ueda}},\
  }\bibfield  {title} {\enquote {\bibinfo {title} {{N}on-{H}er\-mi\-tian
  {M}any-{B}ody {L}ocalization},}\ }\href {\doibase
  10.1103/PhysRevLett.123.090603} {\bibfield  {journal} {\bibinfo  {journal}
  {Phys. Rev. Lett.}\ }\textbf {\bibinfo {volume} {123}},\ \bibinfo {pages}
  {090603} (\bibinfo {year} {2019})}\BibitemShut {NoStop}%
\bibitem [{\citenamefont {Matsumoto}\ \emph {et~al.}(2020)\citenamefont
  {Matsumoto}, \citenamefont {Kawabata}, \citenamefont {Ashida}, \citenamefont
  {Furukawa},\ and\ \citenamefont {Ueda}}]{PhysRevLett.125.260601}%
  \BibitemOpen
  \bibfield  {author} {\bibinfo {author} {\bibfnamefont {N.}~\bibnamefont
  {Matsumoto}}, \bibinfo {author} {\bibfnamefont {K.}~\bibnamefont {Kawabata}},
  \bibinfo {author} {\bibfnamefont {Y.}~\bibnamefont {Ashida}}, \bibinfo
  {author} {\bibfnamefont {S.}~\bibnamefont {Furukawa}}, \ and\ \bibinfo
  {author} {\bibfnamefont {M.}~\bibnamefont {Ueda}},\ }\bibfield  {title}
  {\enquote {\bibinfo {title} {Continuous {P}hase {T}ransition without {G}ap
  {C}losing in {N}on-{H}ermitian {Q}uantum {M}any-{B}ody {S}ystems},}\ }\href
  {\doibase 10.1103/PhysRevLett.125.260601} {\bibfield  {journal} {\bibinfo
  {journal} {Phys. Rev. Lett.}\ }\textbf {\bibinfo {volume} {125}},\ \bibinfo
  {pages} {260601} (\bibinfo {year} {2020})}\BibitemShut {NoStop}%
\bibitem [{\citenamefont {Takasu}\ \emph {et~al.}(2020)\citenamefont {Takasu},
  \citenamefont {Yagami}, \citenamefont {Ashida}, \citenamefont {Hamazaki},
  \citenamefont {Kuno},\ and\ \citenamefont
  {Takahashi}}]{ProgTheorExpPhys_12_2020}%
  \BibitemOpen
  \bibfield  {author} {\bibinfo {author} {\bibfnamefont {Y.}~\bibnamefont
  {Takasu}}, \bibinfo {author} {\bibfnamefont {T.}~\bibnamefont {Yagami}},
  \bibinfo {author} {\bibfnamefont {Y.}~\bibnamefont {Ashida}}, \bibinfo
  {author} {\bibfnamefont {R.}~\bibnamefont {Hamazaki}}, \bibinfo {author}
  {\bibfnamefont {Y.}~\bibnamefont {Kuno}}, \ and\ \bibinfo {author}
  {\bibfnamefont {Y.}~\bibnamefont {Takahashi}},\ }\bibfield  {title} {\enquote
  {\bibinfo {title} {{PT}-symmetric non-{H}ermitian quantum many-body system
  using ultracold atoms in an optical lattice with controlled dissipation},}\
  }\href {\doibase 10.1093/ptep/ptaa094} {\bibfield  {journal} {\bibinfo
  {journal} {Prog. Theor. Exp. Phys.}\ }\textbf {\bibinfo {volume} {2020}},\
  \bibinfo {pages} {12A110} (\bibinfo {year} {2020})}\BibitemShut {NoStop}%
\bibitem [{\citenamefont {Turkeshi}\ and\ \citenamefont
  {Schir\'o}(2023)}]{arXiv:2201.09895}%
  \BibitemOpen
  \bibfield  {author} {\bibinfo {author} {\bibfnamefont {X.}~\bibnamefont
  {Turkeshi}}\ and\ \bibinfo {author} {\bibfnamefont {M.}~\bibnamefont
  {Schir\'o}},\ }\bibfield  {title} {\enquote {\bibinfo {title} {Entanglement
  and correlation spreading in non-{H}ermitian spin chains},}\ }\href {\doibase
  10.1103/PhysRevB.107.L020403} {\bibfield  {journal} {\bibinfo  {journal}
  {Phys. Rev. B}\ }\textbf {\bibinfo {volume} {107}},\ \bibinfo {pages}
  {L020403} (\bibinfo {year} {2023})}\BibitemShut {NoStop}%
\bibitem [{\citenamefont {Gal}\ \emph {et~al.}(2022)\citenamefont {Gal},
  \citenamefont {Turkeshi},\ and\ \citenamefont
  {Schir\'{o}}}]{arXiv:2210.11937}%
  \BibitemOpen
  \bibfield  {author} {\bibinfo {author} {\bibfnamefont {Y.~L.}\ \bibnamefont
  {Gal}}, \bibinfo {author} {\bibfnamefont {X.}~\bibnamefont {Turkeshi}}, \
  and\ \bibinfo {author} {\bibfnamefont {M.}~\bibnamefont {Schir\'{o}}},\
  }\bibfield  {title} {\enquote {\bibinfo {title} {Volume-to-{A}rea {L}aw
  {E}ntanglement {T}ransition in a non-{H}ermitian {F}ree {F}ermionic
  {C}hain},}\ }\href {https://arxiv.org/abs/2210.11937} {\bibfield  {journal}
  {\bibinfo  {journal} {arXiv:2210.11937}\ } (\bibinfo {year}
  {2022})}\BibitemShut {NoStop}%
\bibitem [{\citenamefont {Kawabata}\ \emph {et~al.}(2020)\citenamefont
  {Kawabata}, \citenamefont {Sato},\ and\ \citenamefont
  {Shiozaki}}]{PhysRevB.102.205118}%
  \BibitemOpen
  \bibfield  {author} {\bibinfo {author} {\bibfnamefont {K.}~\bibnamefont
  {Kawabata}}, \bibinfo {author} {\bibfnamefont {M.}~\bibnamefont {Sato}}, \
  and\ \bibinfo {author} {\bibfnamefont {K.}~\bibnamefont {Shiozaki}},\
  }\bibfield  {title} {\enquote {\bibinfo {title} {Higher-order non-{H}ermitian
  skin effect},}\ }\href {\doibase 10.1103/PhysRevB.102.205118} {\bibfield
  {journal} {\bibinfo  {journal} {Phys. Rev. B}\ }\textbf {\bibinfo {volume}
  {102}},\ \bibinfo {pages} {205118} (\bibinfo {year} {2020})}\BibitemShut
  {NoStop}%
\bibitem [{\citenamefont {Peng}\ \emph {et~al.}(2022)\citenamefont {Peng},
  \citenamefont {Jie}, \citenamefont {Yu},\ and\ \citenamefont
  {Wang}}]{arXiv:2201.10318}%
  \BibitemOpen
  \bibfield  {author} {\bibinfo {author} {\bibfnamefont {Y.}~\bibnamefont
  {Peng}}, \bibinfo {author} {\bibfnamefont {J.}~\bibnamefont {Jie}}, \bibinfo
  {author} {\bibfnamefont {D.}~\bibnamefont {Yu}}, \ and\ \bibinfo {author}
  {\bibfnamefont {Y.}~\bibnamefont {Wang}},\ }\bibfield  {title} {\enquote
  {\bibinfo {title} {Manipulating the non-{H}ermitian skin effect via electric
  fields},}\ }\href {\doibase 10.1103/PhysRevB.106.L161402} {\bibfield
  {journal} {\bibinfo  {journal} {Phys. Rev. B}\ }\textbf {\bibinfo {volume}
  {106}},\ \bibinfo {pages} {L161402} (\bibinfo {year} {2022})}\BibitemShut
  {NoStop}%
\bibitem [{\citenamefont {\ifmmode \check{Z}\else
  \v{Z}\fi{}nidari\ifmmode~\check{c}\else
  \v{c}\fi{}}(2022)}]{PhysRevResearch.4.033041}%
  \BibitemOpen
  \bibfield  {author} {\bibinfo {author} {\bibfnamefont {M.}~\bibnamefont
  {\ifmmode \check{Z}\else \v{Z}\fi{}nidari\ifmmode~\check{c}\else
  \v{c}\fi{}}},\ }\bibfield  {title} {\enquote {\bibinfo {title} {Solvable
  non-{H}ermitian skin effect in many-body unitary dynamics},}\ }\href
  {\doibase 10.1103/PhysRevResearch.4.033041} {\bibfield  {journal} {\bibinfo
  {journal} {Phys. Rev. Research}\ }\textbf {\bibinfo {volume} {4}},\ \bibinfo
  {pages} {033041} (\bibinfo {year} {2022})}\BibitemShut {NoStop}%
\bibitem [{\citenamefont {Yao}\ and\ \citenamefont
  {Wang}(2018)}]{PhysRevLett.121.086803}%
  \BibitemOpen
  \bibfield  {author} {\bibinfo {author} {\bibfnamefont {S.}~\bibnamefont
  {Yao}}\ and\ \bibinfo {author} {\bibfnamefont {Z.}~\bibnamefont {Wang}},\
  }\bibfield  {title} {\enquote {\bibinfo {title} {Edge {S}tates and
  {T}opological {I}nvariants of {N}on-{H}ermitian {S}ystems},}\ }\href
  {\doibase 10.1103/PhysRevLett.121.086803} {\bibfield  {journal} {\bibinfo
  {journal} {Phys. Rev. Lett.}\ }\textbf {\bibinfo {volume} {121}},\ \bibinfo
  {pages} {086803} (\bibinfo {year} {2018})}\BibitemShut {NoStop}%
\bibitem [{\citenamefont {Yang}\ \emph {et~al.}(2020)\citenamefont {Yang},
  \citenamefont {Zhang}, \citenamefont {Fang},\ and\ \citenamefont
  {Hu}}]{PhysRevLett.125.226402}%
  \BibitemOpen
  \bibfield  {author} {\bibinfo {author} {\bibfnamefont {Z.}~\bibnamefont
  {Yang}}, \bibinfo {author} {\bibfnamefont {K.}~\bibnamefont {Zhang}},
  \bibinfo {author} {\bibfnamefont {C.}~\bibnamefont {Fang}}, \ and\ \bibinfo
  {author} {\bibfnamefont {J.}~\bibnamefont {Hu}},\ }\bibfield  {title}
  {\enquote {\bibinfo {title} {Non-{H}ermitian {B}ulk-{B}oundary
  {C}orrespondence and {A}uxiliary {G}eneralized {B}rillouin {Z}one
  {T}heory},}\ }\href {\doibase 10.1103/PhysRevLett.125.226402} {\bibfield
  {journal} {\bibinfo  {journal} {Phys. Rev. Lett.}\ }\textbf {\bibinfo
  {volume} {125}},\ \bibinfo {pages} {226402} (\bibinfo {year}
  {2020})}\BibitemShut {NoStop}%
\bibitem [{\citenamefont {Par\'a}\ \emph {et~al.}(2021)\citenamefont {Par\'a},
  \citenamefont {{P}alumbo},\ and\ \citenamefont
  {Macr\`{\i}}}]{PhysRevB.103.155417}%
  \BibitemOpen
  \bibfield  {author} {\bibinfo {author} {\bibfnamefont {Y.}~\bibnamefont
  {Par\'a}}, \bibinfo {author} {\bibfnamefont {G.}~\bibnamefont {{P}alumbo}}, \
  and\ \bibinfo {author} {\bibfnamefont {T.}~\bibnamefont {Macr\`{\i}}},\
  }\bibfield  {title} {\enquote {\bibinfo {title} {Probing non-hermitian phase
  transitions in curved space via quench dynamics},}\ }\href {\doibase
  10.1103/PhysRevB.103.155417} {\bibfield  {journal} {\bibinfo  {journal}
  {Phys. Rev. B}\ }\textbf {\bibinfo {volume} {103}},\ \bibinfo {pages}
  {155417} (\bibinfo {year} {2021})}\BibitemShut {NoStop}%
\bibitem [{\citenamefont {Nakanishi}\ and\ \citenamefont
  {Sasamoto}(2022)}]{PhysRevA.105.022219}%
  \BibitemOpen
  \bibfield  {author} {\bibinfo {author} {\bibfnamefont {Y.}~\bibnamefont
  {Nakanishi}}\ and\ \bibinfo {author} {\bibfnamefont {T.}~\bibnamefont
  {Sasamoto}},\ }\bibfield  {title} {\enquote {\bibinfo {title} {$\mathcal{PT}$
  phase transition in open quantum systems with {L}indblad dynamics},}\ }\href
  {\doibase 10.1103/PhysRevA.105.022219} {\bibfield  {journal} {\bibinfo
  {journal} {Phys. Rev. A}\ }\textbf {\bibinfo {volume} {105}},\ \bibinfo
  {pages} {022219} (\bibinfo {year} {2022})}\BibitemShut {NoStop}%
\bibitem [{\citenamefont {Noronha}\ \emph {et~al.}(2022)\citenamefont
  {Noronha}, \citenamefont {Louren\c{c}o},\ and\ \citenamefont
  {Macr\`{\i}}}]{arXiv:2206.02922}%
  \BibitemOpen
  \bibfield  {author} {\bibinfo {author} {\bibfnamefont {F.}~\bibnamefont
  {Noronha}}, \bibinfo {author} {\bibfnamefont {J.~A.~S.}\ \bibnamefont
  {Louren\c{c}o}}, \ and\ \bibinfo {author} {\bibfnamefont {T.}~\bibnamefont
  {Macr\`{\i}}},\ }\bibfield  {title} {\enquote {\bibinfo {title} {Robust
  quantum boomerang effect in non-{H}ermitian systems},}\ }\href {\doibase
  10.1103/PhysRevB.106.104310} {\bibfield  {journal} {\bibinfo  {journal}
  {Phys. Rev. B}\ }\textbf {\bibinfo {volume} {106}},\ \bibinfo {pages}
  {104310} (\bibinfo {year} {2022})}\BibitemShut {NoStop}%
\bibitem [{\citenamefont {Sajjad}\ \emph {et~al.}(2022)\citenamefont {Sajjad},
  \citenamefont {Tanlimco}, \citenamefont {Mas}, \citenamefont {Cao},
  \citenamefont {Nolasco-Martinez}, \citenamefont {Simmons}, \citenamefont
  {Santos}, \citenamefont {Vignolo}, \citenamefont {Macr\`{\i}},\ and\
  \citenamefont {Weld}}]{PhysRevX.12.011035}%
  \BibitemOpen
  \bibfield  {author} {\bibinfo {author} {\bibfnamefont {R.}~\bibnamefont
  {Sajjad}}, \bibinfo {author} {\bibfnamefont {J.~L.}\ \bibnamefont
  {Tanlimco}}, \bibinfo {author} {\bibfnamefont {H.}~\bibnamefont {Mas}},
  \bibinfo {author} {\bibfnamefont {A.}~\bibnamefont {Cao}}, \bibinfo {author}
  {\bibfnamefont {E.}~\bibnamefont {Nolasco-Martinez}}, \bibinfo {author}
  {\bibfnamefont {E.~Q.}\ \bibnamefont {Simmons}}, \bibinfo {author}
  {\bibfnamefont {F.~L.~N.}\ \bibnamefont {Santos}}, \bibinfo {author}
  {\bibfnamefont {P.}~\bibnamefont {Vignolo}}, \bibinfo {author} {\bibfnamefont
  {T.}~\bibnamefont {Macr\`{\i}}}, \ and\ \bibinfo {author} {\bibfnamefont
  {D.~M.}\ \bibnamefont {Weld}},\ }\bibfield  {title} {\enquote {\bibinfo
  {title} {Observation of the {Q}uantum {B}oomerang {E}ffect},}\ }\href
  {\doibase 10.1103/PhysRevX.12.011035} {\bibfield  {journal} {\bibinfo
  {journal} {Phys. Rev. X}\ }\textbf {\bibinfo {volume} {12}},\ \bibinfo
  {pages} {011035} (\bibinfo {year} {2022})}\BibitemShut {NoStop}%
\bibitem [{\citenamefont {Prat}\ \emph {et~al.}(2019)\citenamefont {Prat},
  \citenamefont {Delande},\ and\ \citenamefont
  {Cherroret}}]{PhysRevA.99.023629}%
  \BibitemOpen
  \bibfield  {author} {\bibinfo {author} {\bibfnamefont {T.}~\bibnamefont
  {Prat}}, \bibinfo {author} {\bibfnamefont {D.}~\bibnamefont {Delande}}, \
  and\ \bibinfo {author} {\bibfnamefont {N.}~\bibnamefont {Cherroret}},\
  }\bibfield  {title} {\enquote {\bibinfo {title} {Quantum boomeranglike effect
  of wave packets in random media},}\ }\href {\doibase
  10.1103/PhysRevA.99.023629} {\bibfield  {journal} {\bibinfo  {journal} {Phys.
  Rev. A}\ }\textbf {\bibinfo {volume} {99}},\ \bibinfo {pages} {023629}
  (\bibinfo {year} {2019})}\BibitemShut {NoStop}%
\bibitem [{\citenamefont {Noronha}\ and\ \citenamefont
  {Macr\`{\i}}(2022)}]{PhysRevB.106.L060301}%
  \BibitemOpen
  \bibfield  {author} {\bibinfo {author} {\bibfnamefont {F.}~\bibnamefont
  {Noronha}}\ and\ \bibinfo {author} {\bibfnamefont {T.}~\bibnamefont
  {Macr\`{\i}}},\ }\bibfield  {title} {\enquote {\bibinfo {title} {Ubiquity of
  the quantum boomerang effect in {H}ermitian {A}nderson-localized systems},}\
  }\href {\doibase 10.1103/PhysRevB.106.L060301} {\bibfield  {journal}
  {\bibinfo  {journal} {Phys. Rev. B}\ }\textbf {\bibinfo {volume} {106}},\
  \bibinfo {pages} {L060301} (\bibinfo {year} {2022})}\BibitemShut {NoStop}%
\bibitem [{\citenamefont {Minganti}\ \emph {et~al.}(2019)\citenamefont
  {Minganti}, \citenamefont {Miranowicz}, \citenamefont {Chhajlany},\ and\
  \citenamefont {Nori}}]{PhysRevA.100.062131}%
  \BibitemOpen
  \bibfield  {author} {\bibinfo {author} {\bibfnamefont {F.}~\bibnamefont
  {Minganti}}, \bibinfo {author} {\bibfnamefont {A.}~\bibnamefont
  {Miranowicz}}, \bibinfo {author} {\bibfnamefont {R.~W.}\ \bibnamefont
  {Chhajlany}}, \ and\ \bibinfo {author} {\bibfnamefont {F.}~\bibnamefont
  {Nori}},\ }\bibfield  {title} {\enquote {\bibinfo {title} {Quantum
  exceptional points of non-{H}ermitian {H}amiltonians and {L}iouvillians:
  {T}he effects of quantum jumps},}\ }\href {\doibase
  10.1103/PhysRevA.100.062131} {\bibfield  {journal} {\bibinfo  {journal}
  {Phys. Rev. A}\ }\textbf {\bibinfo {volume} {100}},\ \bibinfo {pages}
  {062131} (\bibinfo {year} {2019})}\BibitemShut {NoStop}%
\bibitem [{\citenamefont {Minganti}\ \emph {et~al.}(2020)\citenamefont
  {Minganti}, \citenamefont {Miranowicz}, \citenamefont {W.~Chhajlany},
  \citenamefont {I.~Arkhipov},\ and\ \citenamefont
  {Nori}}]{PhysRevA.101.062112}%
  \BibitemOpen
  \bibfield  {author} {\bibinfo {author} {\bibfnamefont {F.}~\bibnamefont
  {Minganti}}, \bibinfo {author} {\bibfnamefont {A.}~\bibnamefont
  {Miranowicz}}, \bibinfo {author} {\bibfnamefont {R.}~\bibnamefont
  {W.~Chhajlany}}, \bibinfo {author} {\bibfnamefont {I.}~\bibnamefont
  {I.~Arkhipov}}, \ and\ \bibinfo {author} {\bibfnamefont {F.}~\bibnamefont
  {Nori}},\ }\bibfield  {title} {\enquote {\bibinfo {title}
  {Hybrid-{L}iouvillian formalism connecting exceptional points of
  non-{H}ermitian {H}amiltonians and {L}iouvillians via postselection of
  quantum trajectories},}\ }\href {\doibase 10.1103/PhysRevA.101.062112}
  {\bibfield  {journal} {\bibinfo  {journal} {Phys. Rev. A}\ }\textbf {\bibinfo
  {volume} {101}},\ \bibinfo {pages} {062112} (\bibinfo {year}
  {2020})}\BibitemShut {NoStop}%
\bibitem [{\citenamefont {Pires}\ and\ \citenamefont
  {Macr\`{\i}}(2021)}]{PhysRevB.104.155141}%
  \BibitemOpen
  \bibfield  {author} {\bibinfo {author} {\bibfnamefont {D.~P.}\ \bibnamefont
  {Pires}}\ and\ \bibinfo {author} {\bibfnamefont {T.}~\bibnamefont
  {Macr\`{\i}}},\ }\bibfield  {title} {\enquote {\bibinfo {title} {Probing
  phase transitions in non-{H}ermitian systems with multiple quantum
  coherences},}\ }\href {\doibase 10.1103/PhysRevB.104.155141} {\bibfield
  {journal} {\bibinfo  {journal} {Phys. Rev. B}\ }\textbf {\bibinfo {volume}
  {104}},\ \bibinfo {pages} {155141} (\bibinfo {year} {2021})}\BibitemShut
  {NoStop}%
\bibitem [{\citenamefont {Huang}\ \emph {et~al.}(2021)\citenamefont {Huang},
  \citenamefont {Zhao},\ and\ \citenamefont {Li}}]{PhysRevA.104.052405}%
  \BibitemOpen
  \bibfield  {author} {\bibinfo {author} {\bibfnamefont {K.-Q.}\ \bibnamefont
  {Huang}}, \bibinfo {author} {\bibfnamefont {W.-L.}\ \bibnamefont {Zhao}}, \
  and\ \bibinfo {author} {\bibfnamefont {Z.}~\bibnamefont {Li}},\ }\bibfield
  {title} {\enquote {\bibinfo {title} {Effective protection of quantum
  coherence by a non-{H}ermitian driving potential},}\ }\href {\doibase
  10.1103/PhysRevA.104.052405} {\bibfield  {journal} {\bibinfo  {journal}
  {Phys. Rev. A}\ }\textbf {\bibinfo {volume} {104}},\ \bibinfo {pages}
  {052405} (\bibinfo {year} {2021})}\BibitemShut {NoStop}%
\bibitem [{\citenamefont {Gopalakrishnan}\ and\ \citenamefont
  {Gullans}(2021)}]{PhysRevLett.126.170503}%
  \BibitemOpen
  \bibfield  {author} {\bibinfo {author} {\bibfnamefont {S.}~\bibnamefont
  {Gopalakrishnan}}\ and\ \bibinfo {author} {\bibfnamefont {M.~J.}\
  \bibnamefont {Gullans}},\ }\bibfield  {title} {\enquote {\bibinfo {title}
  {Entanglement and {P}urification {T}ransitions in {N}on-{H}ermitian {Q}uantum
  {M}echanics},}\ }\href {\doibase 10.1103/PhysRevLett.126.170503} {\bibfield
  {journal} {\bibinfo  {journal} {Phys. Rev. Lett.}\ }\textbf {\bibinfo
  {volume} {126}},\ \bibinfo {pages} {170503} (\bibinfo {year}
  {2021})}\BibitemShut {NoStop}%
\bibitem [{\citenamefont {Chen}\ \emph
  {et~al.}(2022{\natexlab{a}})\citenamefont {Chen}, \citenamefont {Zhou},
  \citenamefont {Chen},\ and\ \citenamefont {Ye}}]{PhysRevB.105.L121115}%
  \BibitemOpen
  \bibfield  {author} {\bibinfo {author} {\bibfnamefont {L.-M.}\ \bibnamefont
  {Chen}}, \bibinfo {author} {\bibfnamefont {Y.}~\bibnamefont {Zhou}}, \bibinfo
  {author} {\bibfnamefont {S.~A.}\ \bibnamefont {Chen}}, \ and\ \bibinfo
  {author} {\bibfnamefont {P.}~\bibnamefont {Ye}},\ }\bibfield  {title}
  {\enquote {\bibinfo {title} {Quantum entanglement of non-{H}ermitian
  quasicrystals},}\ }\href {\doibase 10.1103/PhysRevB.105.L121115} {\bibfield
  {journal} {\bibinfo  {journal} {Phys. Rev. B}\ }\textbf {\bibinfo {volume}
  {105}},\ \bibinfo {pages} {L121115} (\bibinfo {year}
  {2022}{\natexlab{a}})}\BibitemShut {NoStop}%
\bibitem [{\citenamefont {Fang}\ \emph {et~al.}(2022)\citenamefont {Fang},
  \citenamefont {Zhao}, \citenamefont {Chen}, \citenamefont {Zhou},
  \citenamefont {Zhang}, \citenamefont {Wu}, \citenamefont {Yang},\ and\
  \citenamefont {Nori}}]{PhysRevResearch.4.033022}%
  \BibitemOpen
  \bibfield  {author} {\bibinfo {author} {\bibfnamefont {Y.-L.}\ \bibnamefont
  {Fang}}, \bibinfo {author} {\bibfnamefont {J.-L.}\ \bibnamefont {Zhao}},
  \bibinfo {author} {\bibfnamefont {D.-X.}\ \bibnamefont {Chen}}, \bibinfo
  {author} {\bibfnamefont {Y.-H.}\ \bibnamefont {Zhou}}, \bibinfo {author}
  {\bibfnamefont {Y.}~\bibnamefont {Zhang}}, \bibinfo {author} {\bibfnamefont
  {Q.-C.}\ \bibnamefont {Wu}}, \bibinfo {author} {\bibfnamefont {C.-P.}\
  \bibnamefont {Yang}}, \ and\ \bibinfo {author} {\bibfnamefont
  {F.}~\bibnamefont {Nori}},\ }\bibfield  {title} {\enquote {\bibinfo {title}
  {Entanglement dynamics in anti-$\mathcal{PT}$-symmetric systems},}\ }\href
  {\doibase 10.1103/PhysRevResearch.4.033022} {\bibfield  {journal} {\bibinfo
  {journal} {Phys. Rev. Research}\ }\textbf {\bibinfo {volume} {4}},\ \bibinfo
  {pages} {033022} (\bibinfo {year} {2022})}\BibitemShut {NoStop}%
\bibitem [{\citenamefont {Ortega-Taberner}\ \emph {et~al.}(2022)\citenamefont
  {Ortega-Taberner}, \citenamefont {R\o{}dland},\ and\ \citenamefont
  {Hermanns}}]{PhysRevB.105.075103}%
  \BibitemOpen
  \bibfield  {author} {\bibinfo {author} {\bibfnamefont {C.}~\bibnamefont
  {Ortega-Taberner}}, \bibinfo {author} {\bibfnamefont {L.}~\bibnamefont
  {R\o{}dland}}, \ and\ \bibinfo {author} {\bibfnamefont {M.}~\bibnamefont
  {Hermanns}},\ }\bibfield  {title} {\enquote {\bibinfo {title} {Polarization
  and entanglement spectrum in non-{H}ermitian systems},}\ }\href {\doibase
  10.1103/PhysRevB.105.075103} {\bibfield  {journal} {\bibinfo  {journal}
  {Phys. Rev. B}\ }\textbf {\bibinfo {volume} {105}},\ \bibinfo {pages}
  {075103} (\bibinfo {year} {2022})}\BibitemShut {NoStop}%
\bibitem [{\citenamefont {Macieszczak}\ \emph {et~al.}(2019)\citenamefont
  {Macieszczak}, \citenamefont {Levi}, \citenamefont {Macr\`{\i}},
  \citenamefont {Lesanovsky},\ and\ \citenamefont
  {Garrahan}}]{PhysRevA.99.052354}%
  \BibitemOpen
  \bibfield  {author} {\bibinfo {author} {\bibfnamefont {K.}~\bibnamefont
  {Macieszczak}}, \bibinfo {author} {\bibfnamefont {E.}~\bibnamefont {Levi}},
  \bibinfo {author} {\bibfnamefont {T.}~\bibnamefont {Macr\`{\i}}}, \bibinfo
  {author} {\bibfnamefont {I.}~\bibnamefont {Lesanovsky}}, \ and\ \bibinfo
  {author} {\bibfnamefont {J.~P.}\ \bibnamefont {Garrahan}},\ }\bibfield
  {title} {\enquote {\bibinfo {title} {Coherence, entanglement, and quantumness
  in closed and open systems with conserved charge, with an application to
  many-body localization},}\ }\href {\doibase 10.1103/PhysRevA.99.052354}
  {\bibfield  {journal} {\bibinfo  {journal} {Phys. Rev. A}\ }\textbf {\bibinfo
  {volume} {99}},\ \bibinfo {pages} {052354} (\bibinfo {year}
  {2019})}\BibitemShut {NoStop}%
\bibitem [{\citenamefont {Pires}\ \emph {et~al.}(2020)\citenamefont {Pires},
  \citenamefont {Smerzi},\ and\ \citenamefont
  {Macr\`{\i}}}]{PhysRevA.102.012429}%
  \BibitemOpen
  \bibfield  {author} {\bibinfo {author} {\bibfnamefont {D.~P.}\ \bibnamefont
  {Pires}}, \bibinfo {author} {\bibfnamefont {A.}~\bibnamefont {Smerzi}}, \
  and\ \bibinfo {author} {\bibfnamefont {T.}~\bibnamefont {Macr\`{\i}}},\
  }\bibfield  {title} {\enquote {\bibinfo {title} {Relating relative {R}\'enyi
  entropies and {W}igner-{Y}anase-{D}yson skew information to generalized
  multiple quantum coherences},}\ }\href {\doibase 10.1103/PhysRevA.102.012429}
  {\bibfield  {journal} {\bibinfo  {journal} {Phys. Rev. A}\ }\textbf {\bibinfo
  {volume} {102}},\ \bibinfo {pages} {012429} (\bibinfo {year}
  {2020})}\BibitemShut {NoStop}%
\bibitem [{\citenamefont {Sergi}\ and\ \citenamefont
  {Zloshchastiev}(2016)}]{JStatMech_033102_2016}%
  \BibitemOpen
  \bibfield  {author} {\bibinfo {author} {\bibfnamefont {A.}~\bibnamefont
  {Sergi}}\ and\ \bibinfo {author} {\bibfnamefont {K.~G.}\ \bibnamefont
  {Zloshchastiev}},\ }\bibfield  {title} {\enquote {\bibinfo {title} {Quantum
  entropy of systems described by non-{H}ermitian {H}amiltonians},}\ }\href
  {\doibase 10.1088/1742-5468/2016/03/033102} {\bibfield  {journal} {\bibinfo
  {journal} {J. Stat. Mech.}\ }\textbf {\bibinfo {volume} {2016}},\ \bibinfo
  {pages} {033102} (\bibinfo {year} {2016})}\BibitemShut {NoStop}%
\bibitem [{\citenamefont {Zhang}\ \emph {et~al.}(2017)\citenamefont {Zhang},
  \citenamefont {Fang},\ and\ \citenamefont {Xu}}]{QIP_10_234_2017}%
  \BibitemOpen
  \bibfield  {author} {\bibinfo {author} {\bibfnamefont {S.-Y.}\ \bibnamefont
  {Zhang}}, \bibinfo {author} {\bibfnamefont {M.-F.}\ \bibnamefont {Fang}}, \
  and\ \bibinfo {author} {\bibfnamefont {L.}~\bibnamefont {Xu}},\ }\bibfield
  {title} {\enquote {\bibinfo {title} {Quantum entropy of non-{H}ermitian
  entangled systems},}\ }\href {\doibase 10.1007/s11128-017-1685-9} {\bibfield
  {journal} {\bibinfo  {journal} {Quantum Inf. Process.}\ }\textbf {\bibinfo
  {volume} {16}},\ \bibinfo {pages} {234} (\bibinfo {year} {2017})}\BibitemShut
  {NoStop}%
\bibitem [{\citenamefont {Guo}\ \emph {et~al.}(2021)\citenamefont {Guo},
  \citenamefont {Yu}, \citenamefont {Huang}, \citenamefont {Yang},
  \citenamefont {Chi}, \citenamefont {Liao},\ and\ \citenamefont
  {Xiang}}]{arXiv:2105.09793}%
  \BibitemOpen
  \bibfield  {author} {\bibinfo {author} {\bibfnamefont {Y.-B.}\ \bibnamefont
  {Guo}}, \bibinfo {author} {\bibfnamefont {Y.-C.}\ \bibnamefont {Yu}},
  \bibinfo {author} {\bibfnamefont {R.-Z.}\ \bibnamefont {Huang}}, \bibinfo
  {author} {\bibfnamefont {L.-P.}\ \bibnamefont {Yang}}, \bibinfo {author}
  {\bibfnamefont {R.-Z.}\ \bibnamefont {Chi}}, \bibinfo {author} {\bibfnamefont
  {Hai-Jun}\ \bibnamefont {Liao}}, \ and\ \bibinfo {author} {\bibfnamefont
  {T.}~\bibnamefont {Xiang}},\ }\bibfield  {title} {\enquote {\bibinfo {title}
  {Entanglement entropy of non-{H}ermitian free fermions},}\ }\href {\doibase
  10.1088/1361-648X/ac216e} {\bibfield  {journal} {\bibinfo  {journal} {J.
  Phys.: Condens. Matter}\ }\textbf {\bibinfo {volume} {33}},\ \bibinfo {pages}
  {475502} (\bibinfo {year} {2021})}\BibitemShut {NoStop}%
\bibitem [{\citenamefont {Chen}\ \emph
  {et~al.}(2022{\natexlab{b}})\citenamefont {Chen}, \citenamefont {Peng},
  \citenamefont {Lu},\ and\ \citenamefont {Lu}}]{PhysRevB.105.075126}%
  \BibitemOpen
  \bibfield  {author} {\bibinfo {author} {\bibfnamefont {W.}~\bibnamefont
  {Chen}}, \bibinfo {author} {\bibfnamefont {L.}~\bibnamefont {Peng}}, \bibinfo
  {author} {\bibfnamefont {H.}~\bibnamefont {Lu}}, \ and\ \bibinfo {author}
  {\bibfnamefont {X.}~\bibnamefont {Lu}},\ }\bibfield  {title} {\enquote
  {\bibinfo {title} {Characterizing bulk-boundary correspondence of
  one-dimensional non-{H}ermitian interacting systems by edge entanglement
  entropy},}\ }\href {\doibase 10.1103/PhysRevB.105.075126} {\bibfield
  {journal} {\bibinfo  {journal} {Phys. Rev. B}\ }\textbf {\bibinfo {volume}
  {105}},\ \bibinfo {pages} {075126} (\bibinfo {year}
  {2022}{\natexlab{b}})}\BibitemShut {NoStop}%
\bibitem [{\citenamefont {Li}\ \emph {et~al.}(2017)\citenamefont {Li},
  \citenamefont {Fan}, \citenamefont {Wang}, \citenamefont {Ye}, \citenamefont
  {Zeng}, \citenamefont {Zhai}, \citenamefont {Peng},\ and\ \citenamefont
  {Du}}]{PhysRevX.7.031011}%
  \BibitemOpen
  \bibfield  {author} {\bibinfo {author} {\bibfnamefont {J.}~\bibnamefont
  {Li}}, \bibinfo {author} {\bibfnamefont {R.}~\bibnamefont {Fan}}, \bibinfo
  {author} {\bibfnamefont {H.}~\bibnamefont {Wang}}, \bibinfo {author}
  {\bibfnamefont {B.}~\bibnamefont {Ye}}, \bibinfo {author} {\bibfnamefont
  {B.}~\bibnamefont {Zeng}}, \bibinfo {author} {\bibfnamefont {H.}~\bibnamefont
  {Zhai}}, \bibinfo {author} {\bibfnamefont {X.}~\bibnamefont {Peng}}, \ and\
  \bibinfo {author} {\bibfnamefont {J.}~\bibnamefont {Du}},\ }\bibfield
  {title} {\enquote {\bibinfo {title} {Measuring {O}ut-of-{T}ime-{O}rder
  {C}orrelators on a {N}uclear {M}agnetic {R}esonance {Q}uantum {S}imulator},}\
  }\href {\doibase 10.1103/PhysRevX.7.031011} {\bibfield  {journal} {\bibinfo
  {journal} {Phys. Rev. X}\ }\textbf {\bibinfo {volume} {7}},\ \bibinfo {pages}
  {031011} (\bibinfo {year} {2017})}\BibitemShut {NoStop}%
\bibitem [{\citenamefont {Streltsov}\ \emph {et~al.}(2018)\citenamefont
  {Streltsov}, \citenamefont {Kampermann}, \citenamefont {W\"{o}lk},
  \citenamefont {Gessner},\ and\ \citenamefont
  {Bru{\ss}}}]{NJP_20_053058_2018}%
  \BibitemOpen
  \bibfield  {author} {\bibinfo {author} {\bibfnamefont {A.}~\bibnamefont
  {Streltsov}}, \bibinfo {author} {\bibfnamefont {H.}~\bibnamefont
  {Kampermann}}, \bibinfo {author} {\bibfnamefont {S.}~\bibnamefont
  {W\"{o}lk}}, \bibinfo {author} {\bibfnamefont {M.}~\bibnamefont {Gessner}}, \
  and\ \bibinfo {author} {\bibfnamefont {D.}~\bibnamefont {Bru{\ss}}},\
  }\bibfield  {title} {\enquote {\bibinfo {title} {Maximal coherence and the
  resource theory of purity},}\ }\href {\doibase 10.1088/1367-2630/aac484}
  {\bibfield  {journal} {\bibinfo  {journal} {New J. Phys.}\ }\textbf {\bibinfo
  {volume} {20}},\ \bibinfo {pages} {053058} (\bibinfo {year}
  {2018})}\BibitemShut {NoStop}%
\bibitem [{\citenamefont {Islam}\ \emph {et~al.}(2015)\citenamefont {Islam},
  \citenamefont {Ma}, \citenamefont {Preiss}, \citenamefont {Tai},
  \citenamefont {Lukin}, \citenamefont {Rispoli},\ and\ \citenamefont
  {Greiner}}]{20159_arxiv_1509.01160}%
  \BibitemOpen
  \bibfield  {author} {\bibinfo {author} {\bibfnamefont {R.}~\bibnamefont
  {Islam}}, \bibinfo {author} {\bibfnamefont {R.}~\bibnamefont {Ma}}, \bibinfo
  {author} {\bibfnamefont {P.~M.}\ \bibnamefont {Preiss}}, \bibinfo {author}
  {\bibfnamefont {M.~E.}\ \bibnamefont {Tai}}, \bibinfo {author} {\bibfnamefont
  {A.}~\bibnamefont {Lukin}}, \bibinfo {author} {\bibfnamefont
  {M.}~\bibnamefont {Rispoli}}, \ and\ \bibinfo {author} {\bibfnamefont
  {M.}~\bibnamefont {Greiner}},\ }\bibfield  {title} {\enquote {\bibinfo
  {title} {Measuring entanglement entropy in a quantum many-body system},}\
  }\href {\doibase 10.1038/nature15750} {\bibfield  {journal} {\bibinfo
  {journal} {Nature (London)}\ }\textbf {\bibinfo {volume} {528}},\ \bibinfo
  {pages} {77} (\bibinfo {year} {2015})}\BibitemShut {NoStop}%
\bibitem [{\citenamefont {Elben}\ \emph {et~al.}(2018)\citenamefont {Elben},
  \citenamefont {Vermersch}, \citenamefont {Dalmonte}, \citenamefont {Cirac},\
  and\ \citenamefont {Zoller}}]{PhysRevLett.120.050406}%
  \BibitemOpen
  \bibfield  {author} {\bibinfo {author} {\bibfnamefont {A.}~\bibnamefont
  {Elben}}, \bibinfo {author} {\bibfnamefont {B.}~\bibnamefont {Vermersch}},
  \bibinfo {author} {\bibfnamefont {M.}~\bibnamefont {Dalmonte}}, \bibinfo
  {author} {\bibfnamefont {J.~I.}\ \bibnamefont {Cirac}}, \ and\ \bibinfo
  {author} {\bibfnamefont {P.}~\bibnamefont {Zoller}},\ }\bibfield  {title}
  {\enquote {\bibinfo {title} {R\'enyi {E}ntropies from {R}andom {Q}uenches in
  {A}tomic {H}ubbard and {S}pin {M}odels},}\ }\href {\doibase
  10.1103/PhysRevLett.120.050406} {\bibfield  {journal} {\bibinfo  {journal}
  {Phys. Rev. Lett.}\ }\textbf {\bibinfo {volume} {120}},\ \bibinfo {pages}
  {050406} (\bibinfo {year} {2018})}\BibitemShut {NoStop}%
\bibitem [{\citenamefont {Kaufman}\ \emph {et~al.}(2016)\citenamefont
  {Kaufman}, \citenamefont {Tai}, \citenamefont {Lukin}, \citenamefont
  {Rispoli}, \citenamefont {Schittko}, \citenamefont {Preiss},\ and\
  \citenamefont {Greiner}}]{Science_353_6301_794}%
  \BibitemOpen
  \bibfield  {author} {\bibinfo {author} {\bibfnamefont {A.~M.}\ \bibnamefont
  {Kaufman}}, \bibinfo {author} {\bibfnamefont {M.~E.}\ \bibnamefont {Tai}},
  \bibinfo {author} {\bibfnamefont {A.}~\bibnamefont {Lukin}}, \bibinfo
  {author} {\bibfnamefont {M.}~\bibnamefont {Rispoli}}, \bibinfo {author}
  {\bibfnamefont {R.}~\bibnamefont {Schittko}}, \bibinfo {author}
  {\bibfnamefont {P.~M.}\ \bibnamefont {Preiss}}, \ and\ \bibinfo {author}
  {\bibfnamefont {M.}~\bibnamefont {Greiner}},\ }\bibfield  {title} {\enquote
  {\bibinfo {title} {Quantum thermalization through entanglement in an isolated
  many-body system},}\ }\href {\doibase 10.1126/science.aaf6725} {\bibfield
  {journal} {\bibinfo  {journal} {Science}\ }\textbf {\bibinfo {volume}
  {353}},\ \bibinfo {pages} {794} (\bibinfo {year} {2016})}\BibitemShut
  {NoStop}%
\bibitem [{\citenamefont {Linke}\ \emph {et~al.}(2018)\citenamefont {Linke},
  \citenamefont {Johri}, \citenamefont {Figgatt}, \citenamefont {Landsman},
  \citenamefont {Matsuura},\ and\ \citenamefont {Monroe}}]{PhysRevA.98.052334}%
  \BibitemOpen
  \bibfield  {author} {\bibinfo {author} {\bibfnamefont {N.~M.}\ \bibnamefont
  {Linke}}, \bibinfo {author} {\bibfnamefont {S.}~\bibnamefont {Johri}},
  \bibinfo {author} {\bibfnamefont {C.}~\bibnamefont {Figgatt}}, \bibinfo
  {author} {\bibfnamefont {K.~A.}\ \bibnamefont {Landsman}}, \bibinfo {author}
  {\bibfnamefont {A.~Y.}\ \bibnamefont {Matsuura}}, \ and\ \bibinfo {author}
  {\bibfnamefont {C.}~\bibnamefont {Monroe}},\ }\bibfield  {title} {\enquote
  {\bibinfo {title} {Measuring the {R}\'enyi entropy of a two-site
  {F}ermi-{H}ubbard model on a trapped ion quantum computer},}\ }\href
  {\doibase 10.1103/PhysRevA.98.052334} {\bibfield  {journal} {\bibinfo
  {journal} {Phys. Rev. A}\ }\textbf {\bibinfo {volume} {98}},\ \bibinfo
  {pages} {052334} (\bibinfo {year} {2018})}\BibitemShut {NoStop}%
\bibitem [{\citenamefont {Brydges}\ \emph {et~al.}(2019)\citenamefont
  {Brydges}, \citenamefont {Elben}, \citenamefont {Jurcevic}, \citenamefont
  {Vermersch}, \citenamefont {Maier}, \citenamefont {Lanyon}, \citenamefont
  {Zoller}, \citenamefont {Blatt},\ and\ \citenamefont {Roos}}]{Brydges260}%
  \BibitemOpen
  \bibfield  {author} {\bibinfo {author} {\bibfnamefont {T.}~\bibnamefont
  {Brydges}}, \bibinfo {author} {\bibfnamefont {A.}~\bibnamefont {Elben}},
  \bibinfo {author} {\bibfnamefont {P.}~\bibnamefont {Jurcevic}}, \bibinfo
  {author} {\bibfnamefont {B.}~\bibnamefont {Vermersch}}, \bibinfo {author}
  {\bibfnamefont {C.}~\bibnamefont {Maier}}, \bibinfo {author} {\bibfnamefont
  {B.~P.}\ \bibnamefont {Lanyon}}, \bibinfo {author} {\bibfnamefont
  {P.}~\bibnamefont {Zoller}}, \bibinfo {author} {\bibfnamefont
  {R.}~\bibnamefont {Blatt}}, \ and\ \bibinfo {author} {\bibfnamefont {C.~F.}\
  \bibnamefont {Roos}},\ }\bibfield  {title} {\enquote {\bibinfo {title}
  {Probing {R}{\'e}nyi entanglement entropy via randomized measurements},}\
  }\href {\doibase 10.1126/science.aau4963} {\bibfield  {journal} {\bibinfo
  {journal} {Science}\ }\textbf {\bibinfo {volume} {364}},\ \bibinfo {pages}
  {260} (\bibinfo {year} {2019})}\BibitemShut {NoStop}%
\bibitem [{\citenamefont {Yang}(2018)}]{PhysRevD.97.066008}%
  \BibitemOpen
  \bibfield  {author} {\bibinfo {author} {\bibfnamefont {I.-S.}\ \bibnamefont
  {Yang}},\ }\bibfield  {title} {\enquote {\bibinfo {title} {Entanglement
  timescale},}\ }\href {\doibase 10.1103/PhysRevD.97.066008} {\bibfield
  {journal} {\bibinfo  {journal} {Phys. Rev. D}\ }\textbf {\bibinfo {volume}
  {97}},\ \bibinfo {pages} {066008} (\bibinfo {year} {2018})}\BibitemShut
  {NoStop}%
\bibitem [{\citenamefont {Cresswell}(2018)}]{PhysRevA.97.022317}%
  \BibitemOpen
  \bibfield  {author} {\bibinfo {author} {\bibfnamefont {J.~C.}\ \bibnamefont
  {Cresswell}},\ }\bibfield  {title} {\enquote {\bibinfo {title} {Universal
  entanglement timescale for {R}\'enyi entropies},}\ }\href {\doibase
  10.1103/PhysRevA.97.022317} {\bibfield  {journal} {\bibinfo  {journal} {Phys.
  Rev. A}\ }\textbf {\bibinfo {volume} {97}},\ \bibinfo {pages} {022317}
  (\bibinfo {year} {2018})}\BibitemShut {NoStop}%
\bibitem [{\citenamefont {Angelo}\ and\ \citenamefont
  {Furuya}(2005)}]{PhysRevA.71.042321}%
  \BibitemOpen
  \bibfield  {author} {\bibinfo {author} {\bibfnamefont {R.~M.}\ \bibnamefont
  {Angelo}}\ and\ \bibinfo {author} {\bibfnamefont {K.}~\bibnamefont
  {Furuya}},\ }\bibfield  {title} {\enquote {\bibinfo {title} {Semiclassical
  limit of the entanglement in closed pure systems},}\ }\href {\doibase
  10.1103/PhysRevA.71.042321} {\bibfield  {journal} {\bibinfo  {journal} {Phys.
  Rev. A}\ }\textbf {\bibinfo {volume} {71}},\ \bibinfo {pages} {042321}
  (\bibinfo {year} {2005})}\BibitemShut {NoStop}%
\bibitem [{\citenamefont {Nemes}\ and\ \citenamefont {de~{T}oledo
  {P}iza}(1986)}]{NEMES1986367}%
  \BibitemOpen
  \bibfield  {author} {\bibinfo {author} {\bibfnamefont {M.~C.}\ \bibnamefont
  {Nemes}}\ and\ \bibinfo {author} {\bibfnamefont {A.~F.~R.}\ \bibnamefont
  {de~{T}oledo {P}iza}},\ }\bibfield  {title} {\enquote {\bibinfo {title}
  {Effective dynamics of quantum subsystems},}\ }\href {\doibase
  https://doi.org/10.1016/0378-4371(86)90083-X} {\bibfield  {journal} {\bibinfo
   {journal} {Phys. A: Stat. Mech. Appl.}\ }\textbf {\bibinfo {volume} {137}},\
  \bibinfo {pages} {367} (\bibinfo {year} {1986})}\BibitemShut {NoStop}%
\bibitem [{\citenamefont {Zurek}(2003)}]{RevModPhys.75.715}%
  \BibitemOpen
  \bibfield  {author} {\bibinfo {author} {\bibfnamefont {W.~H.}\ \bibnamefont
  {Zurek}},\ }\bibfield  {title} {\enquote {\bibinfo {title} {Decoherence,
  einselection, and the quantum origins of the classical},}\ }\href {\doibase
  10.1103/RevModPhys.75.715} {\bibfield  {journal} {\bibinfo  {journal} {Rev.
  Mod. Phys.}\ }\textbf {\bibinfo {volume} {75}},\ \bibinfo {pages} {715}
  (\bibinfo {year} {2003})}\BibitemShut {NoStop}%
\bibitem [{\citenamefont {Kim}\ \emph {et~al.}(1996)\citenamefont {Kim},
  \citenamefont {Nemes}, \citenamefont {de~Toledo~Piza},\ and\ \citenamefont
  {Borges}}]{PhysRevLett.77.207}%
  \BibitemOpen
  \bibfield  {author} {\bibinfo {author} {\bibfnamefont {J.~II}\ \bibnamefont
  {Kim}}, \bibinfo {author} {\bibfnamefont {M.~C.}\ \bibnamefont {Nemes}},
  \bibinfo {author} {\bibfnamefont {A.~F.~R.}\ \bibnamefont {de~Toledo~Piza}},
  \ and\ \bibinfo {author} {\bibfnamefont {H.~E.}\ \bibnamefont {Borges}},\
  }\bibfield  {title} {\enquote {\bibinfo {title} {Perturbative {E}xpansion for
  {C}oherence {L}oss},}\ }\href {\doibase 10.1103/PhysRevLett.77.207}
  {\bibfield  {journal} {\bibinfo  {journal} {Phys. Rev. Lett.}\ }\textbf
  {\bibinfo {volume} {77}},\ \bibinfo {pages} {207} (\bibinfo {year}
  {1996})}\BibitemShut {NoStop}%
\bibitem [{\citenamefont {Zurek}\ \emph {et~al.}(1993)\citenamefont {Zurek},
  \citenamefont {Habib},\ and\ \citenamefont {Paz}}]{PhysRevLett.70.1187}%
  \BibitemOpen
  \bibfield  {author} {\bibinfo {author} {\bibfnamefont {W.~H.}\ \bibnamefont
  {Zurek}}, \bibinfo {author} {\bibfnamefont {S.}~\bibnamefont {Habib}}, \ and\
  \bibinfo {author} {\bibfnamefont {J.~P.}\ \bibnamefont {Paz}},\ }\bibfield
  {title} {\enquote {\bibinfo {title} {Coherent states via decoherence},}\
  }\href {\doibase 10.1103/PhysRevLett.70.1187} {\bibfield  {journal} {\bibinfo
   {journal} {Phys. Rev. Lett.}\ }\textbf {\bibinfo {volume} {70}},\ \bibinfo
  {pages} {1187} (\bibinfo {year} {1993})}\BibitemShut {NoStop}%
\bibitem [{\citenamefont {Gu}\ and\ \citenamefont
  {Franco}(2017)}]{JPhysChemLett_17_8_2017}%
  \BibitemOpen
  \bibfield  {author} {\bibinfo {author} {\bibfnamefont {B.}~\bibnamefont
  {Gu}}\ and\ \bibinfo {author} {\bibfnamefont {I.}~\bibnamefont {Franco}},\
  }\bibfield  {title} {\enquote {\bibinfo {title} {Quantifying {E}arly {T}ime
  {Q}uantum {D}ecoherence {D}ynamics through {F}luctuations},}\ }\href
  {\doibase 10.1021/acs.jpclett.7b01817} {\bibfield  {journal} {\bibinfo
  {journal} {J. Phys. Chem. Lett.}\ }\textbf {\bibinfo {volume} {8}},\ \bibinfo
  {pages} {4289} (\bibinfo {year} {2017})}\BibitemShut {NoStop}%
\bibitem [{\citenamefont {Cresswell}\ \emph {et~al.}(2019)\citenamefont
  {Cresswell}, \citenamefont {Tzitrin},\ and\ \citenamefont
  {Goldberg}}]{PhysRevA.99.012322}%
  \BibitemOpen
  \bibfield  {author} {\bibinfo {author} {\bibfnamefont {J.~C.}\ \bibnamefont
  {Cresswell}}, \bibinfo {author} {\bibfnamefont {I.}~\bibnamefont {Tzitrin}},
  \ and\ \bibinfo {author} {\bibfnamefont {A.~Z.}\ \bibnamefont {Goldberg}},\
  }\bibfield  {title} {\enquote {\bibinfo {title} {Perturbative expansion of
  entanglement negativity using patterned matrix calculus},}\ }\href {\doibase
  10.1103/PhysRevA.99.012322} {\bibfield  {journal} {\bibinfo  {journal} {Phys.
  Rev. A}\ }\textbf {\bibinfo {volume} {99}},\ \bibinfo {pages} {012322}
  (\bibinfo {year} {2019})}\BibitemShut {NoStop}%
\bibitem [{\citenamefont {Wen}\ and\ \citenamefont
  {Kempf}(2022)}]{arXiv:2209.03976}%
  \BibitemOpen
  \bibfield  {author} {\bibinfo {author} {\bibfnamefont {R.~Y.}\ \bibnamefont
  {Wen}}\ and\ \bibinfo {author} {\bibfnamefont {A.}~\bibnamefont {Kempf}},\
  }\bibfield  {title} {\enquote {\bibinfo {title} {The transfer of entanglement
  negativity at the onset of interactions},}\ }\href {\doibase
  10.1088/1751-8121/aca7a1} {\bibfield  {journal} {\bibinfo  {journal} {J.
  Phys. A: Math. Theor.}\ }\textbf {\bibinfo {volume} {55}},\ \bibinfo {pages}
  {495304} (\bibinfo {year} {2022})}\BibitemShut {NoStop}%
\bibitem [{\citenamefont {Duan}\ and\ \citenamefont
  {Guo}(1997)}]{PhysRevA.56.4466}%
  \BibitemOpen
  \bibfield  {author} {\bibinfo {author} {\bibfnamefont {L.-M.}\ \bibnamefont
  {Duan}}\ and\ \bibinfo {author} {\bibfnamefont {G.-C.}\ \bibnamefont {Guo}},\
  }\bibfield  {title} {\enquote {\bibinfo {title} {Perturbative expansions for
  the fidelities and spatially correlated dissipation of quantum bits},}\
  }\href {\doibase 10.1103/PhysRevA.56.4466} {\bibfield  {journal} {\bibinfo
  {journal} {Phys. Rev. A}\ }\textbf {\bibinfo {volume} {56}},\ \bibinfo
  {pages} {4466} (\bibinfo {year} {1997})}\BibitemShut {NoStop}%
\bibitem [{\citenamefont {Gorin}\ \emph {et~al.}(2006)\citenamefont {Gorin},
  \citenamefont {Prosen}, \citenamefont {Seligman},\ and\ \citenamefont
  {\v{Z}nidari\v{c}}}]{GORIN200633}%
  \BibitemOpen
  \bibfield  {author} {\bibinfo {author} {\bibfnamefont {T.}~\bibnamefont
  {Gorin}}, \bibinfo {author} {\bibfnamefont {T.}~\bibnamefont {Prosen}},
  \bibinfo {author} {\bibfnamefont {T.~H.}\ \bibnamefont {Seligman}}, \ and\
  \bibinfo {author} {\bibfnamefont {M.}~\bibnamefont {\v{Z}nidari\v{c}}},\
  }\bibfield  {title} {\enquote {\bibinfo {title} {Dynamics of {L}oschmidt
  echoes and fidelity decay},}\ }\href {\doibase 10.1016/j.physrep.2006.09.003}
  {\bibfield  {journal} {\bibinfo  {journal} {Phys. Rep.}\ }\textbf {\bibinfo
  {volume} {435}},\ \bibinfo {pages} {33} (\bibinfo {year} {2006})}\BibitemShut
  {NoStop}%
\bibitem [{\citenamefont {Feng}\ \emph {et~al.}(2017)\citenamefont {Feng},
  \citenamefont {El-Ganainy},\ and\ \citenamefont {Ge}}]{Nat_Phot_11_752_2017}%
  \BibitemOpen
  \bibfield  {author} {\bibinfo {author} {\bibfnamefont {L.}~\bibnamefont
  {Feng}}, \bibinfo {author} {\bibfnamefont {R.}~\bibnamefont {El-Ganainy}}, \
  and\ \bibinfo {author} {\bibfnamefont {L.}~\bibnamefont {Ge}},\ }\bibfield
  {title} {\enquote {\bibinfo {title} {Non-{H}ermitian photonics based on
  parity-time symmetry},}\ }\href {\doibase 10.1038/s41566-017-0031-1}
  {\bibfield  {journal} {\bibinfo  {journal} {Nat. Phys.}\ }\textbf {\bibinfo
  {volume} {11}},\ \bibinfo {pages} {752} (\bibinfo {year} {2017})}\BibitemShut
  {NoStop}%
\bibitem [{\citenamefont {Gong}\ \emph {et~al.}(2018)\citenamefont {Gong},
  \citenamefont {Ashida}, \citenamefont {Kawabata}, \citenamefont {Takasan},
  \citenamefont {Higashikawa},\ and\ \citenamefont {Ueda}}]{PhysRevX.8.031079}%
  \BibitemOpen
  \bibfield  {author} {\bibinfo {author} {\bibfnamefont {Z.}~\bibnamefont
  {Gong}}, \bibinfo {author} {\bibfnamefont {Y.}~\bibnamefont {Ashida}},
  \bibinfo {author} {\bibfnamefont {K.}~\bibnamefont {Kawabata}}, \bibinfo
  {author} {\bibfnamefont {K.}~\bibnamefont {Takasan}}, \bibinfo {author}
  {\bibfnamefont {S.}~\bibnamefont {Higashikawa}}, \ and\ \bibinfo {author}
  {\bibfnamefont {M.}~\bibnamefont {Ueda}},\ }\bibfield  {title} {\enquote
  {\bibinfo {title} {Topological {P}hases of {N}on-{H}ermitian {S}ystems},}\
  }\href {\doibase 10.1103/PhysRevX.8.031079} {\bibfield  {journal} {\bibinfo
  {journal} {Phys. Rev. X}\ }\textbf {\bibinfo {volume} {8}},\ \bibinfo {pages}
  {031079} (\bibinfo {year} {2018})}\BibitemShut {NoStop}%
\bibitem [{\citenamefont {del Campo}\ \emph {et~al.}(2013)\citenamefont {del
  Campo}, \citenamefont {Egusquiza}, \citenamefont {Plenio},\ and\
  \citenamefont {Huelga}}]{PhysRevLett.110.050403}%
  \BibitemOpen
  \bibfield  {author} {\bibinfo {author} {\bibfnamefont {A.}~\bibnamefont {del
  Campo}}, \bibinfo {author} {\bibfnamefont {I.~L.}\ \bibnamefont {Egusquiza}},
  \bibinfo {author} {\bibfnamefont {M.~B.}\ \bibnamefont {Plenio}}, \ and\
  \bibinfo {author} {\bibfnamefont {S.~F.}\ \bibnamefont {Huelga}},\ }\bibfield
   {title} {\enquote {\bibinfo {title} {Quantum {S}peed {L}imits in {O}pen
  {S}ystem {D}y\-na\-mics},}\ }\href {\doibase 10.1103/PhysRevLett.110.050403}
  {\bibfield  {journal} {\bibinfo  {journal} {Phys. Rev. Lett.}\ }\textbf
  {\bibinfo {volume} {110}},\ \bibinfo {pages} {050403} (\bibinfo {year}
  {2013})}\BibitemShut {NoStop}%
\bibitem [{\citenamefont {Gessner}\ and\ \citenamefont
  {Smerzi}(2018)}]{PhysRevA.97.022109}%
  \BibitemOpen
  \bibfield  {author} {\bibinfo {author} {\bibfnamefont {M.}~\bibnamefont
  {Gessner}}\ and\ \bibinfo {author} {\bibfnamefont {A.}~\bibnamefont
  {Smerzi}},\ }\bibfield  {title} {\enquote {\bibinfo {title} {Statistical
  speed of quantum states: {G}eneralized quantum {F}isher information and
  {S}chatten speed},}\ }\href {\doibase 10.1103/PhysRevA.97.022109} {\bibfield
  {journal} {\bibinfo  {journal} {Phys. Rev. A}\ }\textbf {\bibinfo {volume}
  {97}},\ \bibinfo {pages} {022109} (\bibinfo {year} {2018})}\BibitemShut
  {NoStop}%
\bibitem [{\citenamefont {Impens}\ \emph {et~al.}(2021)\citenamefont {Impens},
  \citenamefont {D'Angelis}, \citenamefont {Pinheiro},\ and\ \citenamefont
  {Gu\'ery-Odelin}}]{PhysRevA.104.052620}%
  \BibitemOpen
  \bibfield  {author} {\bibinfo {author} {\bibfnamefont {F.}~\bibnamefont
  {Impens}}, \bibinfo {author} {\bibfnamefont {F.~M.}\ \bibnamefont
  {D'Angelis}}, \bibinfo {author} {\bibfnamefont {F.~A.}\ \bibnamefont
  {Pinheiro}}, \ and\ \bibinfo {author} {\bibfnamefont {D.}~\bibnamefont
  {Gu\'ery-Odelin}},\ }\bibfield  {title} {\enquote {\bibinfo {title} {Time
  scaling and quantum speed limit in non-{H}ermitian {H}amiltonians},}\ }\href
  {\doibase 10.1103/PhysRevA.104.052620} {\bibfield  {journal} {\bibinfo
  {journal} {Phys. Rev. A}\ }\textbf {\bibinfo {volume} {104}},\ \bibinfo
  {pages} {052620} (\bibinfo {year} {2021})}\BibitemShut {NoStop}%
\bibitem [{\citenamefont {Pyrialakos}\ \emph {et~al.}(2022)\citenamefont
  {Pyrialakos}, \citenamefont {Ren}, \citenamefont {Jung}, \citenamefont
  {Khajavikhan},\ and\ \citenamefont
  {Christodoulides}}]{PhysRevLett.128.213901}%
  \BibitemOpen
  \bibfield  {author} {\bibinfo {author} {\bibfnamefont {G.~G.}\ \bibnamefont
  {Pyrialakos}}, \bibinfo {author} {\bibfnamefont {H.}~\bibnamefont {Ren}},
  \bibinfo {author} {\bibfnamefont {P.~S.}\ \bibnamefont {Jung}}, \bibinfo
  {author} {\bibfnamefont {M.}~\bibnamefont {Khajavikhan}}, \ and\ \bibinfo
  {author} {\bibfnamefont {D.~N.}\ \bibnamefont {Christodoulides}},\ }\bibfield
   {title} {\enquote {\bibinfo {title} {Thermalization {D}ynamics of
  {N}onlinear non-{H}ermitian {O}ptical {L}attices},}\ }\href {\doibase
  10.1103/PhysRevLett.128.213901} {\bibfield  {journal} {\bibinfo  {journal}
  {Phys. Rev. Lett.}\ }\textbf {\bibinfo {volume} {128}},\ \bibinfo {pages}
  {213901} (\bibinfo {year} {2022})}\BibitemShut {NoStop}%
\bibitem [{\citenamefont {Chen}\ \emph
  {et~al.}(2022{\natexlab{c}})\citenamefont {Chen}, \citenamefont {Chen},\ and\
  \citenamefont {Zhu}}]{arXiv:2202.08638}%
  \BibitemOpen
  \bibfield  {author} {\bibinfo {author} {\bibfnamefont {Q.}~\bibnamefont
  {Chen}}, \bibinfo {author} {\bibfnamefont {S.~A.}\ \bibnamefont {Chen}}, \
  and\ \bibinfo {author} {\bibfnamefont {Z.}~\bibnamefont {Zhu}},\ }\bibfield
  {title} {\enquote {\bibinfo {title} {Weak {E}rgodicity {B}reaking in
  non-{H}ermitian {M}any-body {S}ystems},}\ }\href
  {https://arxiv.org/abs/2202.08638} {\bibfield  {journal} {\bibinfo  {journal}
  {arXiv:2202.08638}\ } (\bibinfo {year} {2022}{\natexlab{c}})}\BibitemShut
  {NoStop}%
\bibitem [{\citenamefont {Schlosshauer}(2007)}]{Schlosshauer_book}%
  \BibitemOpen
  \bibfield  {author} {\bibinfo {author} {\bibfnamefont {M.}~\bibnamefont
  {Schlosshauer}},\ }\href {\doibase 10.1007/978-3-540-35775-9} {\emph
  {\bibinfo {title} {Decoherence and the Quantum-to-Classical Transition}}}\
  (\bibinfo  {publisher} {Springer},\ \bibinfo {address} {Berlin},\ \bibinfo
  {year} {2007})\BibitemShut {NoStop}%
\bibitem [{\citenamefont {G\"{u}hne}\ and\ \citenamefont
  {T\'{o}th}(2009)}]{PhysRep474_1_2009}%
  \BibitemOpen
  \bibfield  {author} {\bibinfo {author} {\bibfnamefont {O.}~\bibnamefont
  {G\"{u}hne}}\ and\ \bibinfo {author} {\bibfnamefont {G.}~\bibnamefont
  {T\'{o}th}},\ }\bibfield  {title} {\enquote {\bibinfo {title} {Entanglement
  detection},}\ }\href {\doibase 10.1016/j.physrep.2009.02.004} {\bibfield
  {journal} {\bibinfo  {journal} {Phys. Rep.}\ }\textbf {\bibinfo {volume}
  {474}},\ \bibinfo {pages} {1} (\bibinfo {year} {2009})}\BibitemShut {NoStop}%
\bibitem [{\citenamefont {Horodecki}\ \emph {et~al.}(2009)\citenamefont
  {Horodecki}, \citenamefont {Horodecki}, \citenamefont {Horodecki},\ and\
  \citenamefont {Horodecki}}]{RevModPhys.81.865}%
  \BibitemOpen
  \bibfield  {author} {\bibinfo {author} {\bibfnamefont {R.}~\bibnamefont
  {Horodecki}}, \bibinfo {author} {\bibfnamefont {P.}~\bibnamefont
  {Horodecki}}, \bibinfo {author} {\bibfnamefont {M.}~\bibnamefont
  {Horodecki}}, \ and\ \bibinfo {author} {\bibfnamefont {K.}~\bibnamefont
  {Horodecki}},\ }\bibfield  {title} {\enquote {\bibinfo {title} {Quantum
  entanglement},}\ }\href {\doibase 10.1103/RevModPhys.81.865} {\bibfield
  {journal} {\bibinfo  {journal} {Rev. Mod. Phys.}\ }\textbf {\bibinfo {volume}
  {81}},\ \bibinfo {pages} {865} (\bibinfo {year} {2009})}\BibitemShut
  {NoStop}%
\bibitem [{\citenamefont {Jaeger}\ \emph {et~al.}(2003)\citenamefont {Jaeger},
  \citenamefont {Sergienko}, \citenamefont {Saleh},\ and\ \citenamefont
  {Teich}}]{PhysRevA.68.022318}%
  \BibitemOpen
  \bibfield  {author} {\bibinfo {author} {\bibfnamefont {G.}~\bibnamefont
  {Jaeger}}, \bibinfo {author} {\bibfnamefont {A.~V.}\ \bibnamefont
  {Sergienko}}, \bibinfo {author} {\bibfnamefont {B.~E.~A.}\ \bibnamefont
  {Saleh}}, \ and\ \bibinfo {author} {\bibfnamefont {M.~C.}\ \bibnamefont
  {Teich}},\ }\bibfield  {title} {\enquote {\bibinfo {title} {Entanglement,
  mixedness, and spin-flip symmetry in multiple-qubit systems},}\ }\href
  {\doibase 10.1103/PhysRevA.68.022318} {\bibfield  {journal} {\bibinfo
  {journal} {Phys. Rev. A}\ }\textbf {\bibinfo {volume} {68}},\ \bibinfo
  {pages} {022318} (\bibinfo {year} {2003})}\BibitemShut {NoStop}%
\bibitem [{\citenamefont {Olaya-Castro}\ \emph {et~al.}(2004)\citenamefont
  {Olaya-Castro}, \citenamefont {Johnson},\ and\ \citenamefont
  {Quiroga}}]{JOptB_6_S730_2004}%
  \BibitemOpen
  \bibfield  {author} {\bibinfo {author} {\bibfnamefont {A.}~\bibnamefont
  {Olaya-Castro}}, \bibinfo {author} {\bibfnamefont {N.~F.}\ \bibnamefont
  {Johnson}}, \ and\ \bibinfo {author} {\bibfnamefont {L.}~\bibnamefont
  {Quiroga}},\ }\bibfield  {title} {\enquote {\bibinfo {title} {Dynamics of
  quantum correlations and linear entropy in a multi-qubit-cavity system},}\
  }\href {\doibase 10.1088/1464-4266/6/8/016} {\bibfield  {journal} {\bibinfo
  {journal} {J. Opt. B: Quantum Semiclass. Opt.}\ }\textbf {\bibinfo {volume}
  {6}},\ \bibinfo {pages} {S730} (\bibinfo {year} {2004})}\BibitemShut
  {NoStop}%
\bibitem [{\citenamefont {Buscemi}\ \emph {et~al.}(2007)\citenamefont
  {Buscemi}, \citenamefont {Bordone},\ and\ \citenamefont
  {Bertoni}}]{PhysRevA.75.032301}%
  \BibitemOpen
  \bibfield  {author} {\bibinfo {author} {\bibfnamefont {F.}~\bibnamefont
  {Buscemi}}, \bibinfo {author} {\bibfnamefont {P.}~\bibnamefont {Bordone}}, \
  and\ \bibinfo {author} {\bibfnamefont {A.}~\bibnamefont {Bertoni}},\
  }\bibfield  {title} {\enquote {\bibinfo {title} {Linear entropy as an
  entanglement measure in two-fermion systems},}\ }\href {\doibase
  10.1103/PhysRevA.75.032301} {\bibfield  {journal} {\bibinfo  {journal} {Phys.
  Rev. A}\ }\textbf {\bibinfo {volume} {75}},\ \bibinfo {pages} {032301}
  (\bibinfo {year} {2007})}\BibitemShut {NoStop}%
\bibitem [{\citenamefont {Alves}\ and\ \citenamefont
  {Jaksch}(2004)}]{PhysRevLett.93.110501}%
  \BibitemOpen
  \bibfield  {author} {\bibinfo {author} {\bibfnamefont {C.~M.}\ \bibnamefont
  {Alves}}\ and\ \bibinfo {author} {\bibfnamefont {D.}~\bibnamefont {Jaksch}},\
  }\bibfield  {title} {\enquote {\bibinfo {title} {Multipartite {E}ntanglement
  {D}etection in {B}osons},}\ }\href {\doibase 10.1103/PhysRevLett.93.110501}
  {\bibfield  {journal} {\bibinfo  {journal} {Phys. Rev. Lett.}\ }\textbf
  {\bibinfo {volume} {93}},\ \bibinfo {pages} {110501} (\bibinfo {year}
  {2004})}\BibitemShut {NoStop}%
\bibitem [{\citenamefont {Horodecki}\ and\ \citenamefont
  {Horodecki}(1996)}]{PhysRevA.54.1838}%
  \BibitemOpen
  \bibfield  {author} {\bibinfo {author} {\bibfnamefont {R.}~\bibnamefont
  {Horodecki}}\ and\ \bibinfo {author} {\bibfnamefont {M.}~\bibnamefont
  {Horodecki}},\ }\bibfield  {title} {\enquote {\bibinfo {title}
  {Information-theoretic aspects of inseparability of mixed states},}\ }\href
  {\doibase 10.1103/PhysRevA.54.1838} {\bibfield  {journal} {\bibinfo
  {journal} {Phys. Rev. A}\ }\textbf {\bibinfo {volume} {54}},\ \bibinfo
  {pages} {1838} (\bibinfo {year} {1996})}\BibitemShut {NoStop}%
\bibitem [{\citenamefont {Vollbrecht}\ and\ \citenamefont
  {Wolf}(2002)}]{doi:10.1063/1.1498490}%
  \BibitemOpen
  \bibfield  {author} {\bibinfo {author} {\bibfnamefont {K.~G.~H.}\
  \bibnamefont {Vollbrecht}}\ and\ \bibinfo {author} {\bibfnamefont {M.~M.}\
  \bibnamefont {Wolf}},\ }\bibfield  {title} {\enquote {\bibinfo {title}
  {Conditional entropies and their relation to entanglement criteria},}\ }\href
  {\doibase 10.1063/1.1498490} {\bibfield  {journal} {\bibinfo  {journal} {J.
  Math. Phys.}\ }\textbf {\bibinfo {volume} {43}},\ \bibinfo {pages} {4299}
  (\bibinfo {year} {2002})}\BibitemShut {NoStop}%
\bibitem [{\citenamefont {Horodecki}\ \emph {et~al.}(1996)\citenamefont
  {Horodecki}, \citenamefont {Horodecki},\ and\ \citenamefont
  {Horodecki}}]{HORODECKI1996377}%
  \BibitemOpen
  \bibfield  {author} {\bibinfo {author} {\bibfnamefont {R.}~\bibnamefont
  {Horodecki}}, \bibinfo {author} {\bibfnamefont {P.}~\bibnamefont
  {Horodecki}}, \ and\ \bibinfo {author} {\bibfnamefont {M.}~\bibnamefont
  {Horodecki}},\ }\bibfield  {title} {\enquote {\bibinfo {title} {Quantum
  entropy inequalities: {I}ndependent condition for local realism?}}\ }\href
  {\doibase https://doi.org/10.1016/0375-9601(95)00930-2} {\bibfield  {journal}
  {\bibinfo  {journal} {Phys. Lett. A}\ }\textbf {\bibinfo {volume} {210}},\
  \bibinfo {pages} {377} (\bibinfo {year} {1996})}\BibitemShut {NoStop}%
\bibitem [{\citenamefont {Cerf}\ and\ \citenamefont
  {Adami}(1997)}]{PhysRevLett.79.5194}%
  \BibitemOpen
  \bibfield  {author} {\bibinfo {author} {\bibfnamefont {N.~J.}\ \bibnamefont
  {Cerf}}\ and\ \bibinfo {author} {\bibfnamefont {C.}~\bibnamefont {Adami}},\
  }\bibfield  {title} {\enquote {\bibinfo {title} {Negative {E}ntropy and
  {I}nformation in {Q}uantum {M}echanics},}\ }\href {\doibase
  10.1103/PhysRevLett.79.5194} {\bibfield  {journal} {\bibinfo  {journal}
  {Phys. Rev. Lett.}\ }\textbf {\bibinfo {volume} {79}},\ \bibinfo {pages}
  {5194} (\bibinfo {year} {1997})}\BibitemShut {NoStop}%
\bibitem [{\citenamefont {Abe}\ and\ \citenamefont
  {Rajagopal}(1999)}]{PhysRevA.60.3461}%
  \BibitemOpen
  \bibfield  {author} {\bibinfo {author} {\bibfnamefont {S.}~\bibnamefont
  {Abe}}\ and\ \bibinfo {author} {\bibfnamefont {A.~K.}\ \bibnamefont
  {Rajagopal}},\ }\bibfield  {title} {\enquote {\bibinfo {title} {Quantum
  entanglement inferred by the principle of maximum nonadditive entropy},}\
  }\href {\doibase 10.1103/PhysRevA.60.3461} {\bibfield  {journal} {\bibinfo
  {journal} {Phys. Rev. A}\ }\textbf {\bibinfo {volume} {60}},\ \bibinfo
  {pages} {3461} (\bibinfo {year} {1999})}\BibitemShut {NoStop}%
\bibitem [{\citenamefont {Vidiella-Barranco}(1999)}]{VIDIELLABARRANCO1999335}%
  \BibitemOpen
  \bibfield  {author} {\bibinfo {author} {\bibfnamefont {A.}~\bibnamefont
  {Vidiella-Barranco}},\ }\bibfield  {title} {\enquote {\bibinfo {title}
  {Entanglement and nonextensive statistics},}\ }\href {\doibase
  https://doi.org/10.1016/S0375-9601(99)00551-4} {\bibfield  {journal}
  {\bibinfo  {journal} {Phys. Lett. A}\ }\textbf {\bibinfo {volume} {260}},\
  \bibinfo {pages} {335} (\bibinfo {year} {1999})}\BibitemShut {NoStop}%
\bibitem [{\citenamefont {Abe}\ and\ \citenamefont
  {Rajagopal}(2001)}]{ABE2001157}%
  \BibitemOpen
  \bibfield  {author} {\bibinfo {author} {\bibfnamefont {S.}~\bibnamefont
  {Abe}}\ and\ \bibinfo {author} {\bibfnamefont {A.~K.}\ \bibnamefont
  {Rajagopal}},\ }\bibfield  {title} {\enquote {\bibinfo {title} {Nonadditive
  conditional entropy and its significance for local realism},}\ }\href
  {\doibase https://doi.org/10.1016/S0378-4371(00)00476-3} {\bibfield
  {journal} {\bibinfo  {journal} {Physica A: Stat. Mech. Appl.}\ }\textbf
  {\bibinfo {volume} {289}},\ \bibinfo {pages} {157} (\bibinfo {year}
  {2001})}\BibitemShut {NoStop}%
\bibitem [{\citenamefont {Tsallis}\ \emph {et~al.}(2001)\citenamefont
  {Tsallis}, \citenamefont {Lloyd},\ and\ \citenamefont
  {Baranger}}]{PhysRevA.63.042104}%
  \BibitemOpen
  \bibfield  {author} {\bibinfo {author} {\bibfnamefont {C.}~\bibnamefont
  {Tsallis}}, \bibinfo {author} {\bibfnamefont {S.}~\bibnamefont {Lloyd}}, \
  and\ \bibinfo {author} {\bibfnamefont {M.}~\bibnamefont {Baranger}},\
  }\bibfield  {title} {\enquote {\bibinfo {title} {Peres criterion for
  separability through nonextensive entropy},}\ }\href {\doibase
  10.1103/PhysRevA.63.042104} {\bibfield  {journal} {\bibinfo  {journal} {Phys.
  Rev. A}\ }\textbf {\bibinfo {volume} {63}},\ \bibinfo {pages} {042104}
  (\bibinfo {year} {2001})}\BibitemShut {NoStop}%
\bibitem [{\citenamefont {Terhal}(2002)}]{TERHAL2002313}%
  \BibitemOpen
  \bibfield  {author} {\bibinfo {author} {\bibfnamefont {B.~M.}\ \bibnamefont
  {Terhal}},\ }\bibfield  {title} {\enquote {\bibinfo {title} {Detecting
  quantum entanglement},}\ }\href {\doibase
  https://doi.org/10.1016/S0304-3975(02)00139-1} {\bibfield  {journal}
  {\bibinfo  {journal} {Theor. Comput. Sci.}\ }\textbf {\bibinfo {volume}
  {287}},\ \bibinfo {pages} {313} (\bibinfo {year} {2002})},\ \bibinfo {note}
  {natural Computing}\BibitemShut {NoStop}%
\bibitem [{\citenamefont {Peters}\ \emph {et~al.}(2004)\citenamefont {Peters},
  \citenamefont {Wei},\ and\ \citenamefont {Kwiat}}]{PhysRevA.70.052309}%
  \BibitemOpen
  \bibfield  {author} {\bibinfo {author} {\bibfnamefont {N.~A.}\ \bibnamefont
  {Peters}}, \bibinfo {author} {\bibfnamefont {T.-C.}\ \bibnamefont {Wei}}, \
  and\ \bibinfo {author} {\bibfnamefont {P.~G.}\ \bibnamefont {Kwiat}},\
  }\bibfield  {title} {\enquote {\bibinfo {title} {Mixed-state sensitivity of
  several quantum-information benchmarks},}\ }\href {\doibase
  10.1103/PhysRevA.70.052309} {\bibfield  {journal} {\bibinfo  {journal} {Phys.
  Rev. A}\ }\textbf {\bibinfo {volume} {70}},\ \bibinfo {pages} {052309}
  (\bibinfo {year} {2004})}\BibitemShut {NoStop}%
\bibitem [{\citenamefont {\ifmmode~\dot{Z}\else \.{Z}\fi{}yczkowski}\ \emph
  {et~al.}(1998)\citenamefont {\ifmmode~\dot{Z}\else \.{Z}\fi{}yczkowski},
  \citenamefont {Horodecki}, \citenamefont {Sanpera},\ and\ \citenamefont
  {Lewenstein}}]{PhysRevA.58.883}%
  \BibitemOpen
  \bibfield  {author} {\bibinfo {author} {\bibfnamefont {K.}~\bibnamefont
  {\ifmmode~\dot{Z}\else \.{Z}\fi{}yczkowski}}, \bibinfo {author}
  {\bibfnamefont {P.}~\bibnamefont {Horodecki}}, \bibinfo {author}
  {\bibfnamefont {A.}~\bibnamefont {Sanpera}}, \ and\ \bibinfo {author}
  {\bibfnamefont {M.}~\bibnamefont {Lewenstein}},\ }\bibfield  {title}
  {\enquote {\bibinfo {title} {Volume of the set of separable states},}\ }\href
  {\doibase 10.1103/PhysRevA.58.883} {\bibfield  {journal} {\bibinfo  {journal}
  {Phys. Rev. A}\ }\textbf {\bibinfo {volume} {58}},\ \bibinfo {pages} {883}
  (\bibinfo {year} {1998})}\BibitemShut {NoStop}%
\bibitem [{\citenamefont {R\'{e}nyi}(1961)}]{Renyi1961}%
  \BibitemOpen
  \bibfield  {author} {\bibinfo {author} {\bibfnamefont {A.}~\bibnamefont
  {R\'{e}nyi}},\ }\bibfield  {title} {\enquote {\bibinfo {title} {{O}n
  {M}easures of {E}ntropy and {I}nformation},}\ }in\ \href
  {https://projecteuclid.org/euclid.bsmsp/1200512181} {\emph {\bibinfo
  {booktitle} {Proceedings of the {F}ourth {B}erkeley {S}ymposium on
  {M}athematical {S}tatistics and Probability, {V}ol. 1: {C}ontributions to the
  {T}heory of {S}tatistics}}},\ \bibinfo {editor} {edited by\ \bibinfo {editor}
  {\bibfnamefont {J.}~\bibnamefont {Neyman}}}\ (\bibinfo  {publisher}
  {University of {C}a\-li\-for\-nia {P}ress},\ \bibinfo {address} {Berkeley},\
  \bibinfo {year} {1961})\ pp.\ \bibinfo {pages} {547--561}\BibitemShut
  {NoStop}%
\bibitem [{\citenamefont {M\"{u}ller-Lennert}\ \emph
  {et~al.}(2013)\citenamefont {M\"{u}ller-Lennert}, \citenamefont {Dupuis},
  \citenamefont {Szehr}, \citenamefont {Fehr},\ and\ \citenamefont
  {Tomamichel}}]{10.1063.1.4838856}%
  \BibitemOpen
  \bibfield  {author} {\bibinfo {author} {\bibfnamefont {M.}~\bibnamefont
  {M\"{u}ller-Lennert}}, \bibinfo {author} {\bibfnamefont {F.}~\bibnamefont
  {Dupuis}}, \bibinfo {author} {\bibfnamefont {O.}~\bibnamefont {Szehr}},
  \bibinfo {author} {\bibfnamefont {S.}~\bibnamefont {Fehr}}, \ and\ \bibinfo
  {author} {\bibfnamefont {M.}~\bibnamefont {Tomamichel}},\ }\bibfield  {title}
  {\enquote {\bibinfo {title} {On quantum {R}\'{e}nyi entropies: {A} new
  generalization and some properties},}\ }\href {\doibase 10.1063/1.4838856}
  {\bibfield  {journal} {\bibinfo  {journal} {J. Math. Phys.}\ }\textbf
  {\bibinfo {volume} {54}},\ \bibinfo {pages} {122203} (\bibinfo {year}
  {2013})}\BibitemShut {NoStop}%
\bibitem [{\citenamefont {Bosyk}\ \emph {et~al.}(2012)\citenamefont {Bosyk},
  \citenamefont {Portesi},\ and\ \citenamefont
  {Plastino}}]{PhysRevA.85.012108}%
  \BibitemOpen
  \bibfield  {author} {\bibinfo {author} {\bibfnamefont {G.~M.}\ \bibnamefont
  {Bosyk}}, \bibinfo {author} {\bibfnamefont {M.}~\bibnamefont {Portesi}}, \
  and\ \bibinfo {author} {\bibfnamefont {A.}~\bibnamefont {Plastino}},\
  }\bibfield  {title} {\enquote {\bibinfo {title} {Collision entropy and
  optimal uncertainty},}\ }\href {\doibase 10.1103/PhysRevA.85.012108}
  {\bibfield  {journal} {\bibinfo  {journal} {Phys. Rev. A}\ }\textbf {\bibinfo
  {volume} {85}},\ \bibinfo {pages} {012108} (\bibinfo {year}
  {2012})}\BibitemShut {NoStop}%
\bibitem [{\citenamefont {Tsallis}(1988)}]{Tsallis1988}%
  \BibitemOpen
  \bibfield  {author} {\bibinfo {author} {\bibfnamefont {C.}~\bibnamefont
  {Tsallis}},\ }\bibfield  {title} {\enquote {\bibinfo {title} {Possible
  generalization of {B}oltzmann-{G}ibbs statistics},}\ }\href {\doibase
  10.1007/BF01016429} {\bibfield  {journal} {\bibinfo  {journal} {J. Stat.
  Phys.}\ }\textbf {\bibinfo {volume} {52}},\ \bibinfo {pages} {479} (\bibinfo
  {year} {1988})}\BibitemShut {NoStop}%
\bibitem [{\citenamefont {Toth}(2017)}]{arXiv:1701.07461}%
  \BibitemOpen
  \bibfield  {author} {\bibinfo {author} {\bibfnamefont {G.}~\bibnamefont
  {Toth}},\ }\bibfield  {title} {\enquote {\bibinfo {title} {Lower bounds on
  the quantum {F}isher information based on the variance and various types of
  entropies},}\ }\href {https://arxiv.org/abs/1701.07461} {\bibfield  {journal}
  {\bibinfo  {journal} {arXiv:1701.07461}\ } (\bibinfo {year}
  {2017})}\BibitemShut {NoStop}%
\bibitem [{\citenamefont {Paris}(2009)}]{IntJQuantInf_125_7_2009}%
  \BibitemOpen
  \bibfield  {author} {\bibinfo {author} {\bibfnamefont {M.~G.~A.}\
  \bibnamefont {Paris}},\ }\bibfield  {title} {\enquote {\bibinfo {title}
  {Quantum estimation for quantum technology},}\ }\href {\doibase
  10.1142/S0219749909004839} {\bibfield  {journal} {\bibinfo  {journal} {Int.
  J. Quantum Inform.}\ }\textbf {\bibinfo {volume} {7}},\ \bibinfo {pages}
  {125} (\bibinfo {year} {2009})}\BibitemShut {NoStop}%
\bibitem [{\citenamefont {Sidhu}\ and\ \citenamefont
  {Kok}(2020)}]{10.1116_1.5119961}%
  \BibitemOpen
  \bibfield  {author} {\bibinfo {author} {\bibfnamefont {J.~S.}\ \bibnamefont
  {Sidhu}}\ and\ \bibinfo {author} {\bibfnamefont {P.}~\bibnamefont {Kok}},\
  }\bibfield  {title} {\enquote {\bibinfo {title} {Geometric perspective on
  quantum parameter estimation},}\ }\href {\doibase 10.1116/1.5119961}
  {\bibfield  {journal} {\bibinfo  {journal} {{A}{V}{S} {Q}uantum {S}ci.}\
  }\textbf {\bibinfo {volume} {2}},\ \bibinfo {pages} {014701} (\bibinfo {year}
  {2020})}\BibitemShut {NoStop}%
\bibitem [{\citenamefont {Sergi}\ and\ \citenamefont
  {Giaquinta}(2016)}]{Entropy_18120451}%
  \BibitemOpen
  \bibfield  {author} {\bibinfo {author} {\bibfnamefont {A.}~\bibnamefont
  {Sergi}}\ and\ \bibinfo {author} {\bibfnamefont {P.~V.}\ \bibnamefont
  {Giaquinta}},\ }\bibfield  {title} {\enquote {\bibinfo {title} {Linear
  {Q}uantum {E}ntropy and {N}on-{H}ermitian {H}amiltonians},}\ }\href {\doibase
  10.3390/e18120451} {\bibfield  {journal} {\bibinfo  {journal} {Entropy}\
  }\textbf {\bibinfo {volume} {18}},\ \bibinfo {pages} {451} (\bibinfo {year}
  {2016})}\BibitemShut {NoStop}%
\bibitem [{\citenamefont {Bender}(2007)}]{Bender2007}%
  \BibitemOpen
  \bibfield  {author} {\bibinfo {author} {\bibfnamefont {C.~M.}\ \bibnamefont
  {Bender}},\ }\bibfield  {title} {\enquote {\bibinfo {title} {Making sense of
  non-{H}ermitian {H}amiltonians},}\ }\href {\doibase
  10.1088/0034-4885/70/6/R03} {\bibfield  {journal} {\bibinfo  {journal} {Rep.
  Prog. Phys.}\ }\textbf {\bibinfo {volume} {70}},\ \bibinfo {pages} {947}
  (\bibinfo {year} {2007})}\BibitemShut {NoStop}%
\bibitem [{\citenamefont {Scolarici}\ and\ \citenamefont
  {Solombrino}(2009)}]{JPhysA_42_5_055303}%
  \BibitemOpen
  \bibfield  {author} {\bibinfo {author} {\bibfnamefont {G.}~\bibnamefont
  {Scolarici}}\ and\ \bibinfo {author} {\bibfnamefont {L.}~\bibnamefont
  {Solombrino}},\ }\bibfield  {title} {\enquote {\bibinfo {title} {Alternative
  descriptions and bipartite compound quantum systems},}\ }\href {\doibase
  10.1088/1751-8113/42/5/055303} {\bibfield  {journal} {\bibinfo  {journal} {J.
  Phys. A: Math. Theor.}\ }\textbf {\bibinfo {volume} {42}},\ \bibinfo {pages}
  {055303} (\bibinfo {year} {2009})}\BibitemShut {NoStop}%
\bibitem [{\citenamefont {Mostafazadeh}(2010)}]{PhysScr_82_038110}%
  \BibitemOpen
  \bibfield  {author} {\bibinfo {author} {\bibfnamefont {A.}~\bibnamefont
  {Mostafazadeh}},\ }\bibfield  {title} {\enquote {\bibinfo {title} {Conceptual
  aspects of $\mathcal{PT}$-symmetry and pseudo-{H}ermiticity: a status
  report},}\ }\href {\doibase 10.1088/0031-8949/82/03/038110} {\bibfield
  {journal} {\bibinfo  {journal} {Phys. Scr.}\ }\textbf {\bibinfo {volume}
  {82}},\ \bibinfo {pages} {038110} (\bibinfo {year} {2010})}\BibitemShut
  {NoStop}%
\bibitem [{\citenamefont {Dattoli}\ \emph {et~al.}(1990)\citenamefont
  {Dattoli}, \citenamefont {Torre},\ and\ \citenamefont
  {Mignani}}]{PhysRevA.42.1467}%
  \BibitemOpen
  \bibfield  {author} {\bibinfo {author} {\bibfnamefont {G.}~\bibnamefont
  {Dattoli}}, \bibinfo {author} {\bibfnamefont {A.}~\bibnamefont {Torre}}, \
  and\ \bibinfo {author} {\bibfnamefont {R.}~\bibnamefont {Mignani}},\
  }\bibfield  {title} {\enquote {\bibinfo {title} {Non-{H}ermitian evolution of
  two-level quantum systems},}\ }\href {\doibase 10.1103/PhysRevA.42.1467}
  {\bibfield  {journal} {\bibinfo  {journal} {Phys. Rev. A}\ }\textbf {\bibinfo
  {volume} {42}},\ \bibinfo {pages} {1467} (\bibinfo {year}
  {1990})}\BibitemShut {NoStop}%
\bibitem [{\citenamefont {Sergi}\ and\ \citenamefont
  {Zloshchastiev}(2013)}]{doi:10.1142}%
  \BibitemOpen
  \bibfield  {author} {\bibinfo {author} {\bibfnamefont {A.}~\bibnamefont
  {Sergi}}\ and\ \bibinfo {author} {\bibfnamefont {K.~G.}\ \bibnamefont
  {Zloshchastiev}},\ }\bibfield  {title} {\enquote {\bibinfo {title}
  {Non-{H}ermitian quantum dynamics of a two-level system and models of
  dissipative environments},}\ }\href {\doibase 10.1142/S0217979213501634}
  {\bibfield  {journal} {\bibinfo  {journal} {Int. J. Mod. Phys. B}\ }\textbf
  {\bibinfo {volume} {27}},\ \bibinfo {pages} {1350163} (\bibinfo {year}
  {2013})}\BibitemShut {NoStop}%
\bibitem [{\citenamefont {Zloshchastiev}(2015)}]{EPJD_253_69_2015}%
  \BibitemOpen
  \bibfield  {author} {\bibinfo {author} {\bibfnamefont {K.~G.}\ \bibnamefont
  {Zloshchastiev}},\ }\bibfield  {title} {\enquote {\bibinfo {title}
  {Non-{H}ermitian {H}amiltonians and stability of pure states},}\ }\href
  {\doibase 10.1140/epjd/e2015-60384-0} {\bibfield  {journal} {\bibinfo
  {journal} {Eur. Phys. J. D}\ }\textbf {\bibinfo {volume} {69}},\ \bibinfo
  {pages} {253} (\bibinfo {year} {2015})}\BibitemShut {NoStop}%
\bibitem [{\citenamefont {Breuer}\ and\ \citenamefont
  {Petruccione}(2002)}]{Breuer_Petruccione_book}%
  \BibitemOpen
  \bibfield  {author} {\bibinfo {author} {\bibfnamefont {H.-P.}\ \bibnamefont
  {Breuer}}\ and\ \bibinfo {author} {\bibfnamefont {F.}~\bibnamefont
  {Petruccione}},\ }\href {\doibase 10.1093/acprof:oso/9780199213900.001.0001}
  {\emph {\bibinfo {title} {The {T}heory of {O}pen {Q}uantum {S}ystems}}}\
  (\bibinfo  {publisher} {Oxford University Press},\ \bibinfo {address} {New
  York},\ \bibinfo {year} {2002})\BibitemShut {NoStop}%
\bibitem [{\citenamefont {Lieb}\ \emph {et~al.}(1961)\citenamefont {Lieb},
  \citenamefont {Schultz},\ and\ \citenamefont {Mattis}}]{LIEB1961407}%
  \BibitemOpen
  \bibfield  {author} {\bibinfo {author} {\bibfnamefont {E.}~\bibnamefont
  {Lieb}}, \bibinfo {author} {\bibfnamefont {T.}~\bibnamefont {Schultz}}, \
  and\ \bibinfo {author} {\bibfnamefont {D.}~\bibnamefont {Mattis}},\
  }\bibfield  {title} {\enquote {\bibinfo {title} {Two soluble models of an
  antiferromagnetic chain},}\ }\href {\doibase 10.1016/0003-4916(61)90115-4}
  {\bibfield  {journal} {\bibinfo  {journal} {Ann. Phys.}\ }\textbf {\bibinfo
  {volume} {16}},\ \bibinfo {pages} {407} (\bibinfo {year} {1961})}\BibitemShut
  {NoStop}%
\bibitem [{\citenamefont {Barouch}\ \emph {et~al.}(1970)\citenamefont
  {Barouch}, \citenamefont {McCoy},\ and\ \citenamefont
  {Dresden}}]{PhysRevA.2.1075}%
  \BibitemOpen
  \bibfield  {author} {\bibinfo {author} {\bibfnamefont {E.}~\bibnamefont
  {Barouch}}, \bibinfo {author} {\bibfnamefont {B.~M.}\ \bibnamefont {McCoy}},
  \ and\ \bibinfo {author} {\bibfnamefont {M.}~\bibnamefont {Dresden}},\
  }\bibfield  {title} {\enquote {\bibinfo {title} {Statistical {M}echanics of
  the $\mathrm{XY}$ {M}odel. {I}},}\ }\href {\doibase 10.1103/PhysRevA.2.1075}
  {\bibfield  {journal} {\bibinfo  {journal} {Phys. Rev. A}\ }\textbf {\bibinfo
  {volume} {2}},\ \bibinfo {pages} {1075} (\bibinfo {year} {1970})}\BibitemShut
  {NoStop}%
\bibitem [{\citenamefont {Barouch}\ and\ \citenamefont
  {McCoy}(1971{\natexlab{a}})}]{PhysRevA.3.786}%
  \BibitemOpen
  \bibfield  {author} {\bibinfo {author} {\bibfnamefont {E.}~\bibnamefont
  {Barouch}}\ and\ \bibinfo {author} {\bibfnamefont {B.~M.}\ \bibnamefont
  {McCoy}},\ }\bibfield  {title} {\enquote {\bibinfo {title} {Statistical
  {M}echanics of the $\mathrm{XY}$ model. {II}. {S}pin-{C}orrelation
  {F}unctions},}\ }\href {\doibase 10.1103/PhysRevA.3.786} {\bibfield
  {journal} {\bibinfo  {journal} {Phys. Rev. A}\ }\textbf {\bibinfo {volume}
  {3}},\ \bibinfo {pages} {786} (\bibinfo {year}
  {1971}{\natexlab{a}})}\BibitemShut {NoStop}%
\bibitem [{\citenamefont {Barouch}\ and\ \citenamefont
  {McCoy}(1971{\natexlab{b}})}]{PhysRevA.3.2137}%
  \BibitemOpen
  \bibfield  {author} {\bibinfo {author} {\bibfnamefont {E.}~\bibnamefont
  {Barouch}}\ and\ \bibinfo {author} {\bibfnamefont {B.~M.}\ \bibnamefont
  {McCoy}},\ }\bibfield  {title} {\enquote {\bibinfo {title} {Statistical
  {M}echanics of the $\mathrm{XY}$ {M}odel. {III}},}\ }\href {\doibase
  10.1103/PhysRevA.3.2137} {\bibfield  {journal} {\bibinfo  {journal} {Phys.
  Rev. A}\ }\textbf {\bibinfo {volume} {3}},\ \bibinfo {pages} {2137} (\bibinfo
  {year} {1971}{\natexlab{b}})}\BibitemShut {NoStop}%
\bibitem [{\citenamefont {McCoy}\ \emph {et~al.}(1971)\citenamefont {McCoy},
  \citenamefont {Barouch},\ and\ \citenamefont {Abraham}}]{PhysRevA.4.2331}%
  \BibitemOpen
  \bibfield  {author} {\bibinfo {author} {\bibfnamefont {B.~M.}\ \bibnamefont
  {McCoy}}, \bibinfo {author} {\bibfnamefont {E.}~\bibnamefont {Barouch}}, \
  and\ \bibinfo {author} {\bibfnamefont {D.~B.}\ \bibnamefont {Abraham}},\
  }\bibfield  {title} {\enquote {\bibinfo {title} {Statistical {M}echanics of
  the $\mathrm{XY}$ {M}odel. {IV}. {T}ime-{D}ependent {S}pin-{C}orrelation
  {F}unctions},}\ }\href {\doibase 10.1103/PhysRevA.4.2331} {\bibfield
  {journal} {\bibinfo  {journal} {Phys. Rev. A}\ }\textbf {\bibinfo {volume}
  {4}},\ \bibinfo {pages} {2331} (\bibinfo {year} {1971})}\BibitemShut
  {NoStop}%
\bibitem [{\citenamefont {Pires}(2022)}]{PhysRevA.106.012403}%
  \BibitemOpen
  \bibfield  {author} {\bibinfo {author} {\bibfnamefont {D.~P.}\ \bibnamefont
  {Pires}},\ }\bibfield  {title} {\enquote {\bibinfo {title} {Unified entropies
  and quantum speed limits for nonunitary dynamics},}\ }\href {\doibase
  10.1103/PhysRevA.106.012403} {\bibfield  {journal} {\bibinfo  {journal}
  {Phys. Rev. A}\ }\textbf {\bibinfo {volume} {106}},\ \bibinfo {pages}
  {012403} (\bibinfo {year} {2022})}\BibitemShut {NoStop}%
\bibitem [{\citenamefont {Pires}\ and\ \citenamefont
  {de~Oliveira}(2021)}]{PhysRevA.104.052223}%
  \BibitemOpen
  \bibfield  {author} {\bibinfo {author} {\bibfnamefont {D.~P.}\ \bibnamefont
  {Pires}}\ and\ \bibinfo {author} {\bibfnamefont {T.~R.}\ \bibnamefont
  {de~Oliveira}},\ }\bibfield  {title} {\enquote {\bibinfo {title} {Relative
  purity, speed of fluctuations, and bounds on equilibration times},}\ }\href
  {\doibase 10.1103/PhysRevA.104.052223} {\bibfield  {journal} {\bibinfo
  {journal} {Phys. Rev. A}\ }\textbf {\bibinfo {volume} {104}},\ \bibinfo
  {pages} {052223} (\bibinfo {year} {2021})}\BibitemShut {NoStop}%
\bibitem [{\citenamefont {Lau}\ and\ \citenamefont {Clerk}(2018)}]{Lau2018}%
  \BibitemOpen
  \bibfield  {author} {\bibinfo {author} {\bibfnamefont {H.-K.}\ \bibnamefont
  {Lau}}\ and\ \bibinfo {author} {\bibfnamefont {A.~A.}\ \bibnamefont
  {Clerk}},\ }\bibfield  {title} {\enquote {\bibinfo {title} {Fundamental
  limits and non-reciprocal approaches in non-{H}ermitian quantum sensing},}\
  }\href {\doibase 10.1038/s41467-018-06477-7} {\bibfield  {journal} {\bibinfo
  {journal} {Nat. Commun.}\ }\textbf {\bibinfo {volume} {9}},\ \bibinfo {pages}
  {4320} (\bibinfo {year} {2018})}\BibitemShut {NoStop}%
\bibitem [{\citenamefont {Bao}\ \emph {et~al.}(2021)\citenamefont {Bao},
  \citenamefont {Qi}, \citenamefont {Dong},\ and\ \citenamefont
  {Nori}}]{PhysRevA.103.042418}%
  \BibitemOpen
  \bibfield  {author} {\bibinfo {author} {\bibfnamefont {L.}~\bibnamefont
  {Bao}}, \bibinfo {author} {\bibfnamefont {B.}~\bibnamefont {Qi}}, \bibinfo
  {author} {\bibfnamefont {D.}~\bibnamefont {Dong}}, \ and\ \bibinfo {author}
  {\bibfnamefont {F.}~\bibnamefont {Nori}},\ }\bibfield  {title} {\enquote
  {\bibinfo {title} {Fundamental limits for reciprocal and nonreciprocal
  non-{H}ermitian quantum sensing},}\ }\href {\doibase
  10.1103/PhysRevA.103.042418} {\bibfield  {journal} {\bibinfo  {journal}
  {Phys. Rev. A}\ }\textbf {\bibinfo {volume} {103}},\ \bibinfo {pages}
  {042418} (\bibinfo {year} {2021})}\BibitemShut {NoStop}%
\bibitem [{\citenamefont {Gardas}\ \emph {et~al.}(2016)\citenamefont {Gardas},
  \citenamefont {Deffner},\ and\ \citenamefont
  {Saxena}}]{deffner_scirep_23408_2016}%
  \BibitemOpen
  \bibfield  {author} {\bibinfo {author} {\bibfnamefont {B.}~\bibnamefont
  {Gardas}}, \bibinfo {author} {\bibfnamefont {S.}~\bibnamefont {Deffner}}, \
  and\ \bibinfo {author} {\bibfnamefont {A.}~\bibnamefont {Saxena}},\
  }\bibfield  {title} {\enquote {\bibinfo {title} {Non-{H}ermitian quantum
  thermodynamics},}\ }\href {\doibase 10.1038/srep23408} {\bibfield  {journal}
  {\bibinfo  {journal} {Sci. Rep.}\ }\textbf {\bibinfo {volume} {6}},\ \bibinfo
  {pages} {23408} (\bibinfo {year} {2016})}\BibitemShut {NoStop}%
\bibitem [{\citenamefont {Defenu}\ \emph {et~al.}(2021)\citenamefont {Defenu},
  \citenamefont {Donner}, \citenamefont {Macr\`{i}}, \citenamefont {Pagano},
  \citenamefont {Ruffo},\ and\ \citenamefont {Trombettoni}}]{arXiv:2109.01063}%
  \BibitemOpen
  \bibfield  {author} {\bibinfo {author} {\bibfnamefont {N.}~\bibnamefont
  {Defenu}}, \bibinfo {author} {\bibfnamefont {T.}~\bibnamefont {Donner}},
  \bibinfo {author} {\bibfnamefont {T.}~\bibnamefont {Macr\`{i}}}, \bibinfo
  {author} {\bibfnamefont {G.}~\bibnamefont {Pagano}}, \bibinfo {author}
  {\bibfnamefont {S.}~\bibnamefont {Ruffo}}, \ and\ \bibinfo {author}
  {\bibfnamefont {A.}~\bibnamefont {Trombettoni}},\ }\bibfield  {title}
  {\enquote {\bibinfo {title} {Long-range interacting quantum systems},}\
  }\href {https://arxiv.org/abs/2109.01063} {\bibfield  {journal} {\bibinfo
  {journal} {arXiv:2109.01063}\ } (\bibinfo {year} {2021})}\BibitemShut
  {NoStop}%
\bibitem [{\citenamefont {Bender}(2020)}]{Bender_2020}%
  \BibitemOpen
  \bibfield  {author} {\bibinfo {author} {\bibfnamefont {C.~M.}\ \bibnamefont
  {Bender}},\ }\bibfield  {title} {\enquote {\bibinfo {title} {{PT}-symmetric
  quantum field theory},}\ }\href {\doibase 10.1088/1742-6596/1586/1/012004}
  {\bibfield  {journal} {\bibinfo  {journal} {J. Phys.: Conf. Ser.}\ }\textbf
  {\bibinfo {volume} {1586}},\ \bibinfo {pages} {012004} (\bibinfo {year}
  {2020})}\BibitemShut {NoStop}%
\bibitem [{\citenamefont {Brody}(2013)}]{Brody_2013}%
  \BibitemOpen
  \bibfield  {author} {\bibinfo {author} {\bibfnamefont {D.~C.}\ \bibnamefont
  {Brody}},\ }\bibfield  {title} {\enquote {\bibinfo {title} {Biorthogonal
  quantum mechanics},}\ }\href {\doibase 10.1088/1751-8113/47/3/035305}
  {\bibfield  {journal} {\bibinfo  {journal} {J. Phys. A: Math. Theor.}\
  }\textbf {\bibinfo {volume} {47}},\ \bibinfo {pages} {035305} (\bibinfo
  {year} {2013})}\BibitemShut {NoStop}%
\end{thebibliography}

%

\end{document}